\documentclass[11pt,a4paper,twoside]{article} 
\usepackage{fmrep}
\usepackage{amsmath,amssymb,amsfonts}
\usepackage{graphicx}
\usepackage{algpseudocode}
\usepackage{rotating}
\usepackage{multirow}
\usepackage{multicol}
\usepackage{booktabs}
\usepackage{color}
\usepackage{subfig}
\usepackage{listings}
\usepackage{float}
\usepackage{enumitem}
\definecolor{bluep}{rgb}{0.2, 0.2, 0.6}
\definecolor{fgreen}{rgb}{0.0, 0.27, 0.13}
\definecolor{maroon}{rgb}{0.76, 0.13, 0.28}
\usepackage[ruled,vlined]{algorithm2e}
\usepackage{placeins}

\usepackage[dvipsnames]{xcolor}
\usepackage{pagecolor}
\usepackage{afterpage}
\usepackage[dvipsnames]{xcolor}

\definecolor{mygreen}{rgb}{0,0.6,0}
\definecolor{mygray}{rgb}{0.5,0.5,0.5}
\definecolor{mymauve}{rgb}{0.58,0,0.82}
\definecolor{numbercolor}{HTML}{A1A60F}
\newcommand\myemptypage{
    \null
    \thispagestyle{empty}
    \addtocounter{page}{-1}
    \newpage
}
\renewcommand{\theequation}{\arabic{equation}}
\FLD{01}{2024}
\title{GPU Accelerated Implicit Kinetic Meshfree Method based on Modified LU-SGS }
\shorttitle{GPU Accelerated Implicit LSKUM Solver  }
\author{Mayuri Verma and Anil Nemili\affil{ Department of Mathematics\\
BITS Pilani - Hyderabad Campus \\ Hyderabad 500078, India \\{\em Email:} {\tt anil@hyderabad.bits-pilani.ac.in }}
Nischay Ram Mamidi\affil{Rutgers, The State University of New Jersey, New Jersey, 08102, USA. \\
{\em Email:} {\tt nischay.mamidi@rutgers.edu }}
}
\shortauthor{Mayuri, Nischay, and Anil }
\date{Feb 29, 2024 \\
Revised Feb 29, 2024}

\begin{document} 

\begin{titlepage}
\definecolor{titlepage-color}{HTML}{fdeed4}
\newpagecolor{titlepage-color}\afterpage{\restorepagecolor}
\newcommand{\colorRule}[3][black]{\textcolor[HTML]{#1}{\rule{#2}{#3}}}
\begin{flushleft}
\noindent
\\[-1em]
\color[HTML]{000000}
\makebox[0pt][l]{\colorRule[000000]{1.3\textwidth}{2pt}}
\par
\noindent

{
  \vfill
  \noindent {\fontfamily{SourceSansPro-LF}\fontsize{16}{15}\selectfont \textbf{GPU Accelerated Implicit Kinetic Meshfree Method based on Modified LU-SGS}}
    \vskip 2em
  {\fontfamily{SourceSansPro-LF}\fontsize{14}{16}\selectfont \textbf{Scientific Computing Report - 01:2024}}
    \vskip 2em
   \noindent {\fontfamily{SourceSansPro-LF}\fontsize{12}{15}\selectfont \textbf{Mayuri Verma, Nischay Ram Mamidi and Anil Nemili}}
  \vfill
   \includegraphics[height=1.8cm]{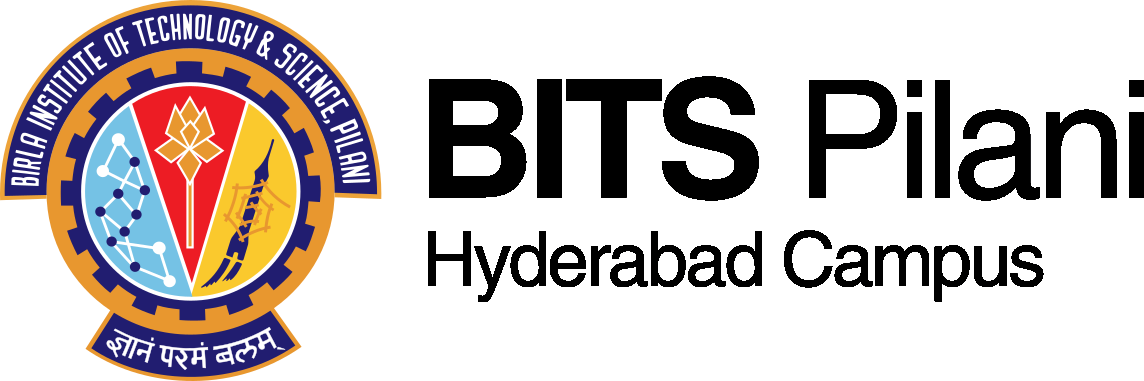}
}
%
%
\end{flushleft}
\end{titlepage}
\myemptypage
\maketitle
\begin{abstract}
This report presents the GPU acceleration of implicit kinetic meshfree methods using modified LU-SGS algorithms. The meshfree scheme is based on the least squares kinetic upwind method (LSKUM). In the existing matrix-free LU-SGS approaches for kinetic meshfree methods, the products of split flux Jacobians and increments in conserved vectors are approximated by increments in the split fluxes. In our modified LU-SGS approach, the Jacobian vector products are computed exactly using algorithmic differentiation (AD). The implicit GPU solvers with exact and approximate computation of the Jacobian vector products are applied to the standard test cases for two-dimensional inviscid flows. Numerical results have shown that the GPU solvers with the exact computation of the Jacobian vector products are computationally more efficient and yield better convergence rates than the solvers with approximations to the Jacobian vector products. Benchmarks are presented to assess the performance of implicit GPU solvers compared to the explicit GPU solver and the implicit serial LSKUM solver. 
\end{abstract}
\section{Introduction}
Numerical simulation of flow fields around complex geometries of specific interest in aerodynamics is computationally intensive. Traditionally, computational fluid dynamics (CFD) researchers used CPU parallel codes for such simulations \cite{openfoam,su2}. However, with significant technological advancements, GPUs have evolved from computer graphics application devices to truly general-purpose parallel processors \cite{gpus-general-purpose-processors}. With superior floating-point operation rates than multicore CPUs, GPUs are more suited for highly parallel numerical computations. In recent years, many research groups have developed finite volume or finite element based GPU parallel solvers, and some of them applied their codes to large-scale applications in CFD  \cite{pyfr,srikanth-bernardini-gpus}. \\ \\
In contrast to the mesh-based methods, the meshfree schemes \cite{deshpande-meshfree-lskum,batina-meshfree,meshfree-lohner,meshfree-morinishi,meshfree-sridar,meshfree-katz,meshfree-Chiu} provide an alternative approach for the numerical solution of Euler or Navier Stokes equations that govern the fluid flow. These schemes operate on a distribution of points, known as a point cloud. The point cloud can be obtained from structured, unstructured, chimera grids or quadtree algorithms. A meshfree scheme of particular interest to us is the Least Squares Kinetic Upwind Method (LSKUM) \cite{deshpande-meshfree-lskum,lskum-ghosh-aiaa,lskum-ghosh-journal}. Over the past two decades, this method has been successfully used for computing flows around realistic configurations, moving boundaries, flutter prediction, and other applications \cite{qlskum,meshfree-unsteady-Ramesh,mahendra-lskum-2011,anandh-lskum-sepdynamic-jaircraft,anandh-lskum-viscous-aiaa-journal,qlskum-flutter}.\\ \\
It is well-known that the explicit solvers for large-scale problems require more iterations to converge the numerical solutions. A convergence acceleration technique that is widely used to reduce the computational time is the lower upper symmetric Gauss-Seidel (LU-SGS) algorithm \cite{lusgs-jameson}. This algorithm requires the evaluation of split flux Jacobians in the implicit operator, which is computationally expensive. To mitigate these costs, a matrix-free version of the LU-SGS scheme has been developed \cite{lusgs-Sharov-1998}. In this approach, the products of split flux Jacobians with incremental conserved vectors are approximated by incremental changes in the split flux vectors. Note that these Jacobian vector products are part of the off-diagonal terms of the implicit operator. Furthermore, the split flux Jacobians that appear in the diagonal matrix of the implicit operator are approximated by a function of unsplit flux Jacobians and their spectral radii \cite{lusgs-jameson}. Typically, in finite difference and finite volume methods, this approximation transforms the block diagonal matrix into a scalar diagonal matrix. However, when implemented in the meshfree LSKUM, it may not lead to a scalar diagonal matrix due to the non-symmetric nature of the split stencils \cite{lusgs-anandh-icas-2004}. In order to extend the LU-SGS algorithm to kinetic meshfree methods, Anandhanarayanan et al. \cite{lusgs-anandh-icas-2004,anandh-lskum-viscous-aiaa-journal} developed a modified approximation for the split flux Jacobians. In another work, Manish et al. \cite{lusgs-NAL-Manish-2015} used the original approximation in \cite{lusgs-jameson} but moved the unsplit Jacobians that result in the block diagonal matrix to the explicit operator. \\ \\ 
This research presents the modified versions of meshfree LU-SGS schemes proposed by Anandhanarayanan et al. and Manish et al. In the modified LU-SGS, instead of approximating the Jacobian vector products, they are computed exactly using algorithmic differentiation (AD) techniques \cite{ad-book-andreas}. In the numerical results, we show that the LSKUM solvers with modified LU-SGS approaches yield a faster residual convergence rate than the solvers with approximate Jacobian vector products. \\ \\
In the works mentioned earlier on kinetic meshfree schemes, numerical simulations were performed using explicit or implicit CPU parallel solvers. Although an effort has been made to develop explicit kinetic meshfree GPU solvers \cite{hipc-2022-nischay}, to our knowledge, research on implicit kinetic meshfree solvers on GPUs is yet to be pursued. This research presents the development of implicit LSKUM solvers on GPUs using original and modified LU-SGS approaches. The programming model {\tt CUDA} is used to develop the GPU solvers. Benchmark simulations are performed to show that the implicit GPU solvers with AD are computationally more efficient than the solvers without AD. \\ \\
%
%
In the numerical simulation of fluid flows, the changes in the flow field occur at the wall boundary points first and then propagate inside the whole computational domain. Therefore, the wall points are updated first in the forward and backward sweeps of LU-SGS, followed by interior and outer boundary points. We analyse the effect of such an implementation on the convergence and run-time of an implicit LSKUM GPU solver and compare it with looping over all points in the cloud. \\ \\
%
%
This report is organised as follows. Section \ref{sec-lskum} presents the basic theory of the meshfree LSKUM. Section \ref{sec-implicit-lusgs-lskum} presents implicit schemes for LSKUM based on modified LU-SGS algorithms. Details on developing a GPU accelerated implicit LSKUM solver are given in Section \ref{section-gpu-accelerated-solver}. In Section \ref{sec-num-results}, residual convergence histories are shown for the standard test cases for $2D$ inviscid flow. Section \ref{sec-performance-analysis} presents benchmarks to assess the computational efficiency of the implicit GPU solvers. Finally, conclusions are drawn in Section \ref{sec-conclusions}.
\section{Basic Theory of LSKUM }
\label{sec-lskum}
In this section, we present the basic theory of the Least Squares Kinetic Upwind Method (LSKUM) with reference to $2D$ Euler equations, 
\begin{equation}
\frac{\partial \boldsymbol{U}}{\partial t} + \frac{\partial \boldsymbol{Gx}}{\partial x} + \frac{\partial \boldsymbol{Gy}}{\partial y}  = 0 
\label{ee-conservation-laws}
\end{equation}
where $\boldsymbol{U}$ is the conserved vector, $\boldsymbol{Gx}$ and $\boldsymbol{Gy}$ are the flux vectors along the coordinates $x$ and $y$, respectively. These vectors are defined by 
\begin{equation}
    \boldsymbol{U} = \begin{bmatrix}
        \rho \\ \rho u_1 \\ \rho u_2 \\ \rho e
    \end{bmatrix}, \; \; \boldsymbol{Gx} = \begin{bmatrix}
        \rho u_1 \\ p + \rho u_1^2 \\ \rho u_1 u_2 \\ \left( p + \rho e \right) u_1 
    \end{bmatrix}, \; \; \; \; \boldsymbol{Gy} = \begin{bmatrix}
        \rho u_2 \\ \rho u_1 u_2 \\ p + \rho u_2^2 \\ \left( p + \rho e \right) u_2 
    \end{bmatrix}
\end{equation}
Here, $\rho$ is the fluid density, $p$ is the pressure, $e$ is the internal energy per unit mass, $u_1$ and $u_2$ are the fluid velocities along $x$ and $y$ coordinate directions, respectively. The above conservation laws can be obtained as the moments of the $2D$ Boltzmann equation with the velocity distribution function being Maxwellian, 
\begin{equation}
\left< \boldsymbol{\Psi}, \frac{\partial F}{\partial t} + v_1 \frac{\partial F}{\partial x} + v_2 \frac{\partial F}{\partial y} \right> = 0, \; \; \left< \boldsymbol{\Psi},v_iF \right> = \int\limits_{\mathcal{R^+} \times \mathcal{R}^2} v_i\boldsymbol{\Psi} F dv_1 dv_2 dI, \; i = 1,2
\label{ee-be-me}
\end{equation}
where, $F$ is the Maxwellian velocity distribution function and $\boldsymbol{\Psi}$ is the moment function vector, defined by
\begin{equation}
    F = \frac{\rho}{I_0}\frac{\beta}{\pi} e^{-\beta\left\lbrace \left(v_1 - u_1 \right)^2 + \left(v_2 - u_2 \right)^2 \right\rbrace - \frac{I}{I_0}},\; \; \; \boldsymbol{\Psi} = \left[1, v_1, v_2, I + \frac{v_1^2+v_2^2}{2} \right] 
\end{equation}
Here, $v_1$ and $v_2$ are the molecular velocities along the coordinates $x$ and $y$, respectively. $\beta = \rho/{2p}$, $I$ is the internal energy variable, and $I_0$ is the internal energy due to non-translational degrees of freedom. \\ 
Using the Courant-Issacson-Rees (CIR) splitting \cite{cir-splitting} of the molecular velocities, an upwind scheme for the $2D$ Boltzmann equation in eq. (\ref{ee-be-me}) is given by 
\begin{equation}
    \frac{\partial F}{\partial t} + \frac{v_1 + \left| v_1 \right| }{2} \frac{\partial F}{\partial x} + \frac{v_1 - \left| v_1 \right| }{2} \frac{\partial F}{\partial x} + \frac{v_2 + \left| v_2 \right| }{2} \frac{\partial F}{\partial x} \frac{v_2 - \left| v_2 \right| }{2} \frac{\partial F}{\partial x} = 0
\label{cir-be}    
\end{equation}
Taking $\boldsymbol{\Psi}$-moments of the above equation results in the upwind kinetic flux vector split (KFVS) Euler equations \cite{kfvs},
\begin{equation}
\frac{\partial \boldsymbol{U}}{\partial t} + \frac{\partial \boldsymbol{Gx}^+}{\partial x } + \frac{\partial \boldsymbol{Gx}^-}{\partial x } + \frac{\partial \boldsymbol{Gy}^+}{\partial y } + \frac{\partial \boldsymbol{Gy}^-}{\partial y }  = 0 
\label{kfvs-ee}
\end{equation}
In order to find a meshfree upwind scheme for the above equation, the spatial derivatives in eq. (\ref{cir-be}) are first discretised using the least-squares (LS) principle. Suitable moments of the LS formulae with appropriate stencils lead to the least squares kinetic upwind method (LSKUM). We first derive a second-order accurate approximation for the spatial derivatives $F_x$ and $F_y$ at a point $P_0$ using the defect correction procedure \cite{lskum-ghosh-journal}. Consider the Taylor series expansion of $F$ up to quadratic terms at a point $P_i \in  N\left(P_0\right)$,
\begin{equation}
\begin{split}
\Delta F_i & =  \Delta x_i F_{x_0} + \Delta y_i F_{y_0} 
+  \frac{\Delta x_i}{2} \left( \Delta x_i F_{xx_0} + \Delta y_i F_{xy_0}  \right)  \\
 & +  \frac{\Delta y_i}{2} \left( \Delta x_i F_{xy_0} 
 +  \Delta y_i F_{yy_0}  \right) + 
  O \left( \Delta x_i, \Delta y_i \right)^3, \; 
 i= 1, \hdots, n
\end{split}
\label{taylor-series-exp-F-second-order-rearranged}
\end{equation}
where $\Delta x_i = x_i - x_0$, $\Delta y_i = y_i - y_0$, $\Delta F_i = F_i - F_0$. $N\left(P_0\right)$ is the set of neighbours or the stencil of $P_0$. Here, $n$ denotes the number of neighbours of the point $P_0$. Replacing the second-order terms in the above equation using the Taylor series expansions of $F_x$ and $F_y$ to linear terms at a point $P_i$, we get 
\begin{equation}
\Delta F_i - \frac{1}{2} \left\lbrace \Delta x_i  \Delta F_{x_i}   +   \Delta y_i \Delta F_{y_i} \right\rbrace =  \Delta x_i F_{x_0} + \Delta y_i F_{y_0} +   +  O \left( \Delta x_i, \Delta y_i \right)^3, \; \; i  = 1, \hdots, n 
\label{taylor-series-second-order-rearranged}
\end{equation}
Denoting the left hand side as $\Delta \widetilde{F}_i=  \widetilde{  F}_i - \widetilde{F}_0 $, the above equation reduces to 
\begin{equation}
\Delta \widetilde{ F}_i   =  \Delta x_i F_{x_0} +  \Delta y_i F_{y_0}  +  O \left( \Delta x_i, \Delta y_i \right)^3, \; i  = 1, \hdots, n 
\label{taylor-series-exp-F-second-order-modified-final}
\end{equation}
For $n \geq 3$, eq. (\ref{taylor-series-exp-F-second-order-modified-final}) leads to an over-determined linear system of equations. Using the least-squares principle, the second-order approximations to $F_x$ and $F_y$ at the point $P_0$ are given by
\begin{equation}
\begin{bmatrix}
F_x \\ F_y \end{bmatrix}_{P_0}
 = 
{\begin{bmatrix}
\sum \Delta x_i^2 &  \sum \Delta x_i \Delta y_i \\
\sum \Delta x_i \Delta y_i & \sum \Delta y_i^2
\end{bmatrix}}^{-1} 
\begin{bmatrix}
\sum \Delta x_i \Delta \widetilde{ F}_i \\
\sum \Delta y_i \Delta \widetilde{ F}_i \\
\end{bmatrix}_{P_i \in N \left( P_0 \right) }
\label{ls-formulae-fx-fy-second-order}
\end{equation}
Taking $ \boldsymbol{\Psi} $ - moments of eq. (\ref{cir-be}) along with the above least-squares formulae, we get the semi-discrete second-order upwind meshfree scheme based on LSKUM for $2D$ Euler equations. The least-squares formulae for the spatial derivatives of $\boldsymbol{Gx}^{\pm}$ at a point $P_0$ are given by 
\begin{equation}
\frac{\partial \boldsymbol{Gx}^{\pm}}{\partial x } 
 = \left[ \frac{\sum \Delta y_i^2 \sum \Delta x_i \Delta \widetilde{\boldsymbol{Gx}}_i^{\pm} - \sum \Delta x_i \Delta y_i \sum \Delta y_i \Delta \widetilde{\boldsymbol{Gx}}_i^{\pm} }{\sum \Delta x_i^2 \sum \Delta y_i^2  - \left(\sum \Delta x_i \Delta y_i\right)^2 }  \right]
\label{ls-formulae-second-order-gxpm}
\end{equation}
Here, the perturbations $\Delta \widetilde{\boldsymbol{Gx}}_i^{\pm}$ are defined by
\begin{equation}
\Delta \widetilde{\boldsymbol{Gx}}_i^{\pm}=  \Delta {\boldsymbol{Gx}}^{\pm}_i - \frac{1}{2} \left\lbrace   \Delta x_i \frac{\partial}{\partial x} \Delta \boldsymbol{Gx}_i^{\pm} +  \Delta y_i \frac{\partial}{\partial y} \Delta \boldsymbol{Gx}_i^{\pm} \right\rbrace  
\label{pert-mod-split-fluxes}
\end{equation}
The split flux derivatives in eq. (\ref{ls-formulae-second-order-gxpm}) are approximated using the stencils 
\begin{equation}
 N_x^{\pm} \left( P_0 \right)  =   \left\lbrace P_i \in N \left( P_0 \right) \mid \Delta x_i \lessgtr 0 \right\rbrace    
\end{equation}
Similarly, using appropriate split stencils, we can write the formulae for the spatial derivatives of $\boldsymbol{Gy}^{\pm}$. A drawback of the defect correction approach using the Maxwellian distributions is that the numerical solution may not be positive as $\Delta \widetilde{F}_i$ is not the difference between two Maxwellians. Instead, it is the difference between two perturbed Maxwellian distributions $\widetilde{F}_i$ and $\widetilde{F}_0$ \cite{lskum-ghosh-journal,q-lskum}.  For preserving the positivity of the solution, instead of the Maxwellians, $\boldsymbol{q}$-variables \cite{smd-nasa-report-entropy-variables} can be used in the defect correction procedure \cite{q-lskum}. Note that $\boldsymbol{q}$-variables can represent the fluid flow at the macroscopic level as the transformations $\boldsymbol{U} \longleftrightarrow \boldsymbol{q} $ and $F\longleftrightarrow \boldsymbol{q}$ are unique. The $\boldsymbol{q}$-variables in $2D$ are defined by
\begin{equation}
\boldsymbol{q} = \begin{bmatrix}
\ln \rho + \frac{\ln \beta}{\gamma - 1} - \beta \left(u_1^2 + u_2^2 \right), \; 2 \beta u_1, \; 2 \beta u_2, \; -2\beta 
\end{bmatrix}, \; \beta = \frac{\rho}{2 p}    
\label{q_variables}
\end{equation}
Where $\gamma$ is the ratio of the specific heats. Using $\boldsymbol{q}$-variables, second-order accuracy can be achieved by replacing $\Delta \widetilde{\boldsymbol{Gx}}_i^{\pm}$ in eq. (\ref{pert-mod-split-fluxes}) with $\Delta \boldsymbol{Gx}_i^{\pm} \left(\widetilde{\boldsymbol{q}}\right) =  \boldsymbol{Gx}^{\pm} \left(\widetilde{\boldsymbol{q}}_i\right) - \boldsymbol{Gx}^{\pm} \left(\widetilde{\boldsymbol{q}}_0\right) 
$. Here, $\widetilde{\boldsymbol{q}}_i$ and $\widetilde{\boldsymbol{q}}_0$ are the modified $\boldsymbol{q}$-variables, defined by
\begin{equation}
\widetilde{\boldsymbol{q}}_{i}  =  \boldsymbol{q}_{i} - \frac{1}{2} \left( \Delta x_i {\boldsymbol{q}_x}_{i} +  \Delta y_i {\boldsymbol{q}_y}_{i}  \right), \; \; \widetilde{\boldsymbol{q}}_{0}  =  \boldsymbol{q}_{0} - \frac{1}{2} \left( \Delta x_i {\boldsymbol{q}_x}_{0} +  \Delta y_i {\boldsymbol{q}_y}_{0}  \right) 
\label{q-tilde-variables}
\end{equation}
Here $\boldsymbol{q}_x$ and $\boldsymbol{q}_y$ are evaluated to second-order using the least-squares formuale
\begin{equation}
    \begin{bmatrix}
    \boldsymbol{q}_x \\ \boldsymbol{q}_y
    \end{bmatrix} = {\begin{bmatrix}
\sum \Delta x_i^2 &  \sum \Delta x_i \Delta y_i \\
\sum \Delta x_i \Delta y_i & \sum \Delta y_i^2  
\end{bmatrix}}^{-1} 
\begin{bmatrix}
\sum \Delta x_i  \Delta \widetilde{ \boldsymbol{q}}_i  \\
\sum \Delta y_i \Delta \widetilde{\boldsymbol{q}}_i  \\
\end{bmatrix}_{P_i \in N \left( P_0 \right) } 
\label{ls-formulae-q-derivatives}
\end{equation}
The above formulae for $\boldsymbol{q}_x$ and $\boldsymbol{q}_y$ are implicit and must be solved iteratively. These iterations are known as inner iterations. 
\section{Implicit LSKUM based on Modified LU-SGS}
\label{sec-implicit-lusgs-lskum}
Consider the kinetic flux vector split $2D$ Euler equations in the implicit form.
\begin{equation}
\left( \frac{\partial \boldsymbol{U}}{\partial t} \right)^n + \frac{\partial}{\partial x} \left( \boldsymbol{Gx}^+\right)^{n+1} + \frac{\partial}{\partial x} \left( \boldsymbol{Gx}^-\right)^{n+1} + \frac{\partial}{\partial y} \left( \boldsymbol{Gy}^+\right)^{n+1} + \frac{\partial}{\partial y} \left( \boldsymbol{Gy}^-\right)^{n+1} = 0
\label{2d-implicit-kfvs-ee}
\end{equation}
Linearising the split fluxes $\left(\boldsymbol{Gx}^{\pm}\right)^{n+1}$ and $\left( \boldsymbol{Gy}^{\pm}\right)^{n+1}$ in time, we get 
\begin{equation}
\left(\boldsymbol{Gx}^{\pm}\right)^{n+1} = \left(\boldsymbol{Gx}^{\pm}\right)^{n} + \left(\boldsymbol{Ax}^{\pm}\right)^{n} \delta \boldsymbol{U}^n, \; \;
\left(\boldsymbol{Gy}^{\pm}\right)^{n+1} = \left(\boldsymbol{Gy}^{\pm}\right)^{n} + \left(\boldsymbol{Ay}^{\pm}\right)^{n} \delta \boldsymbol{U}^n 
\label{linearised-split-fluxes}
\end{equation}
where, $\delta \boldsymbol{U}^n = \boldsymbol{U}^{n+1} - \boldsymbol{U}^{n}$ is the solution increment vector. $\boldsymbol{Ax}^{\pm}$ and $\boldsymbol{Ay}^{\pm}$ are the split flux Jacobians along the coordinate directions $x$ and $y$ respectively. Using the above expressions in eq. (\ref{2d-implicit-kfvs-ee}), we obtain
\begin{equation}
\begin{split}
\left( \frac{\partial \boldsymbol{U}}{\partial t} \right)^n + & \frac{\partial}{\partial x} \left( \boldsymbol{Ax}^{+} \delta \boldsymbol{U}\right)^{n} + 
\frac{\partial}{\partial x} \left( \boldsymbol{Ax}^{-} \delta \boldsymbol{U}\right)^{n} + 
\frac{\partial}{\partial y} \left( \boldsymbol{Ay}^{+} \delta \boldsymbol{U}\right)^{n} + 
\frac{\partial}{\partial y} \left( \boldsymbol{Ay}^{-} \delta \boldsymbol{U}\right)^{n}   \\
 = & - \left(\frac{\partial \boldsymbol{Gx}^+}{\partial x} + \frac{\partial \boldsymbol{Gx}^-}{\partial x} + \frac{\partial \boldsymbol{Gy}^+}{\partial y} + \frac{\partial \boldsymbol{Gy}^-}{\partial y} \right)^n  \\
= & - \boldsymbol{R}^n
\end{split}
\label{implicit-lskum-semidiscrete}
\end{equation}
Here, $\boldsymbol{R}^n$ is the second-order accurate flux residual based on meshfree LSKUM. In the above equation, the time derivative is approximated using a first-order forward difference formula. The spatial derivatives on the left-hand side are discretised using the first-order least-squares approximations with appropriate split stencils. \\ \\
To simplify the left-hand side of the above equation, 
consider the least-squares formulae for the spatial derivatives of $\left( \boldsymbol{Ax}^{\pm} \delta \boldsymbol{U}\right)^{n}$ at the point $P_0$
\begin{equation}
\frac{\partial}{\partial x} \left( \boldsymbol{Ax}^{\pm} \delta \boldsymbol{U}\right)^{n}_0 = \left\lbrace \frac{\sum \Delta y_i^2 \sum \Delta x_i \Delta \left( \boldsymbol{Ax}^{\pm} \delta \boldsymbol{U}\right)^n_i - \sum \Delta x_i \Delta y_i \sum \Delta y_i \Delta \left( \boldsymbol{Ax}^{\pm} \delta \boldsymbol{U}\right)^n_i}{\sum \Delta x_i^2 \sum \Delta y_i^2 - \left( \sum \Delta x_i \Delta y_i \right)^2 } \right\rbrace
\end{equation}
Using the expression $ \Delta \left( \boldsymbol{Ax}^{\pm} \delta \boldsymbol{U}\right)^n_i =  \left( \boldsymbol{Ax}^{\pm} \delta \boldsymbol{U}\right)^n_i  -  \left( \boldsymbol{Ax}^{\pm} \delta \boldsymbol{U}\right)^n_0 $, the above equation reduces to
\begin{equation}
\begin{split}
\frac{\partial}{\partial x} \left( \boldsymbol{Ax}^{\pm} \delta \boldsymbol{U}\right)^{n}_{0} & = \left\lbrace \frac{\sum \Delta y_i^2 \sum \Delta x_i \left( \boldsymbol{Ax}^{\pm} \delta \boldsymbol{U}\right)^n_i - \sum \Delta x_i \Delta y_i \sum \Delta y_i \left( \boldsymbol{Ax}^{\pm} \delta \boldsymbol{U}\right)^n_i}{\sum \Delta x_i^2 \sum \Delta y_i^2 - \left( \sum \Delta x_i \Delta y_i \right)^2 } \right\rbrace \\ 
& - \left\lbrace \frac{\sum \Delta y_i^2 \sum \Delta x_i \left( \boldsymbol{Ax}^{\pm} \delta \boldsymbol{U}\right)^n_0 - \sum \Delta x_i \Delta y_i \sum \Delta y_i \left( \boldsymbol{Ax}^{\pm} \delta \boldsymbol{U}\right)^n_0}{\sum \Delta x_i^2 \sum \Delta y_i^2 - \left( \sum \Delta x_i \Delta y_i \right)^2 } \right\rbrace 
\end{split}
\end{equation}
Regrouping the first term on the right-hand side of the above equation based on $P_i< P_0$ and $P_i > P_0$, we arrive at
\begin{equation}
\begin{split}
 \frac{\partial}{\partial x} \left( \boldsymbol{Ax}^{\pm} \delta \boldsymbol{U}\right)^{n}_{0} & = \left\lbrace \frac{\sum \Delta y_i^2 \sum\limits_{i<0} \Delta x_i \left( \boldsymbol{Ax}^{\pm} \delta \boldsymbol{U}\right)^n_i - \sum \Delta x_i \Delta y_i \sum\limits_{i<0} \Delta y_i \left( \boldsymbol{Ax}^{\pm} \delta \boldsymbol{U}\right)^n_i}{\sum \Delta x_i^2 \sum \Delta y_i^2 - \left( \sum \Delta x_i \Delta y_i \right)^2 } \right\rbrace \\ 
& +  \left\lbrace \frac{\sum \Delta y_i^2 \sum\limits_{i>0} \Delta x_i \left( \boldsymbol{Ax}^{\pm} \delta \boldsymbol{U}\right)^n_i - \sum \Delta x_i \Delta y_i \sum\limits_{i>0} \Delta y_i \left( \boldsymbol{Ax}^{\pm} \delta \boldsymbol{U}\right)^n_i}{\sum \Delta x_i^2 \sum \Delta y_i^2 - \left( \sum \Delta x_i \Delta y_i \right)^2 } \right\rbrace \\ 
& - \left\lbrace \frac{\sum \Delta y_i^2 \sum \Delta x_i \left( \boldsymbol{Ax}^{\pm} \delta \boldsymbol{U}\right)^n_0 - \sum \Delta x_i \Delta y_i \sum \Delta y_i \left( \boldsymbol{Ax}^{\pm} \delta \boldsymbol{U}\right)^n_0}{\sum \Delta x_i^2 \sum \Delta y_i^2 - \left( \sum \Delta x_i \Delta y_i \right)^2 } \right\rbrace 
\end{split}
\end{equation}
Denoting the first, second and third terms on the right-hand side of the above equation as $LSx\left( \boldsymbol{Ax}^{\pm} \delta \boldsymbol{U}\right)^n_{i, i<0}$, $LSx\left( \boldsymbol{Ax}^{\pm} \delta \boldsymbol{U}\right)^n_{i, i>0}$ and $LSx\left( \boldsymbol{Ax}^{\pm} \delta \boldsymbol{U}\right)^n_0$ respectively, we get
\begin{equation}
 \frac{\partial}{\partial x} \left( \boldsymbol{Ax}^{\pm} \delta \boldsymbol{U}\right)^{n}_{0} = LSx\left( \boldsymbol{Ax}^{\pm} \delta \boldsymbol{U}\right)^n_{i, i<0} + LSx\left( \boldsymbol{Ax}^{\pm} \delta \boldsymbol{U}\right)^n_{i, i>0} - LSx\left( \boldsymbol{Ax}^{\pm} \right)^n_0  \delta \boldsymbol{U}^n_0    
\label{ls-approx-axdelu}
\end{equation}
Similarly, the spatial approximations of $\left( \boldsymbol{Ay}^{\pm} \delta \boldsymbol{U}\right)^{n}_0$, evaluated using the split stencils $Ny^{\pm}_0 = \left\lbrace P_i \in N \left( P_0 \right) \mid \Delta y_i \lessgtr 0 \right\rbrace$ are given by

\begin{equation}
\begin{split}
\frac{\partial}{\partial y} \left( \boldsymbol{Ay}^{\pm} \delta \boldsymbol{U}\right)^{n}_{0} & = \left\lbrace \frac{\sum \Delta x_i^2 \sum\limits_{i<0} \Delta y_i \left( \boldsymbol{Ay}^{\pm} \delta \boldsymbol{U}\right)^n_i - \sum \Delta x_i \Delta y_i \sum\limits_{i<0} \Delta x_i \left( \boldsymbol{Ay}^{\pm} \delta \boldsymbol{U}\right)^n_i}{\sum \Delta x_i^2 \sum \Delta y_i^2 - \left( \sum \Delta x_i \Delta y_i \right)^2 } \right\rbrace \\ 
& +  \left\lbrace \frac{\sum \Delta x_i^2 \sum\limits_{i>0} \Delta y_i \left( \boldsymbol{Ay}^{\pm} \delta \boldsymbol{U}\right)^n_i - \sum \Delta x_i \Delta y_i \sum\limits_{i>0} \Delta x_i \left( \boldsymbol{Ay}^{\pm} \delta \boldsymbol{U}\right)^n_i}{\sum \Delta x_i^2 \sum \Delta y_i^2 - \left( \sum \Delta x_i \Delta y_i \right)^2 } \right\rbrace \\ 
& - \left\lbrace \frac{\sum \Delta x_i^2 \sum \Delta y_i \left( \boldsymbol{Ay}^{\pm} \delta \boldsymbol{U}\right)^n_0 - \sum \Delta x_i \Delta y_i \sum \Delta x_i \left( \boldsymbol{Ay}^{\pm} \delta \boldsymbol{U}\right)^n_0}{\sum \Delta x_i^2 \sum \Delta y_i^2 - \left( \sum \Delta x_i \Delta y_i \right)^2 } \right\rbrace \\
 & = LSy\left( \boldsymbol{Ay}^{\pm} \delta \boldsymbol{U}\right)^n_{i, i<0} + LSy\left( \boldsymbol{Ay}^{\pm} \delta \boldsymbol{U}\right)^n_{i, i>0} - LSy\left( \boldsymbol{Ay}^{\pm} \right)^n_0  \delta \boldsymbol{U}^n_0    
\end{split}
\label{ls-approx-aydelu}
\end{equation}
Using the expressions obtained in eqs. (\ref{ls-approx-axdelu}) and (\ref{ls-approx-aydelu}), eq. (\ref{implicit-lskum-semidiscrete}) reduces to 
\begin{equation}
\begin{split}
&\left\lbrace \frac{\mathcal{I}}{{\Delta t}_0}  -   LSx \left( \boldsymbol{Ax}^{+}\right)^n_0  - LSx \left( \boldsymbol{Ax}^{-}\right)^n_0  - LSy \left( \boldsymbol{Ay}^{+}\right)^n_0 - LSy \left( \boldsymbol{Ay}^{-}\right)^n_0 \right\rbrace \delta \boldsymbol{U}^{n}_{0}
 \\ 
+ & \left\lbrace LSx\left( \boldsymbol{Ax}^{+} \delta \boldsymbol{U}\right)^n_{i, i<0} + LSx\left( \boldsymbol{Ax}^{-} \delta \boldsymbol{U}\right)^n_{i, i<0} + LSy\left( \boldsymbol{Ay}^{+} \delta \boldsymbol{U}\right)^n_{i, i<0} + LSy\left( \boldsymbol{Ay}^{-} \delta \boldsymbol{U}\right)^n_{i, i<0} \right\rbrace \\
+ & \left\lbrace LSx\left( \boldsymbol{Ax}^{+} \delta \boldsymbol{U}\right)^n_{i, i>0} + LSx\left( \boldsymbol{Ax}^{-} \delta \boldsymbol{U}\right)^n_{i, i>0} + LSy\left( \boldsymbol{Ay}^{+} \delta \boldsymbol{U}\right)^n_{i, i>0} + LSy\left( \boldsymbol{Ay}^{-} \delta \boldsymbol{U}\right)^n_{i, i>0} \right\rbrace \\
= & - \boldsymbol{R}^n_{0}
\end{split}
\label{full-ldu-equation}
\end{equation}
Denote the first, second, and third terms on the left-hand side of the above equation as $\widetilde{\mathcal{D}}_0$, $\mathcal{L}_0$ and $\mathcal{U}_0$ respectively. Here, $\widetilde{\mathcal{D}}$ is the block diagonal matrix, $\mathcal{L}$ and $\mathcal{U}$ are strictly the lower and upper triangular matrices. For all points in the given computational domain, eq. (\ref{full-ldu-equation}) leads to a linear system of equations of the form
\begin{equation}
    \left( \mathcal{L}^n_i + \widetilde{\mathcal{D}}^n_i + \mathcal{U}^n_i\right) \delta \boldsymbol{U}^n_i = - \boldsymbol{R}^n_i, \; \; i = 1 \hdots N
    \label{short-ldu-equation}
\end{equation}
Here $N$ is the number of points in the domain. Using the implicit lower upper symmetric Gauss-Seidel (LU-SGS) algorithm \cite{lusgs-jameson}, the linear system can be solved using a two-step procedure as
\begin{equation}
\begin{split}
\delta \boldsymbol{U}^{*}_i & =  - {\widetilde{\mathcal{D}}}^{-1}_i \left( \boldsymbol{R}^n_i + \mathcal{ L}_i \delta \boldsymbol{U}^{*}_i \right),\; \; i=1\hdots N \; \; \; \text{(Forward sweep)} \\
 \delta \boldsymbol{U}^{n}_i & =  \delta \boldsymbol{U}^{*}_i - \widetilde{\mathcal{D}}^{-1}_i \left(\mathcal{U}\delta \boldsymbol{U}^{n} \right),\; \; i=N \hdots 1 \; \; \; \text{(Reverse sweep)}
\end{split}    
\label{lusgs-initial-forward-reverse-sweeps}
\end{equation}
We observe that the computation of $\mathcal{L} \delta \boldsymbol{U}^{*}$ and $\mathcal{U} \delta \boldsymbol{U}^{n}$ in the forward and reverse sweeps involves the evaluation of split flux Jacobian matrices and their products with the solution increment vectors, $\boldsymbol{Ax}^{\pm} \delta \boldsymbol{U}$ and $\boldsymbol{Ay}^{\pm} \delta \boldsymbol{U}$. The computation of expensive Jacobians can be avoided by approximating the Jacobian vector products with the increments in the numerical flux functions, 
\begin{equation}
\begin{split}
\boldsymbol{Ax}^{\pm} \delta \boldsymbol{U} & = \boldsymbol{Gx}^{\pm} \left(\boldsymbol{U} + \delta \boldsymbol{U} \right) - \boldsymbol{Gx}^{\pm} \left(\boldsymbol{U}\right) = \delta \boldsymbol{Gx}^{\pm} \\
\boldsymbol{Ay}^{\pm} \delta \boldsymbol{U} & = \boldsymbol{Gy}^{\pm} \left(\boldsymbol{U} + \delta \boldsymbol{U} \right) - \boldsymbol{Gy}^{\pm} \left(\boldsymbol{U} \right) = \delta \boldsymbol{Gy}^{\pm} 
\end{split}
\label{jacobian-vector-product-approximations}
\end{equation}
This matrix-free approach proposed for finite volume methods \cite{lusgs-Sharov-1998} is followed in the earlier works on implicit meshfree solvers based on LSKUM \cite{lusgs-anandh-icas-2004,lusgs-NAL-Manish-2015,anandh-lskum-viscous-aiaa-journal}. The LU-SGS algorithm also requires the inverse of split flux Jacobians to compute $\widetilde{\mathcal{D}}^{-1}$. In order to ease the computation, the split flux Jacobians are approximated using spectral radii \cite{lusgs-jameson} as 
\begin{equation}
\boldsymbol{Ax}^{\pm} = \frac{1}{2} \left( \boldsymbol{Ax} \pm \rho \left( \boldsymbol{Ax} \right) \boldsymbol{I} \right), \; \;  \boldsymbol{Ay}^{\pm} = \frac{1}{2} \left( \boldsymbol{Ay} \pm \rho \left( \boldsymbol{Ay} \right) \boldsymbol{I} \right)
\label{jameson-split-flux-jacobian-approxs}
\end{equation}
Here, $\boldsymbol{Ax}$ and $\boldsymbol{Ay}$ are the unsplit flux Jacobians. $\rho \left( \boldsymbol{Ax} \right)$ and $\rho \left( \boldsymbol{Ay} \right)$ are the spectral radii of the flux Jacobians, given by
\begin{equation}
    \rho \left( \boldsymbol{Ax} \right) = \left | u_1 \right | + a,\; \;   \rho \left( \boldsymbol{Ay} \right) = \left | u_2 \right | + a
\end{equation}
where $a$ is the speed of sound. Using these approximations the expression for $LSx\left(\boldsymbol{Ax^+}\right)^n_0\delta \boldsymbol{U}^n_0$ in eq. (\ref{full-ldu-equation}) reduces to
\begin{equation}
\begin{split}
    LSx\left(\boldsymbol{Ax^+}\right)^n_0 \delta \boldsymbol{U}^n_0 & = \frac{1}{2}\left\lbrace \frac{\sum \Delta y_i^2 \sum \Delta x_i \left( \boldsymbol{Ax} \delta \boldsymbol{U}\right)^n_0 - \sum \Delta x_i \Delta y_i \sum \Delta y_i \left( \boldsymbol{Ax} \delta \boldsymbol{U}\right)^n_0}{\sum \Delta x_i^2 \sum \Delta y_i^2 - \left( \sum \Delta x_i \Delta y_i \right)^2 } \right\rbrace \\
    & + \frac{1}{2} \rho \left(\boldsymbol{Ax}\right) \left\lbrace \frac{\sum \Delta y_i^2 \sum \Delta x_i  - \sum \Delta x_i \Delta y_i \sum \Delta y_i }{\sum \Delta x_i^2 \sum \Delta y_i^2 - \left( \sum \Delta x_i \Delta y_i \right)^2 } \right\rbrace \delta \boldsymbol{U}^n_0 \\ 
    & = \frac{1}{2} LSxp\left( \boldsymbol{Ax} \delta \boldsymbol{U}\right)^n_0 + \frac{1}{2} \rho \left(\boldsymbol{Ax}\right) LSxp(1) \delta \boldsymbol{U}^n_0 
\end{split}
\label{lsxp-final-short-expression}
\end{equation}
Similarly, the expressions for $LSx\left(\boldsymbol{Ax^-}\right)^n_0 \delta \boldsymbol{U}^n_0$, and $LSy\left(\boldsymbol{Ay^{\pm}}\right)^n_0 \delta \boldsymbol{U}^n_0$ can be written as 
\begin{equation}
    \begin{split}
 LSx\left(\boldsymbol{Ax^-}\right)^n_0 \delta \boldsymbol{U}^n_0 & = \frac{1}{2} LSxn\left( \boldsymbol{Ax} \delta \boldsymbol{U}\right)^n_0 - \frac{1}{2} \rho \left(\boldsymbol{Ax}\right) LSxn(1) \delta \boldsymbol{U}^n_0    \\
 LSy\left(\boldsymbol{Ay^+}\right)^n_0 \delta \boldsymbol{U}^n_0 & = \frac{1}{2} LSyp\left( \boldsymbol{Ax} \delta \boldsymbol{U}\right)^n_0 + \frac{1}{2} \rho \left(\boldsymbol{Ay}\right) LSyp(1) \delta \boldsymbol{U}^n_0    \\
LSy\left(\boldsymbol{Ay^-}\right)^n_0 \delta \boldsymbol{U}^n_0 & = \frac{1}{2} LSyn\left( \boldsymbol{Ax} \delta \boldsymbol{U}\right)^n_0 - \frac{1}{2} \rho \left(\boldsymbol{Ay}\right) LSyn(1) \delta \boldsymbol{U}^n_0   \\
\end{split}
\label{lsxypm-final-short-expressions}
\end{equation}
From eqs. (\ref{lsxp-final-short-expression}) and (\ref{lsxypm-final-short-expressions}), $\widetilde{\mathcal{D}}\delta \boldsymbol{U}^n_0$ in eq. (\ref{full-ldu-equation}) reduces to 
\begin{equation}
\begin{split}
\widetilde{\mathcal{D}}\delta \boldsymbol{U}^n_0 & = 
 -\frac{1}{2} \left\lbrace LSxp\left( \boldsymbol{Ax} \delta \boldsymbol{U}\right)^n_0 + LSxn\left( \boldsymbol{Ax} \delta \boldsymbol{U}\right)^n_0 + LSyp\left( \boldsymbol{Ay} \delta \boldsymbol{U}\right)^n_0 + LSyn\left( \boldsymbol{Ay} \delta \boldsymbol{U}\right)^n_0\right\rbrace \\ 
& + \left\lbrace \frac{\mathcal{I}}{\Delta t} - \frac{1}{2}\rho \left(\boldsymbol{Ax}\right) \left\lbrace LSxp(1) - LSxn(1) \right\rbrace  - \frac{1}{2}\rho \left(\boldsymbol{Ay}\right) \left\lbrace LSyp(1) - LSyn(1) \right\rbrace \right\rbrace \delta \boldsymbol{U}^n_0 \\ 
& = \mathcal{S}_0^n + \mathcal{D} \delta \boldsymbol{U}^n_0
\end{split}
\end{equation}
In the above equation, $\mathcal{D}$ is a scalar diagonal matrix. The term $\mathcal{S}$ on the right-hand side vanishes when the point distribution around the point $P_0$ is symmetric, thus transforming the block diagonal matrix $\widetilde{\mathcal{D}}$ to a scalar matrix. However, for an arbitrary distribution of points, ensuring a symmetric stencil for each point may not always be possible. To avoid the term $\mathcal{S}$, instead of eq. (\ref{jameson-split-flux-jacobian-approxs}), Anandhanarayanan et al. \cite{lusgs-anandh-icas-2004}  approximated the split flux Jacobians using the spectral radii, 
\begin{equation}
\boldsymbol{Ax}^{\pm} = \pm \rho \left(\boldsymbol{Ax}^{\pm} \right) \boldsymbol{I}, \; \; \boldsymbol{Ay}^{\pm} = \pm \rho \left(\boldsymbol{Ay}^{\pm} \right) \boldsymbol{I}     
\label{spectral-radii-anandh}
\end{equation}
where $\rho \left(\boldsymbol{Ax}^{\pm}\right)$ and $\rho \left(\boldsymbol{Ay}^{\pm}\right)$ are the spectral radii of $\boldsymbol{Ax}^{\pm}$ and $\boldsymbol{Ay}^{\pm}$ respectively. They are defined by 
\begin{equation}
\rho \left(\boldsymbol{Ax}^{\pm}\right) = \frac{1}{2} \left| \left(u_1 \pm a \right) \pm \left| u_1 \pm a \right | \right |, \; \; \rho \left(\boldsymbol{Ay}^{\pm}\right) = \frac{1}{2} \left| \left(u_2 \pm a \right) \pm \left| u_2 \pm a \right | \right |    
\end{equation}
In a later research work, Manish et al. \cite{lusgs-NAL-Manish-2015} retained the term $\mathcal{S}$ and moved it to the explicit part of eq. (\ref{full-ldu-equation}). It was done to increase the diagonal dominance of the linear system of equations and thus increase the convergence rate. The linear system in eq. (\ref{short-ldu-equation}) then reduces to 
\begin{equation}
    \left(\mathcal{L}^n_i + \mathcal{D}^n_i + \mathcal{U}^n_i \right) \delta \boldsymbol{U}^n_i = - \boldsymbol{R}^n_i + \mathcal{S}^n_i, \; \; i = 1, \hdots, N
\end{equation}
The LU-SGS algorithm is then given by 
\begin{equation}
\begin{split}
\delta \boldsymbol{U}^{*}_i & =  - {\mathcal{D}}^{-1}_i \left( \boldsymbol{R}^n_i - \mathcal{S}^n_i + \mathcal{ L}_i \delta \boldsymbol{U}^{*}_i \right),\; \; i=1,\hdots, N \; \; \; \text{(Forward sweep)} \\
 \delta \boldsymbol{U}^{n}_i & =  \delta \boldsymbol{U}^{*}_i - \mathcal{D}^{-1}_i \left(\mathcal{U}\delta \boldsymbol{U}^{n} \right),\; \; i=N, \hdots, 1 \; \; \; \text{(Reverse sweep)}
\end{split}    
\label{lusgs-revised-forward-reverse-sweeps}
\end{equation}
Since $\delta \boldsymbol{U}^n$ is the unknown vector, the term $\mathcal{S}$ can be computed using the available flow solution at the $(n-1)^{th}$ iteration. The Jacobian vector products $\boldsymbol{Ax}\delta \boldsymbol{U}$ and $\boldsymbol{Ay}\delta \boldsymbol{U}$ present in $\mathcal{S}$ are evaluated using the matrix-free approximations,
\begin{equation}
\begin{split}
\boldsymbol{Ax} \delta \boldsymbol{U} & = \boldsymbol{Gx} \left(\boldsymbol{U} + \delta \boldsymbol{U} \right) - \boldsymbol{Gx} \left(\boldsymbol{U}\right) = \delta \boldsymbol{Gx} \\
\boldsymbol{Ay} \delta \boldsymbol{U} & = \boldsymbol{Gy} \left(\boldsymbol{U} + \delta \boldsymbol{U} \right) - \boldsymbol{Gy}\left(\boldsymbol{U} \right) = \delta \boldsymbol{Gy}
\end{split}
\label{full-jacobian-vector-product-approximations}
\end{equation}
In the numerical results, we show that the implicit LSKUM solver with the LU-SGS algorithm in eq. (\ref{lusgs-revised-forward-reverse-sweeps}) does not improve the residual convergence rate over the LU-SGS scheme in eq. (\ref{lusgs-initial-forward-reverse-sweeps}). \\ \\ 
In this research, the Jacobian vector products present in the term $\mathcal{S}$ of eq. (\ref{lusgs-revised-forward-reverse-sweeps}), and also in the computation of $\mathcal{L}\delta \boldsymbol{U}^*$ and $\mathcal{U}\delta \boldsymbol{U}^n$ in eqs. (\ref{lusgs-initial-forward-reverse-sweeps}) and (\ref{lusgs-revised-forward-reverse-sweeps}) are evaluated exactly using the algorithmic differentiation (AD) techniques \cite{ad-book-andreas}. In the numerical results, we show that the exact computation of the Jacobian vector products notably enhances the convergence rate. \\ \\ 
We illustrate the exact computation of the flux Jacobian vector product $\boldsymbol{Ax}\delta \boldsymbol{U}$ using AD. Listing \ref{routine-for-flux-Gx} shows a {\tt Fortran} subroutine, $\tt flux\_Gx$ for computing the full flux vector $\boldsymbol{Gx}$ along the coordinate direction $x$. Given the conserved vector $\boldsymbol{U}$ as the input, this routine yields the flux vector $\boldsymbol{Gx}$ as the output. Listing \ref{blackbox-routine-for-flux-Gxd} shows the tangent linear code generated by algorithmically differentiating the routine $\tt flux\_Gx$ in a black-box fashion using the AD tool Tapenade \cite{tapenade}. Here, the derivative routine is named by appending $\tt\_d$ to the primal routine $\tt flux\_Gx$. By specifying the input vectors $\tt U$ and $\tt Ud$, the tangent code computes the flux vector $\tt Gx$ and its derivative $\tt Gxd$ in the direction of $\tt Ud$. Here, $\tt Gxd$ is the Jacobian vector product $\boldsymbol{Ax}\delta \boldsymbol{U}$. Since we are interested in finding $\tt Gxd$ alone, the black-box tangent code can be modified to omit the computation of the flux vector $\tt Gx$. Listing \ref{optimised-routine-for-flux-Gxd} shows the optimised version of the tangent code. Given the input vectors $\tt U$ and $\tt Ud$, this code gives the desired Jacobian vector product $\tt AxdU$ as the output. Following the same approach, the split flux Jacobian vector products, $\boldsymbol{Ax}^{\pm}\delta \boldsymbol{U}$ and $\boldsymbol{Ay}^{\pm}\delta \boldsymbol{U}$ too are computed exactly using AD. Note that the computation of a split flux Jacobian conserved vector product using incremental fluxes requires two function calls, one with $\boldsymbol{U} + \delta \boldsymbol{U}$ and the other with $\boldsymbol{U}$ as input vectors. Since the expressions for kinetic split fluxes involve an {\tt error} function, invoking two split flux functions doubled the calls to the error function. On the other hand, computing a split flux Jacobian vector product using AD involves only one call to the {\tt error} function. In Section \ref{sec-performance-analysis}, we show that the implicit meshfree GPU solvers using AD are computationally more efficient than those employing matrix-free approximations for the flux Jacobian conserved vector products. 
\section{GPU Accelerated Implicit LSKUM Solver}
\label{section-gpu-accelerated-solver}
This section presents the development of a GPU accelerated implicit meshfree LSKUM solver based on LU-SGS. Algorithm \ref{algo-lskum-lusgs-solver} presents the general structure of a serial implicit solver for steady-state flows. The solver consists of a fixed point iterative scheme, represented by the $\tt for$ loop. The subroutine $\tt q\_variables()$ computes the $\boldsymbol{q}$-variables defined in eq. (\ref{q_variables}). $\tt q\_derivatives()$ performs the inner iterations in the defect correction step to evaluate $\boldsymbol{q}_x$ and $\boldsymbol{q}_y$ using the formulae in eq. (\ref{ls-formulae-q-derivatives}). $\tt flux\_residual()$ computes the second-order accurate approximation of the kinetic split flux derivatives in eq. (\ref{implicit-lskum-semidiscrete}) using least-squares formulae. The routine $\tt LUSGS()$ performs 
the forward and backward sweeps of the implicit scheme. Since the solution develops at the wall boundary, the wall points are first updated in the forward and backward sweeps, followed by interior and outer boundary points. Algorithm \ref{algo-forward-backward-sweeps} presents the corresponding forward and backward sweep subroutines. Finally, $\tt state\_update()$ updates the flow solution. All input and output operations are performed in the subroutines $\tt preprocessor()$ and $\tt postprocessor()$, respectively. The parameter $N$ represents the fixed point iterations required to achieve the desired convergence in the flow solution.   \\ \\
\scalebox{0.9}{ 
 \begin{algorithm}[H]
    \footnotesize
    \DontPrintSemicolon
    \SetAlgoLined
    \vspace{1mm}
    \SetKwFunction{FMain}{LSKUM()}
    \SetKwProg{Fn}{subroutine}{}{}
        \Fn{\FMain}
    {
		call {\tt{ preprocessor() }}\\
        \For{$n \leftarrow 1$ \KwTo $n \leq N$}
        {
                call {\tt  q\_variables() } \\ 
                call {\tt  q\_derivatives() } \\
                call {\tt  flux\_residual() } \\
                call {\tt  LUSGS() } \\
                call  {\tt state\_update() } \\
        }
		call {\tt{ postprocessor() }}
    }
    \textbf{end subroutine}
    \vspace{1mm}
    \caption{Implicit LSKUM.}
            \label{algo-lskum-lusgs-solver}
\end{algorithm}} \\ \\
 \begin{algorithm}[H]
 \begin{multicols}{2}
    \footnotesize
    \DontPrintSemicolon
    \SetAlgoLined
    \vspace{1mm}
    \SetKwFunction{FMain}{LUSGS\_forward\_sweep()}
    \SetKwProg{Fn}{subroutine}{}{}
        \Fn{\FMain}
    {
		\For{$i \leftarrow 1$ \KwTo {\tt wall\_points} }
        {
        $p = {\tt wall\_point(i)}$ \\
        call {\tt{forward\_sweep(p) }}}
        \For{$i \leftarrow 1$ \KwTo {\tt interior\_points} }
        {
        $p = {\tt interior\_point(i)}$ \\
        call {\tt{forward\_sweep(p) }}}
        \For{$i \leftarrow 1$ \KwTo {\tt outer\_points} }
        {
        $p = {\tt outer\_point(i)}$ \\
        call {\tt{forward\_sweep(p) }}}
    }
	\textbf{end subroutine} \\
    \vspace{1mm}
    \SetAlgoLined
    \vspace{1mm}
    \SetKwFunction{FMain}{LUSGS\_backward\_sweep()}
    \SetKwProg{Fn}{subroutine}{}{}
        \Fn{\FMain}
    {
		\For{$i \leftarrow  {\tt wall\_points}$ \KwTo $i \ge 1$ }
        {
        $p = {\tt wall\_point(i)}$ \\
        call {\tt{backward\_sweep(p) }}}
        \For{$i \leftarrow  {\tt interior\_points} $ \KwTo $i\ge 1$ }
        {
        $p = {\tt interior\_point(i)}$ \\
        call {\tt{backward\_sweep(p) }}}
        \For{$i \leftarrow {\tt outer\_points} $ \KwTo $i \ge 1$ }
        {
        $p = {\tt outer\_point(i)}$ \\
        call {\tt{backward\_sweep(p) }}}
    }
	\textbf{end subroutine}
    \vspace{1mm}
\end{multicols}
\vspace{1mm}
    \caption{Forward and Backward sweeps of the implicit LU-SGS algorithm.}
        \label{algo-forward-backward-sweeps}
\end{algorithm}
Algorithm \ref{algo-lskum-lusgs-gpu-solver} presents the structure of a GPU accelerated implicit LSKUM solver. The GPU solver mainly consists of the following sequence of operations: transfer the input meshfree data structure from host to device, perform the fixed-point iterations on the GPU, and finally transfer the converged flow solution from device to host. In the baseline GPU code, for each subroutine in Algorithm \ref{algo-lskum-lusgs-solver}, equivalent {\tt CUDA} kernels are constructed. The GPU code is then profiled using NVIDIA Nsight Compute \cite{nvidia-documentation}. The profiler metrics have shown that the register usage of the {\tt flux\_residual} and {\tt LUSGS} kernels is very high as these kernels are too large. When the kernel requires more registers than what is available on the GPU, it causes register spilling. This results in performance degradation as the excess data is stored in the slower global memory, leading to an increase in memory latency \cite{hipc-2022-nischay}. In order to reduce the register pressure, these kernels are split into smaller kernels. The {\tt flux\_residual} kernel is split into four kernels that compute the spatial derivatives of the split fluxes $\boldsymbol{Gx}^{\pm}$ and $\boldsymbol{Gy}^{\pm}$. From eq. (\ref{implicit-lskum-semidiscrete}), the sum of these spatial derivatives gives the flux residual $\boldsymbol{R}^n$. Similarly, the {\tt LUSGS} kernel has been split into two kernels that perform the forward and backward sweeps of the implicit scheme. These changes have reduced the register pressure and improved the performance of the GPU solver. \\ \\ 
From Algorithm \ref{algo-forward-backward-sweeps}, we observe that the {\tt i-loops} in the forward and backward sweeps must be executed sequentially. Numerical experiments have shown that the implicit GPU solver based on Algorithm \ref{algo-forward-backward-sweeps} yields a marginal speedup of around three compared to the serial solver. This is due to the inherent structure of the LU-SGS algorithm, which prevents the full utilisation of GPU resources. On the other hand, launching a {\tt CUDA} kernel by ignoring the loop structure of the forward sweep would update $\delta \boldsymbol{U}^*$ at a point $P_i$. However, it could lead to an undesirable scenario where threads operating simultaneously on other points might access $\delta \boldsymbol{U}^*_i$, which has not been updated. This race condition is caused by the presence of the point $P_i$ in the connectivity set of other points.  \\ \\
\scalebox{0.9}{ 
 \begin{algorithm}[H]
    \footnotesize
    \DontPrintSemicolon
    \SetAlgoLined
    \vspace{1mm}
    \SetKwFunction{FMain}{LSKUM\_cuda()}
    \SetKwProg{Fn}{subroutine}{}{}
        \Fn{\FMain}
    {
		call {\tt{ preprocessor() }}\\
        \textbf{cudaHostToDevice}(CPU\_data, GPU\_data) \\
        \For{$n \leftarrow 1$ \KwTo $n \leq N$}
        {
                \textbf{kernel} $<<<$ grid, block $>>>$  {\tt  q\_variables\_cuda() } \\ 
                \textbf{kernel} $<<<$ grid, block $>>>$  {\tt  q\_derivatives\_cuda() } \\
                \textbf{kernel} $<<<$ grid, block $>>>$ {\tt  spatial\_derivative\_Gx\_positive\_cuda() } \\
                \textbf{kernel} $<<<$ grid, block $>>>$ {\tt  spatial\_derivative\_Gx\_negative\_cuda() } \\
                \textbf{kernel} $<<<$ grid, block $>>>$ {\tt  spatial\_derivative\_Gy\_positive\_cuda() } \\
                \textbf{kernel} $<<<$ grid, block $>>>$ {\tt  spatial\_derivative\_Gy\_negative\_cuda() } \\
                \textbf{kernel} $<<<$ grid, block $>>>$ {\tt  LUSGS\_forward\_sweep\_cuda() } \\
                \textbf{kernel} $<<<$ grid, block $>>>$  {\tt  LUSGS\_backward\_sweep\_cuda() } \\
                \textbf{kernel} $<<<$ grid, block $>>>$  {\tt state\_update\_cuda() } \\
        }
        \textbf{cudaDeviceToHost}(GPU\_data, CPU\_data) \\
		call {\tt{ postprocessor() }} \\
        \vspace{1mm}
    }
    \textbf{end subroutine}
    \vspace{1mm}
    \caption{GPU accelerated implicit LSKUM solver.}
            \label{algo-lskum-lusgs-gpu-solver}
\end{algorithm} } \\ \\ 
To address this challenge and to extract parallelism, modifications must be made to the forward and reverse sweep algorithms without affecting the connectivity related data dependency. It can be achieved by dividing the points in the computational domain into several disjoint sets. Here, each set consists of points that do not lie in the connectivity of other points. To distinguish these sets, a unique color is assigned to the points in each set. Typically, the point colors are represented by positive integers. An advantage of this approach is that the forward and reverse sweeps of the points with the same color can be executed in parallel using {\tt CUDA} threads. \\ \\ 
The coloring scheme employed in the present work is shown in Algorithm \ref{algo-coloring-scheme}. The basic idea of this scheme is that a point and its neighbours should have different colors. A necessary condition for the coloring scheme is that if a point $P_i$ is in the neighbourhood of $P_0$, then $C\left(P_i\right) \cap C\left(P_0\right) \supseteq \left\lbrace P_i, P_0 \right\rbrace$. Here, $C\left(P_i\right)$ and $C\left(P_0\right)$ are the connectivity sets of the points $P_i$ and $P_0$, respectively. Violating this criterion may result in a point and some of its neighbours having the same color. However, this condition on the connectivity sets is not required for the least-squares approximation of the spatial derivatives in the meshfree numerical algorithm.\\ \\

%
%
%
\scalebox{0.9}{ 
 \begin{algorithm}[H]
    \footnotesize
    \DontPrintSemicolon
    \SetAlgoLined
    \vspace{1mm}
    \SetKwFunction{FMain}{point\_coloring\_scheme()}
    \SetKwProg{Fn}{subroutine}{}{}
        \Fn{\FMain}
    {
        \vspace{1mm}
        \textbf{Comment:} Initialise the colors of all points. \\
        \For{$i \leftarrow 1$ \KwTo $ {\tt max\_points} $}
        {   $color(i) = 0$
        }
        $color(1) = 1$ \\
        \textbf{Comment:} Find the color of subsequent points. \\
        \For{$i \leftarrow 1$ \KwTo $\tt max\_points$}
        {
        \For{$j \leftarrow 1$ \KwTo $\tt neighbours(i)$}
        {
            $p$ =  {\tt connectivity(i,j)} \\
          \If{$\left(color(p) = 0\right)$} 
   {$S = \text {set of colors of the neighbours of } p$ \\
            $color(p) = \text{smallest integer } k>0 {\text{ and }}  k \notin S $} 
          }}}
    \textbf{end subroutine}
    \vspace{1mm}
    \caption{\small Coloring scheme for the point distribution.}
            \label{algo-coloring-scheme}
\end{algorithm}} \\ \\ 
Algorithm \ref{algo-forward-backward-sweeps-modified} shows the modified forward and backward sweeps using the point colors. Here, the {\tt i-loops}, which run over the point colors, must be executed sequentially. On the other hand, the {\tt j-loops} correspond to the points with the same color. For these points, the forward sweeps can be performed simultaneously on a GPU using {\tt CUDA} threads. The more the number of points in a {\tt j-loop}, the higher the number of {\tt CUDA} threads executed parallelly, leading to an increase in the speedup. The performance of the GPU solver can be enhanced further if only one {\tt i-loop} exists in the forward sweep. Note that the three {\tt i-loops} in this algorithm are due to the grouping of the point distribution into wall, interior, and outer boundary points. \\ \\
Algorithm \ref{algo-forward-backward-sweeps-efficient} shows a computationally more efficient implementation of forward and backward sweeps. Here, the given point distribution is treated as one group, which results in only one {\tt i-loop} per sweep. Since the number of points with a given color is more than the number of wall, interior, or outer boundary points with the same color, it allows for more thread parallelism on a GPU. Although this approach may result in a slower convergence rate, it is well compensated by a significant speedup over Algorithm \ref{algo-forward-backward-sweeps-modified}. The corresponding {\tt CUDA} kernels are shown in Algorithm \ref{algo-cuda-kernels-efficient-forward-backward-sweeps}. The computational efficiency of this approach over Algorithm \ref{algo-forward-backward-sweeps-modified} is demonstrated in the numerical results. \\ 
 \begin{algorithm}[H]
 \begin{multicols}{2}
    \footnotesize
    \DontPrintSemicolon
    \SetAlgoLined
    \vspace{1mm}
    \SetKwFunction{FMain}{LUSGS\_forward\_sweep()}
    \SetKwProg{Fn}{subroutine}{}{}
        \Fn{\FMain}
    {
		\For{$i \leftarrow 1$ \KwTo {\tt wall\_point\_colors} }
        {
        $n_w = {\tt wall\_points\_with\_color(i)}$ \\ 
        \If{$\left(n_w \neq 0 \right)$} 
        { 
        \For{$j \leftarrow 1$ \KwTo $n_w$ }
        {
        $p$ = {\tt wall\_point(i,j)} \\
        call {\tt{forward\_sweep(p) }}} }}
        \For{$i \leftarrow 1$ \KwTo {\tt interior\_point\_colors} }
        {
        $n_i = {\tt interior\_points\_with\_color(i)}$ \\ 
        \If{$\left(n_i \neq 0 \right)$} 
        { 
        \For{$j \leftarrow 1$ \KwTo $n_i$ }
        {
        $p$ = {\tt interior\_point(i,j)} \\
        call {\tt{forward\_sweep(p) }}} }}
        \For{$i \leftarrow 1$ \KwTo {\tt outer\_point\_colors} }
        {
        $n_o = {\tt outer\_points\_with\_color(i)}$ \\ 
          \If{$\left(n_o \neq 0 \right)$} 
        { 
        \For{$j \leftarrow 1$ \KwTo $n_o$ }
        {
        $p$ = {\tt outer\_point(i,j)} \\
        call {\tt{forward\_sweep(p) }}} }}
    }
	\textbf{end subroutine} 
    \\
    \vspace{1mm}
    \SetKwFunction{FMain}{LUSGS\_backward\_sweep()}
    \SetKwProg{Fn}{subroutine}{}{}
        \Fn{\FMain}
    {
		\For{$i \leftarrow $ {\tt wall\_point\_colors} \KwTo $i \geq 1 $ }
        {
        $n_w = {\tt wall\_points\_with\_color(i)}$ \\ 
        \If{$\left(n_w \neq 0 \right)$} 
        { 
        \For{$j \leftarrow 1$ \KwTo $n_w$ }
        {
        $p$ = {\tt wall\_point(i,j)} \\
        call {\tt{backward\_sweep(p) }}} }}
        \For{$i \leftarrow $ {\tt interior\_point\_colors} \KwTo $i\geq 1$}
        {
        $n_i = {\tt interior\_points\_with\_color(i)}$ \\ 
        \If{$\left(n_i \neq 0 \right)$} 
        { 
        \For{$j \leftarrow 1$ \KwTo $n_i$ }
        {
        $p$ = {\tt interior\_point(i,j)} \\
        call {\tt{backward\_sweep(p) }}} }}
        \For{$i \leftarrow $ {\tt outer\_point\_colors} \KwTo $i\geq 1$}
        {
        $n_o = {\tt outer\_points\_with\_color(i)}$ \\ 
          \If{$\left(n_o \neq 0 \right)$} 
        { 
        \For{$j \leftarrow 1$ \KwTo $n_o$ }
        {
        $p$ = {\tt outer\_point(i,j)} \\
        call {\tt{backward\_sweep(p) }}} }}
    }
	\textbf{end subroutine}
    \vspace{1mm}
    \DontPrintSemicolon
    \SetAlgoLined
            \label{algo-forward-backward-sweeps-modified}
            \end{multicols}
            \vspace{1mm}
    \caption{\small Modified forward and backward sweeps.}
\end{algorithm}
 \begin{algorithm}[H]
 \begin{multicols}{2}
    \footnotesize
    \DontPrintSemicolon
    \SetAlgoLined
    \vspace{1mm}
    \SetKwFunction{FMain}{LUSGS\_forward\_sweep()}
    \SetKwProg{Fn}{subroutine}{}{}
        \Fn{\FMain}
    {
		\For{$i \leftarrow 1$ \KwTo {\tt point\_colors} }
        {
        $n_p = {\tt points\_with\_color(i)}$ \\
        \For{$j \leftarrow 1$ \KwTo $n_p$ }
        {
        $p$ = {\tt point(i,j)} \\
        call {\tt{forward\_sweep(p) }}} 
        }
        
    }
	\textbf{end subroutine} \\
    \vspace{1mm}
    \SetKwFunction{FMain}{LUSGS\_backward\_sweep()}
    \SetKwProg{Fn}{subroutine}{}{}
        \Fn{\FMain}
    {
		\For{$i \leftarrow {\tt point\_colors}$ \KwTo $i\ge 1$ }
        {
        $n_p = {\tt points\_with\_color(i)}$ \\
        \For{$j \leftarrow 1$ \KwTo $n_p$ }
        {
        $p$ = {\tt point(i,j)} \\
        call {\tt{backward\_sweep(p) }}} 
        }
        
    }
	\textbf{end subroutine}
 \end{multicols}
    \vspace{1mm}
    \caption{\small Efficient forward and backward sweeps.}
        \label{algo-forward-backward-sweeps-efficient}
\end{algorithm}
 \begin{algorithm}[H]
 \begin{multicols}{2}
    \footnotesize
    \DontPrintSemicolon
    \SetAlgoLined
    \vspace{1mm}
    \SetKwFunction{FForward}{LUSGS\_forward\_sweep\_cuda()}
    \SetKwFunction{FBackward}{LUSGS\_backward\_sweep\_cuda()}
    \SetKwProg{Fn}{subroutine}{}{}
    
    \Fn{\FForward}
    {
        \For{$i \leftarrow 1$ \KwTo {\tt point\_colors}}
        {
            $n_p = {\tt points\_with\_color(i)}$ \\
            \textbf{kernel} $<<<$ grid, block $>>>$  {\tt forward\_sweep\_cuda($n_p$)} \\
        }
    }
    \textbf{end subroutine}
    
    \Fn{\FBackward}
    {
        \For{$i \leftarrow {\tt point\_colors}$ \KwTo $1$ }
        {
            $n_p = {\tt points\_with\_color(i)}$ \\
            \textbf{kernel} $<<<$ grid, block $>>>$  {\tt backward\_sweep\_cuda($n_p$)} \\
        }
    }
    \textbf{end subroutine}
    \end{multicols}
    \vspace{1mm}
        \caption{CUDA kernels for efficient forward and backward sweeps.}
    \label{algo-cuda-kernels-efficient-forward-backward-sweeps}
\end{algorithm}
\section{Numerical Results}
\label{sec-num-results}
In this section, we illustrate the performance of the GPU accelerated implicit meshfree LSKUM solvers with exact computation of the flux Jacobian conserved vector products in LU-SGS algorithms presented in Section \ref{sec-implicit-lusgs-lskum}. The test cases considered are the inviscid subsonic, transonic, and supersonic flows over the NACA 0012 airfoil and the subsonic flow over a three-element airfoil. For residual comparisons, the following GPU solvers are considered. 
    \begin{enumerate}[label=(\alph*)]
    \item Implicit GPU solver based on the LU-SGS scheme in eq. (\ref{lusgs-revised-forward-reverse-sweeps}) with incremental flux approximations for the flux Jacobian conserved vector products. We refer to it as the implicit solver based on Manish et al.'s approach. 
    \item Implicit GPU solver based on the LU-SGS scheme in eq. (\ref{lusgs-revised-forward-reverse-sweeps}) with the exact computation of the flux Jacobian conserved vector products. We denote this approach as the implicit solver based on Manish et al.+AD. 
   \item Implicit GPU solver based on the LU-SGS scheme in eq. (\ref{lusgs-initial-forward-reverse-sweeps}) with incremental flux approximations for the flux Jacobian conserved vector products. We refer to it as the implicit solver based on Anandhanarayanan et al.'s approach.
   \item Implicit GPU solver based on the LU-SGS scheme in eq. (\ref{lusgs-initial-forward-reverse-sweeps}) with the exact computation of the flux Jacobian conserved vector products. We refer to this approach as Anandhanarayanan et al.+AD.
    \item Explicit GPU LSKUM solver. 
\end{enumerate}
All the GPU solvers are written in CUDA {\tt Fortran}. These codes are compiled with \texttt{nvcc 23.3} using the flags: \texttt{-O3} and \texttt{-mcmodel=large} \cite{cuda}. Table \ref{workstation-config} shows the hardware configuration of the GPU node used for the simulations. 
\begin{table}[htbp]
\centering
\begin{tabular}{lcc}
\toprule
& CPU   & GPU   \\
\midrule
Model & AMD EPYC\textsuperscript{TM} $ 7532 $   & Nvidia Ampere $A100$ SXM4  \\ [0.3em]
Cores & $64$ $\left( 2 \times 32 \right)$ & $5120$ \\ [0.3em]
Core Frequency & $2.40$ GHz & $1.23$ GHz \\ [0.3em]
Global Memory & $1$ TiB & $80$ GiB \\ [0.3em]
$L2$ Cache & $16$ MiB & $40$ MiB \\ [0.1em]
%
\bottomrule
\end{tabular}
\caption{Hardware configuration used to perform numerical simulations. }
\label{workstation-config}
\end{table}
\subsection{Subsonic flow over the NACA 0012 airfoil}
\begin{figure}[t]
\centering
    \subfloat[\centering Coarse distribution. ] {{\includegraphics[width=0.48\textwidth,trim={5mm 0mm 5mm 5mm},angle=0,clip]{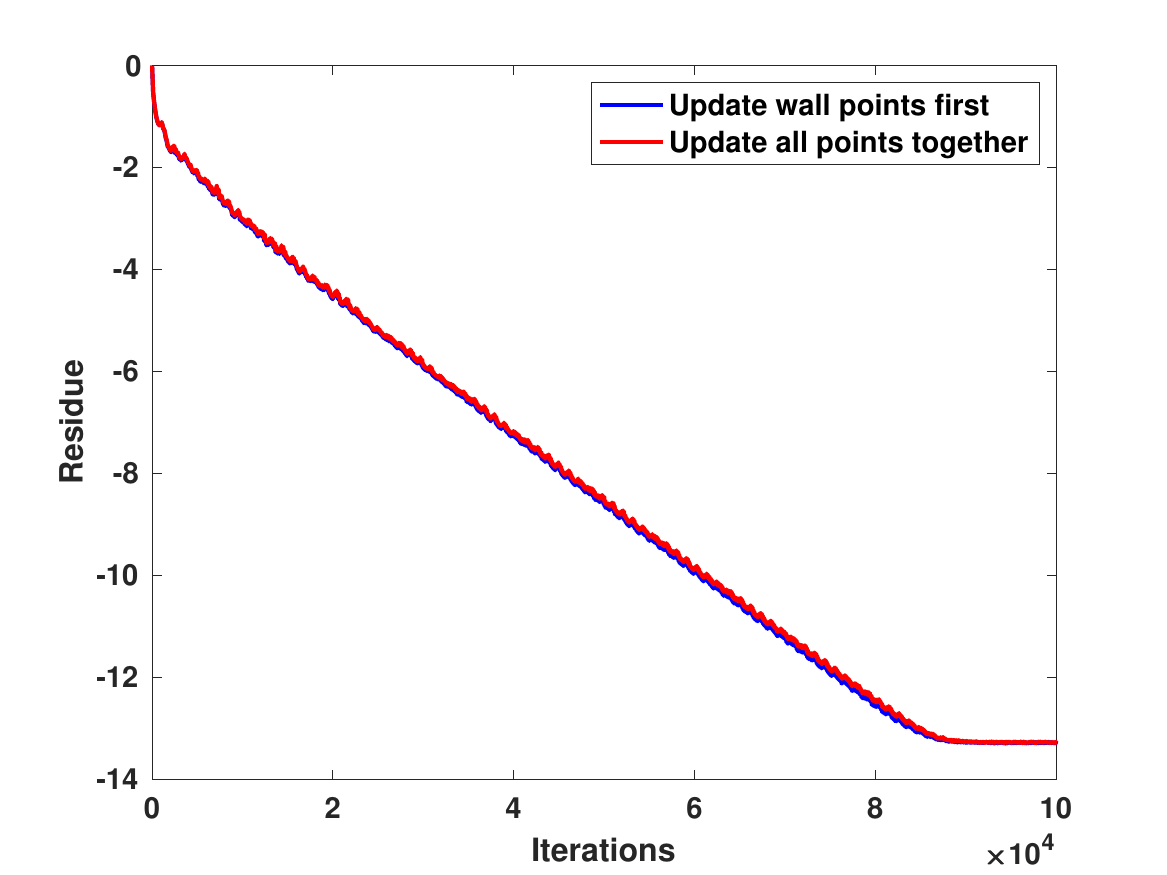}} }\label{subsonic-naca-0012-coarse-distribution-wall-first-and-update-all-residues}
        \subfloat[\centering Fine distribution. ]{{\includegraphics[width=0.48\textwidth,trim={5mm 0mm 5mm 5mm},angle=0,clip]{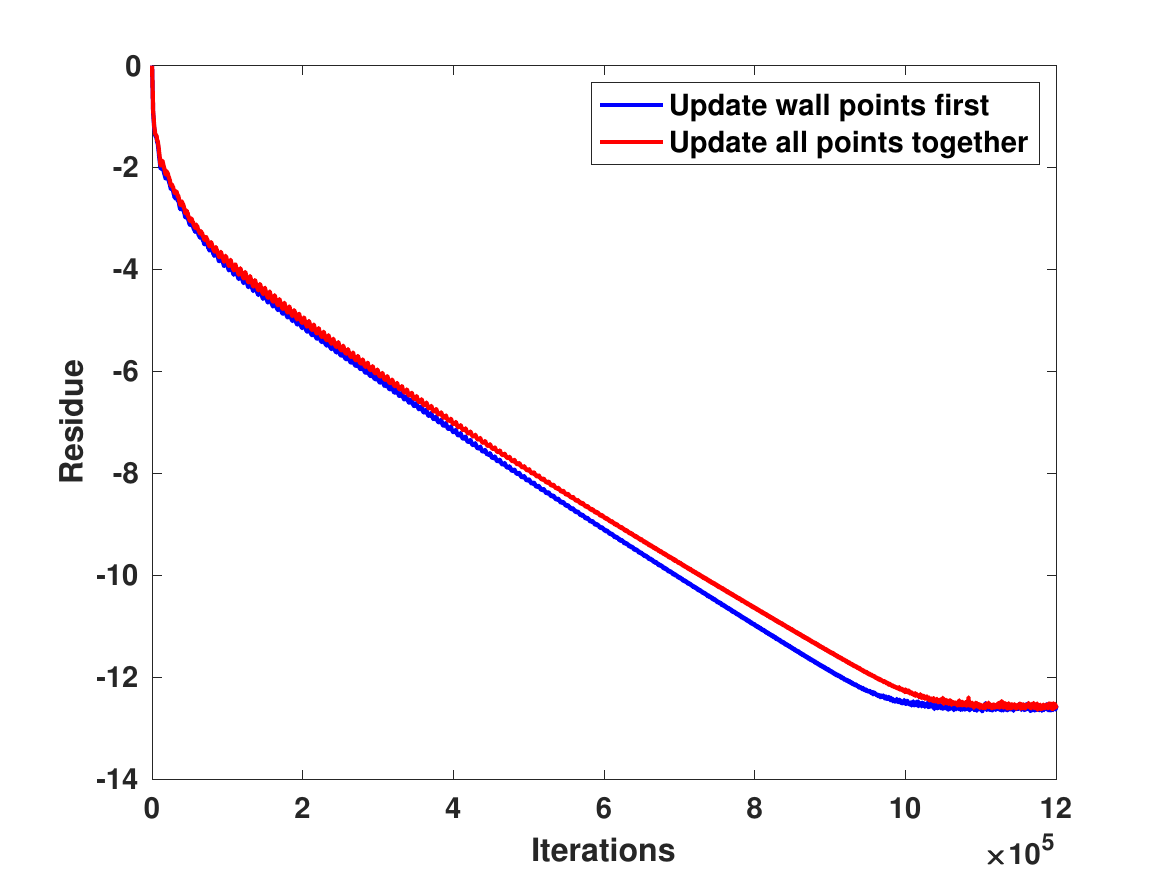}} }\label{subsonic-naca-0012-fine-distribution-wall-first-and-update-all-residues} 
\caption{Subsonic flow over the NACA 0012 airfoil at $M_{\infty} = 0.63$ and $AoA = 2^o$. Comparison of residues based on the implicit GPU solvers with the forward and reverse sweep strategies in Algorithm \ref{algo-forward-backward-sweeps-modified} and \ref{algo-forward-backward-sweeps-efficient}.}
\label{subsonic-naca-0012-coarse-fine-distributions-wall-first-and-update-all-residues}
 \end{figure}

\begin{figure}[htbp]
\centering
    \subfloat[\centering  ]{{\includegraphics[width=0.48\textwidth,trim={5mm 0mm 5mm 5mm},angle=0,clip]{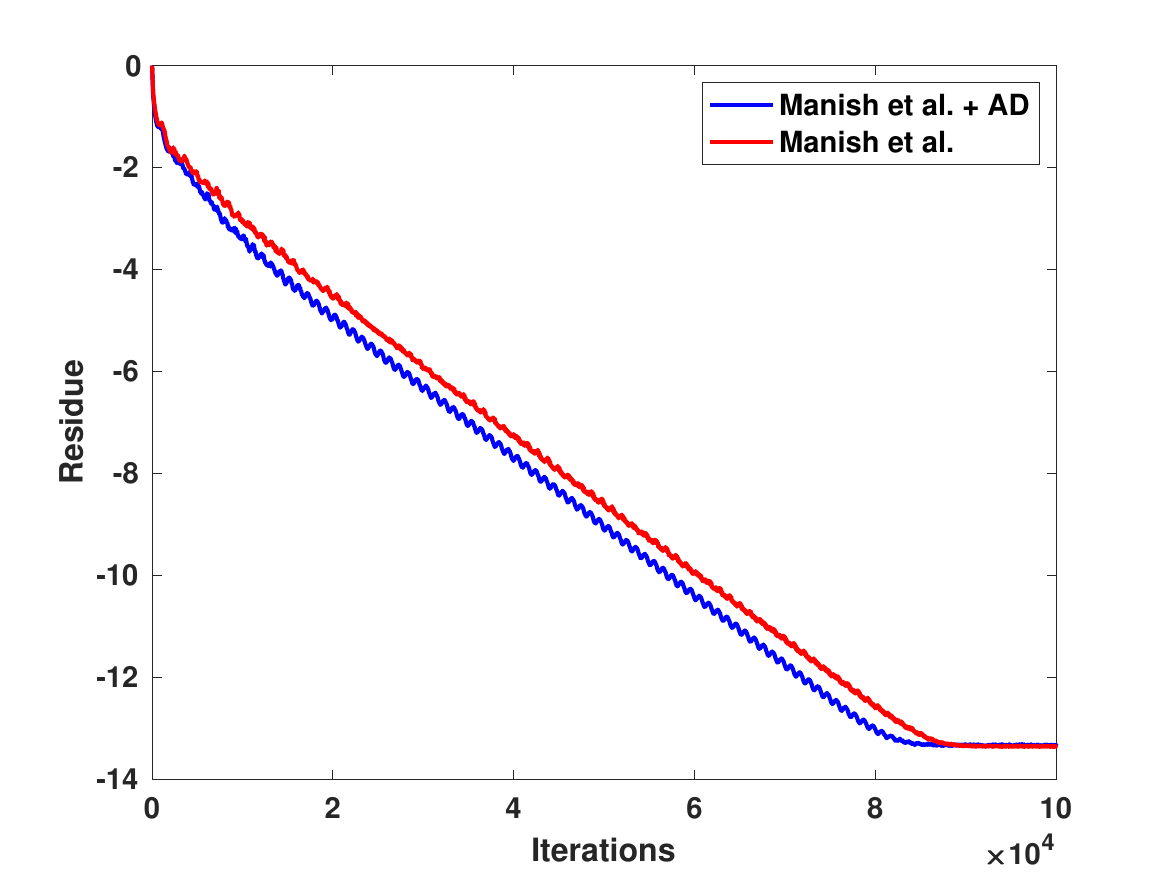} }\label{subsonic-naca-0012-coarse-distribution-residues-plot-a}}
            \subfloat[\centering ]{{\includegraphics[width=0.48\textwidth,trim={5mm 0mm 5mm 5mm},angle=0,clip]{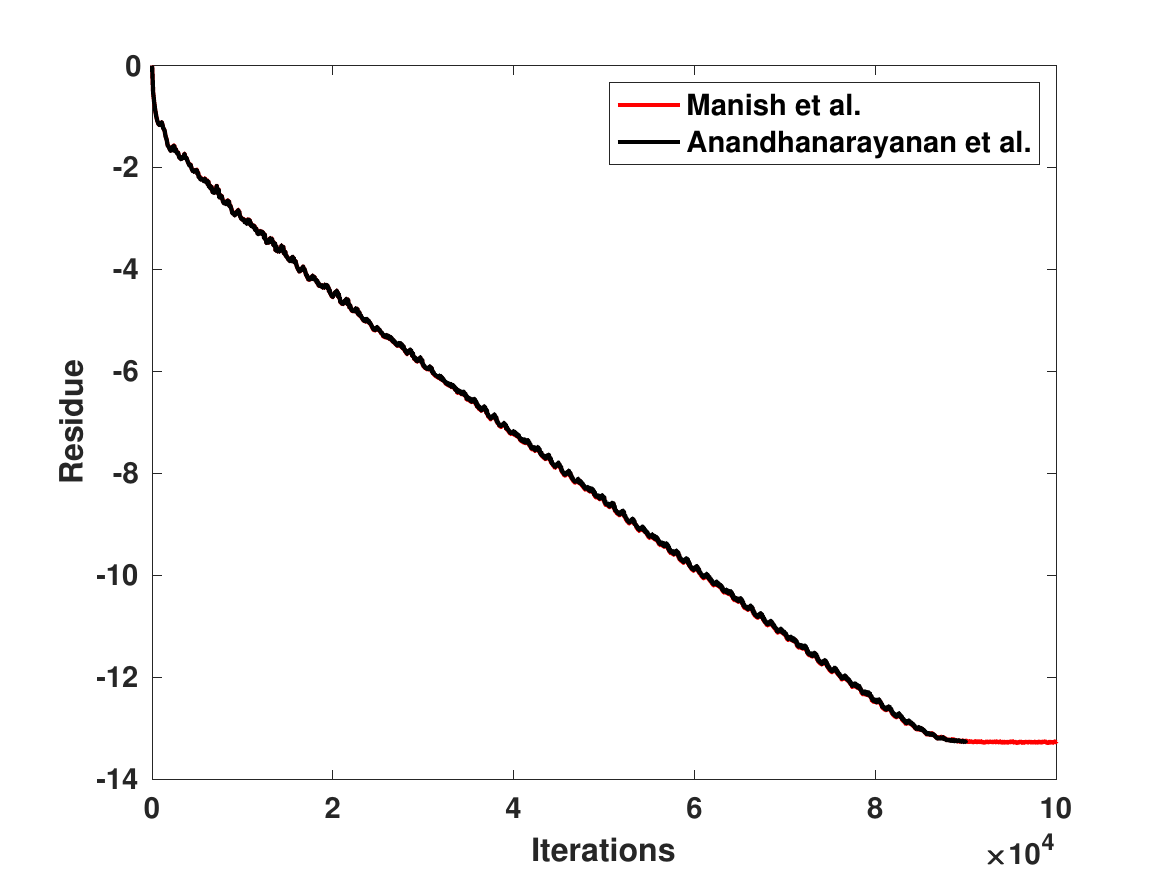} }\label{subsonic-naca-0012-coarse-distribution-residues-plot-b}} \\
         \subfloat[\centering ]{{\includegraphics[width=0.48\textwidth,trim={5mm 0mm 5mm 5mm},angle=0,clip]{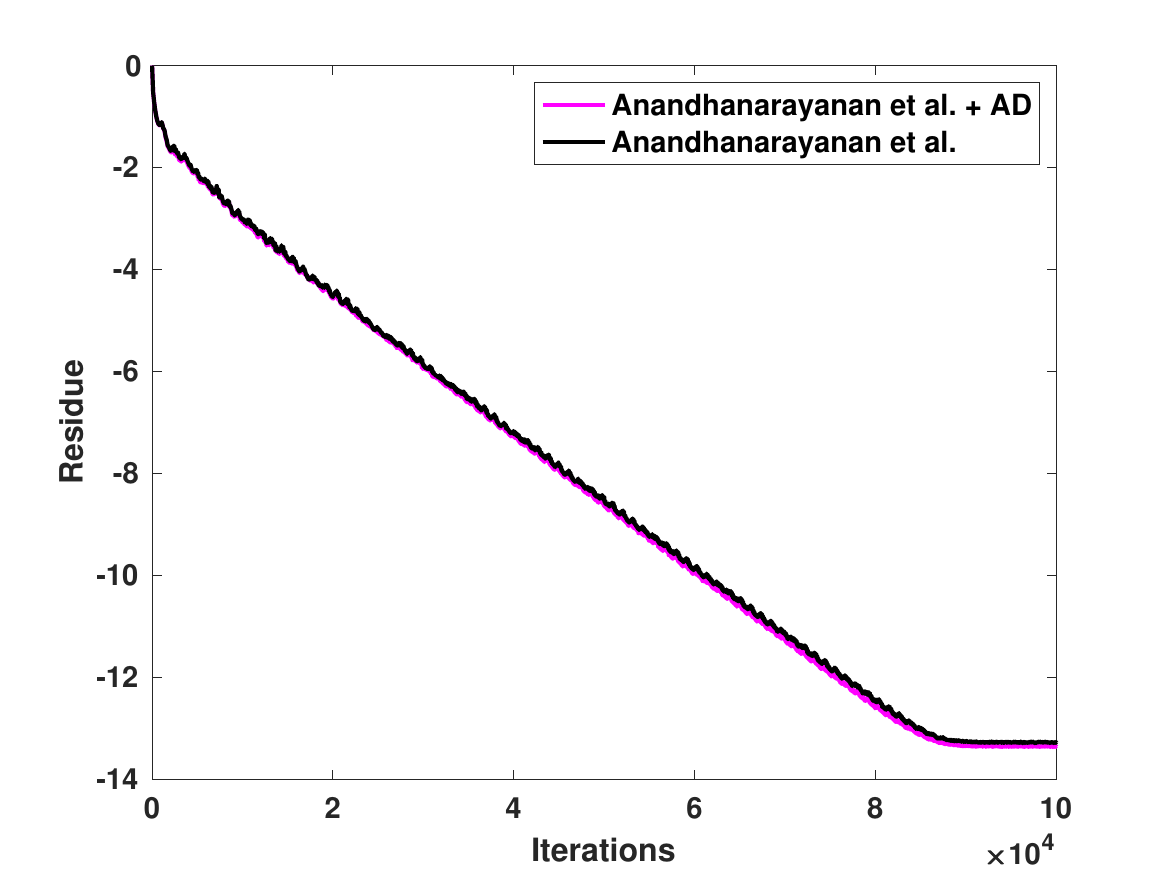} }
         \label{subsonic-naca-0012-coarse-distribution-residues-plot-c}}
        \subfloat[\centering ]{{\includegraphics[width=0.48\textwidth,trim={5mm 0mm 5mm 5mm},angle=0,clip]{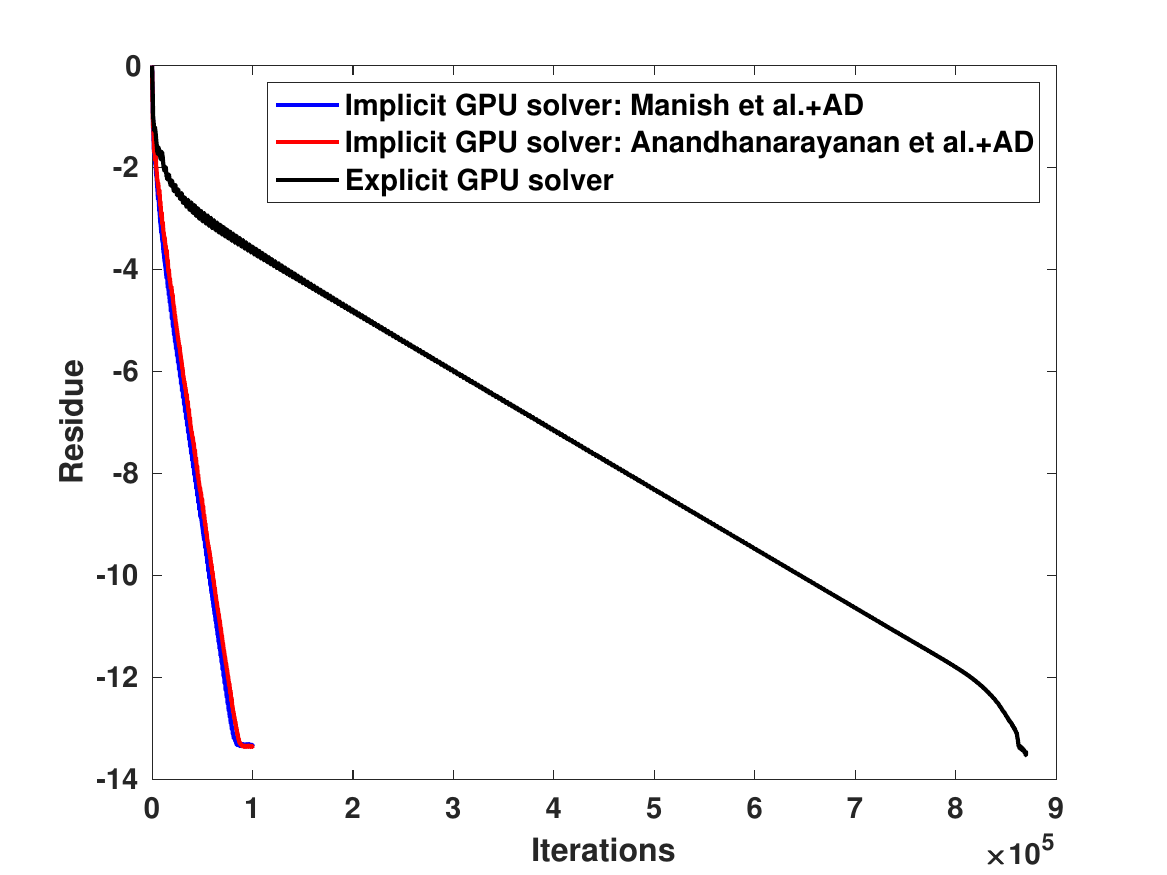} }\label{subsonic-naca-0012-coarse-distribution-residues-plot-d}}
\caption{Subsonic flow over the NACA 0012 airfoil at $M_{\infty} = 0.63$ and $AoA = 2^o$. Comparison of residues on the coarse distribution with $38,400$ points. }
\label{subsonic-naca-0012-coarse-distribution-residues}
 \end{figure}

\begin{figure}[htbp]
\centering
    \subfloat[\centering  ]{{\includegraphics[width=0.48\textwidth,trim={5mm 0mm 5mm 5mm},angle=0,clip]{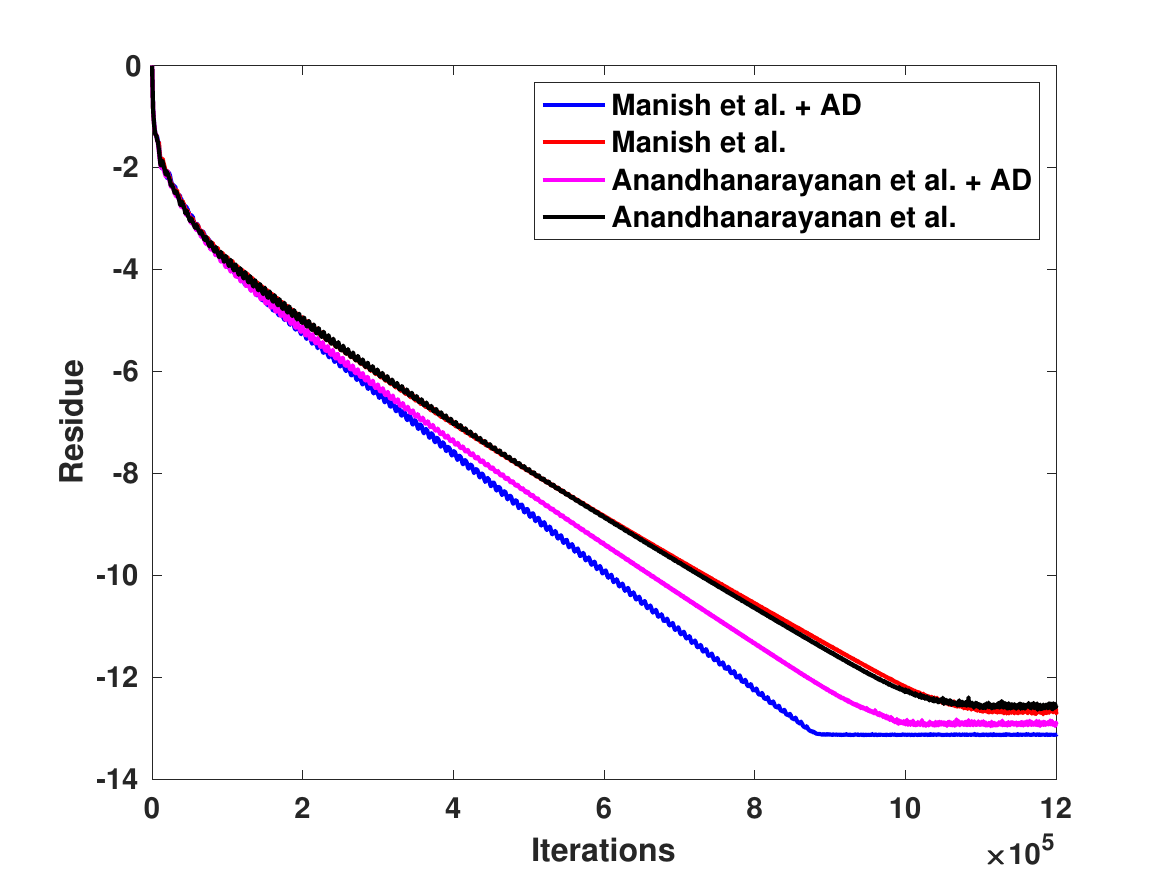} }\label{subsonic-naca-0012-fine-distribution-residues-plot-a}}
        \subfloat[\centering ]{{\includegraphics[width=0.48\textwidth,trim={5mm 0mm 5mm 5mm},angle=0,clip]{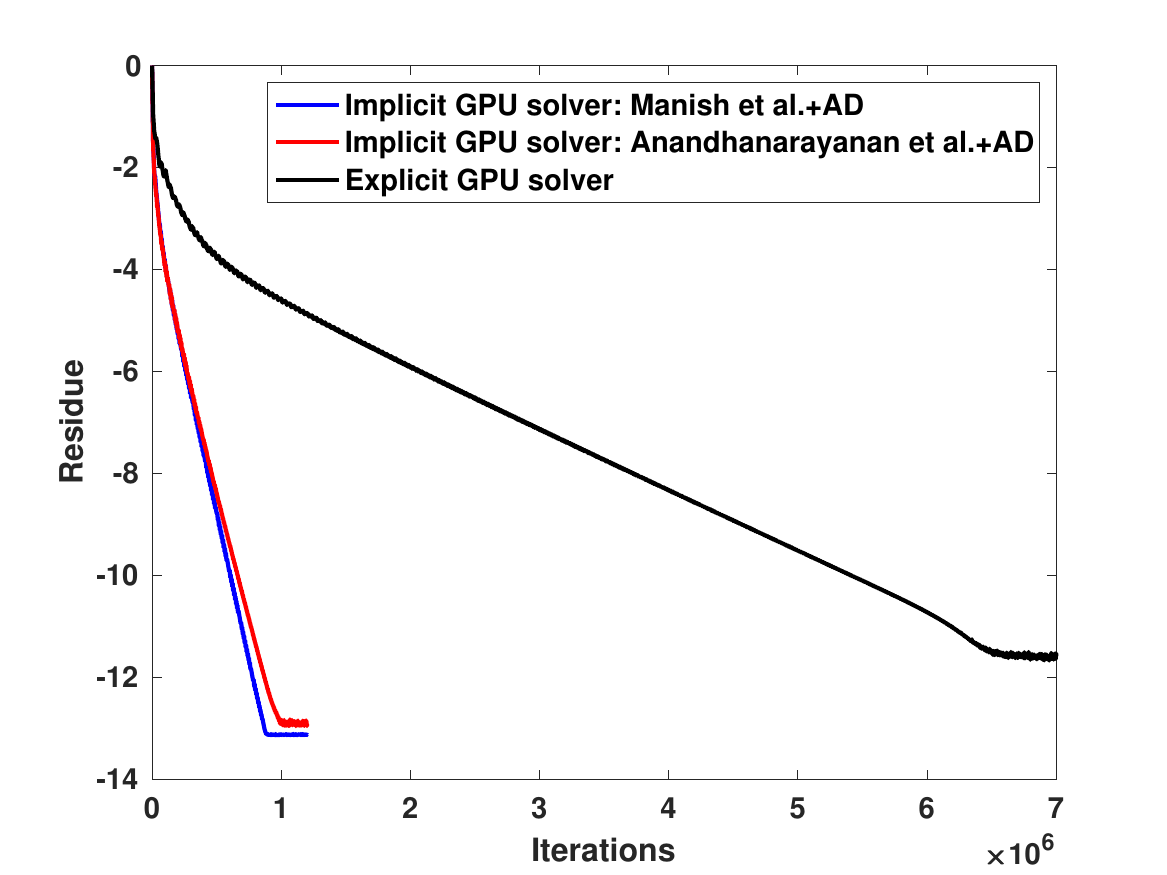} }\label{subsonic-naca-0012-fine-distribution-residues-plot-b}} 
\caption{Subsonic flow over the NACA 0012 airfoil at $M_{\infty} = 0.63$ and $AoA = 2^o$. Comparison of residues on the fine distribution with $1,228,800$ points. }
\label{subsonic-naca-0012-fine-distribution-residues}
 \end{figure}

\begin{figure}[htbp]
\centering
    \subfloat[\centering  Surface pressure distribution]{{\includegraphics[width=0.48\textwidth,trim={5mm 0mm 5mm 5mm},angle=0,clip]{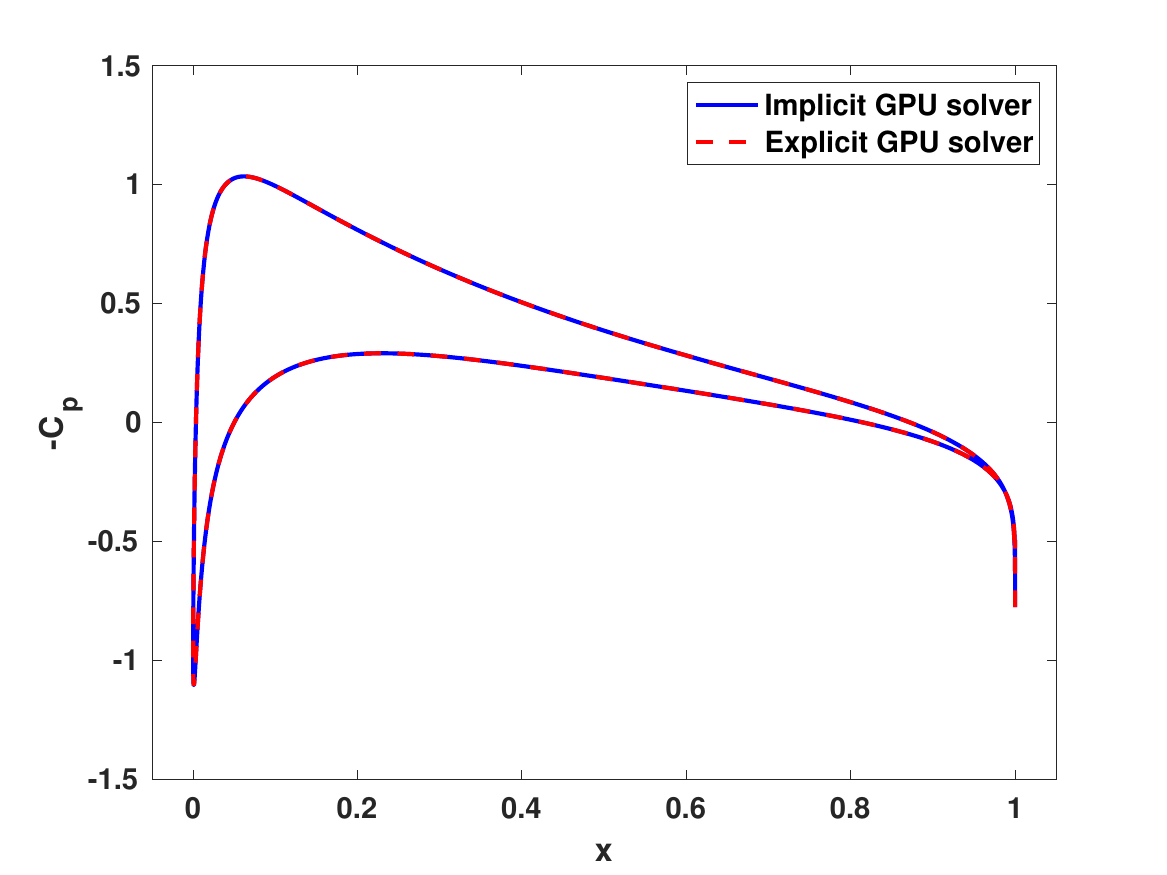}} \label{subsonic-naca-0012-fine-distribution-cp-plot}}
        \subfloat[\centering Mach contours]{{\includegraphics[width=0.48\textwidth,trim={15mm 50mm 0mm 50mm},angle=0,clip]{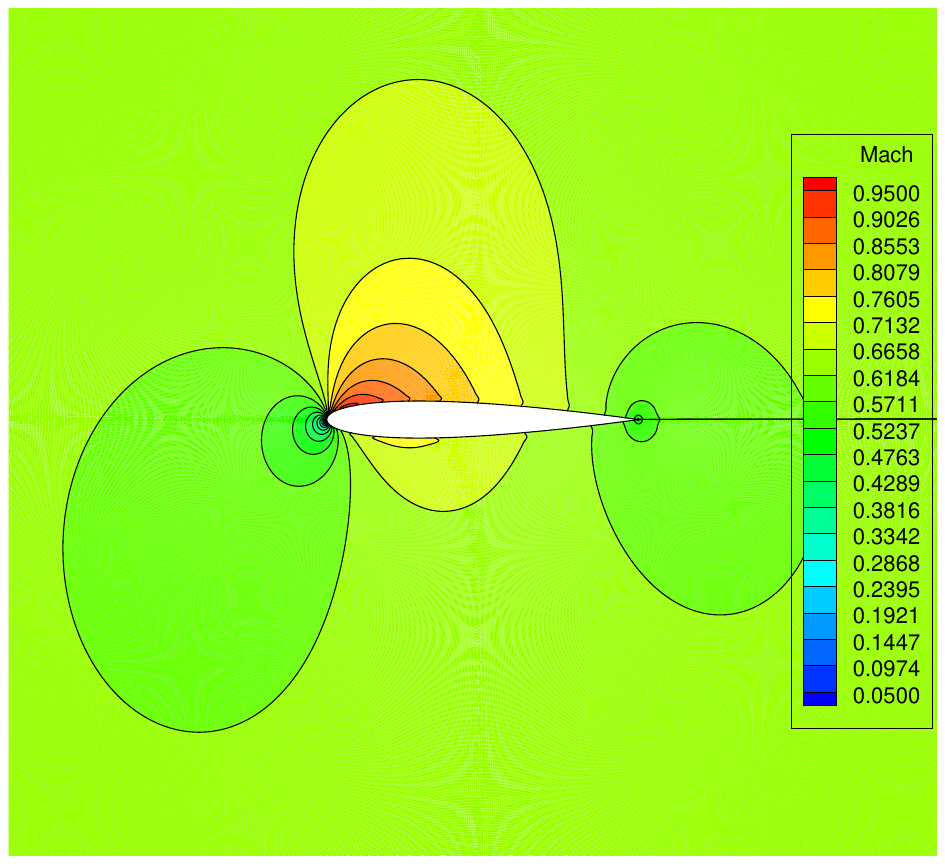}} \label{subsonic-naca-0012-fine-distribution-mach-contours} }
\caption{Subsonic flow over the NACA 0012 airfoil at $M_{\infty} = 0.63$ and $AoA = 2^o$. Surface pressure distribution and Mach contours on the fine distribution with $1,228,800$ points. }
\label{subsonic-naca-0012-fine-distribution-cp-mach-contours}
 \end{figure}

The test case under investigation is the subsonic flow over the NACA 0012 airfoil with a freestream Mach number, $M_{\infty}=0.63$ and angle of attack, $AoA=2^{o}$. Numerical simulations are performed on coarse and very fine point clouds, with $38,400$ and $1,228,800$ points. \\ \\ 
We first investigate the performance of the forward and reverse sweeps strategies presented in Algorithms \ref{algo-forward-backward-sweeps-modified} and \ref{algo-forward-backward-sweeps-efficient}. For this purpose, these algorithms are integrated into the implicit GPU solver based on Anandhanarayanan et al.'s approach. Note that in the forward and backward sweeps of Algorithm \ref{algo-forward-backward-sweeps-modified}, the wall points with a given color are first updated, followed by interior and outer boundary points. On the other hand, in Algorithm \ref{algo-forward-backward-sweeps-efficient}, all points with the same color are updated together. Figure \ref{subsonic-naca-0012-coarse-fine-distributions-wall-first-and-update-all-residues} shows the density residual history of the implicit GPU solvers on the coarse and fine point distributions. We observe that the implicit solver with wall points first update in the LU-SGS sweeps yields a slightly faster rate of convergence compared to the solver with all points updated together. However, numerical experiments have shown that the run-time per iteration of the implicit GPU solver with Algorithm \ref{algo-forward-backward-sweeps-modified} is found to be higher compared to the solver using Algorithm \ref{algo-forward-backward-sweeps-efficient}. On the coarse distribution, the run-time is increased by a factor of $1.61$ times, while on the fine distribution, it is increased by $1.12$ times. \\ \\
This increase in runtime can be attributed to two main factors, one of which is the overhead costs. In executing the implicit GPU solver, Algorithm \ref{algo-forward-backward-sweeps-modified} has six {\tt i-loops} in forward and backward sweeps together compared to two {\tt i-loops} in Algorithm \ref{algo-forward-backward-sweeps-efficient}. Correspondingly, Algorithm \ref{algo-forward-backward-sweeps-modified} requires six small {\tt CUDA} kernels while Algorithm \ref{algo-forward-backward-sweeps-efficient} uses only two kernels. Launching a {\tt CUDA} kernel incurs overhead, which is notably higher than calling the same function on a CPU. It is important to highlight that the overhead costs of loop execution and kernel launches are fixed and do not depend on the point cloud size. \\ \\ 
Another source for an increase in the run-time is the number of points with a given color. Table \ref{coarse-fine-point-color-groups} shows the color groups and the number of points in each group for both the coarse and fine distributions. Note that the number of wall, interior, or outer boundary points with a given color is less than the total number of points with the same color. On the coarse distribution, the number of points in each color group of Algorithm \ref{algo-forward-backward-sweeps-modified} is small. Due to this, the implicit solver could not fully utilise the available GPU resources. Furthermore, the overhead costs are also significant compared to the cost of floating point operations within a kernel. Both these factors resulted in a higher run-time factor on the coarse distribution. However, on the fine distribution, with more points in each color group, the overheads are relatively smaller than the floating point operations cost. The efficient utilisation of GPU resources on the fine distribution reduced the run-time factor significantly. Given these findings, the computational efficiency of Algorithm \ref{algo-forward-backward-sweeps-efficient} makes it the preferred choice for implementation in all versions of the implicit GPU meshfree LSKUM solver. \\ \\
%
%
\begin{table}[t]
\centering
\begin{center}
\scalebox{0.85}{
\begin{tabular}{lccccccccc}
\toprule
& \multicolumn{7}{c}{\centering{Color group}}  & \\ [0.2em]
\cline{2-9}
 & $1$ & $2$ & $3$ & $4$ & $5$  & $6$ & $7$ & $8$ \\[0.2em]
\midrule
Level & \multicolumn{7}{c}{\centering{Number of points in each color group}}  & \\ [0.2em]
\cline{1-9}
Coarse & $9441$ & $9383$ & $9441$ & $9382$ & $433$ & $316$ & $3$ & $1$\\[0.5em]
Fine & $303491$ & $302393$ & $300101$ & $296968$ & $22526$ & $2912$ & $396$ & $13$ \\[0.5em]
\bottomrule
\end{tabular}
}
\end{center}
\caption{{\small Number of points in each color group for the coarse and fine point distributions.  }}
\label{coarse-fine-point-color-groups}
\end{table} 
Figures \ref{subsonic-naca-0012-coarse-distribution-residues} and \ref{subsonic-naca-0012-fine-distribution-residues} compare the residual histories for all the four versions of the implicit GPU solver and the explicit GPU solver. Figures \ref{subsonic-naca-0012-coarse-distribution-residues-plot-a} and \ref{subsonic-naca-0012-fine-distribution-residues-plot-a} show that the rate of residue fall is higher for the implicit GPU solver based on Manish et al.+AD approach. From Figures \ref{subsonic-naca-0012-coarse-distribution-residues-plot-b} and \ref{subsonic-naca-0012-fine-distribution-residues-plot-a}, we observe that the LU-SGS algorithm of Manish et al. do not affect the rate of convergence as the residue plots are almost the same as those from Anandhanarayanan et al.'s approach. Figure \ref{subsonic-naca-0012-coarse-distribution-residues-plot-c} shows that the Anandhanarayanan et al.+AD approach has a very minimal effect on the convergence rate on the coarse distribution. From Figure \ref{subsonic-naca-0012-coarse-distribution-residues-plot-a}, convergence acceleration on the coarse distribution can be attributed to the diagonal dominance induced by the exact computation of the flux Jacobian conserved vector products in the LU-SGS algorithm of eq. (\ref{lusgs-revised-forward-reverse-sweeps}). Figure \ref{subsonic-naca-0012-fine-distribution-residues-plot-a} shows the exact computation of Jacobian vector products in eqs. (\ref{lusgs-initial-forward-reverse-sweeps}) and (\ref{lusgs-revised-forward-reverse-sweeps}) has significantly improved the convergence rate on the fine distribution. Figures \ref{subsonic-naca-0012-coarse-distribution-residues-plot-d} and \ref{subsonic-naca-0012-fine-distribution-residues-plot-b} show that a very good speedup in convergence is achieved using the implicit GPU solver over the explicit solver. \\ \\
Figure \ref{subsonic-naca-0012-fine-distribution-cp-mach-contours} shows the surface pressure distribution $\left(C_p\right)$ and the Mach contours on the finest distribution. The suction peak in the $C_p$ distribution is accurately captured, and the Mach contours are smooth. Furthermore, the $C_p$ plots for the implicit and explicit GPU solvers are identical. %
\subsection{Transonic flow over the NACA 0012 airfoil}
\label{subsec-num-results-transonic}
\begin{figure}[htbp]
\centering
    \subfloat[\centering  ]{{\includegraphics[width=0.48\textwidth,trim={5mm 0mm 5mm 5mm},angle=0,clip]{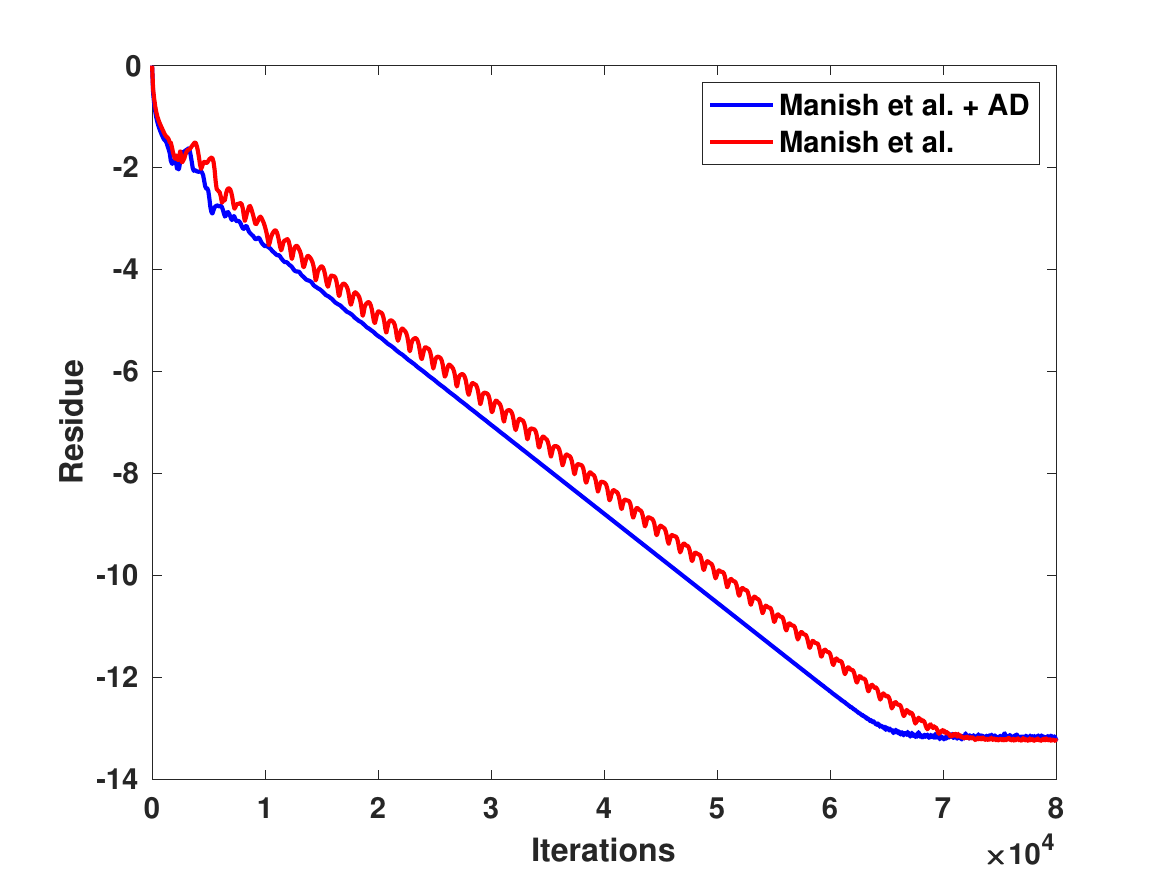} }\label{transonic-naca-0012-coarse-distribution-residues-plot-a}}
            \subfloat[\centering ]{{\includegraphics[width=0.48\textwidth,trim={5mm 0mm 5mm 5mm},angle=0,clip]{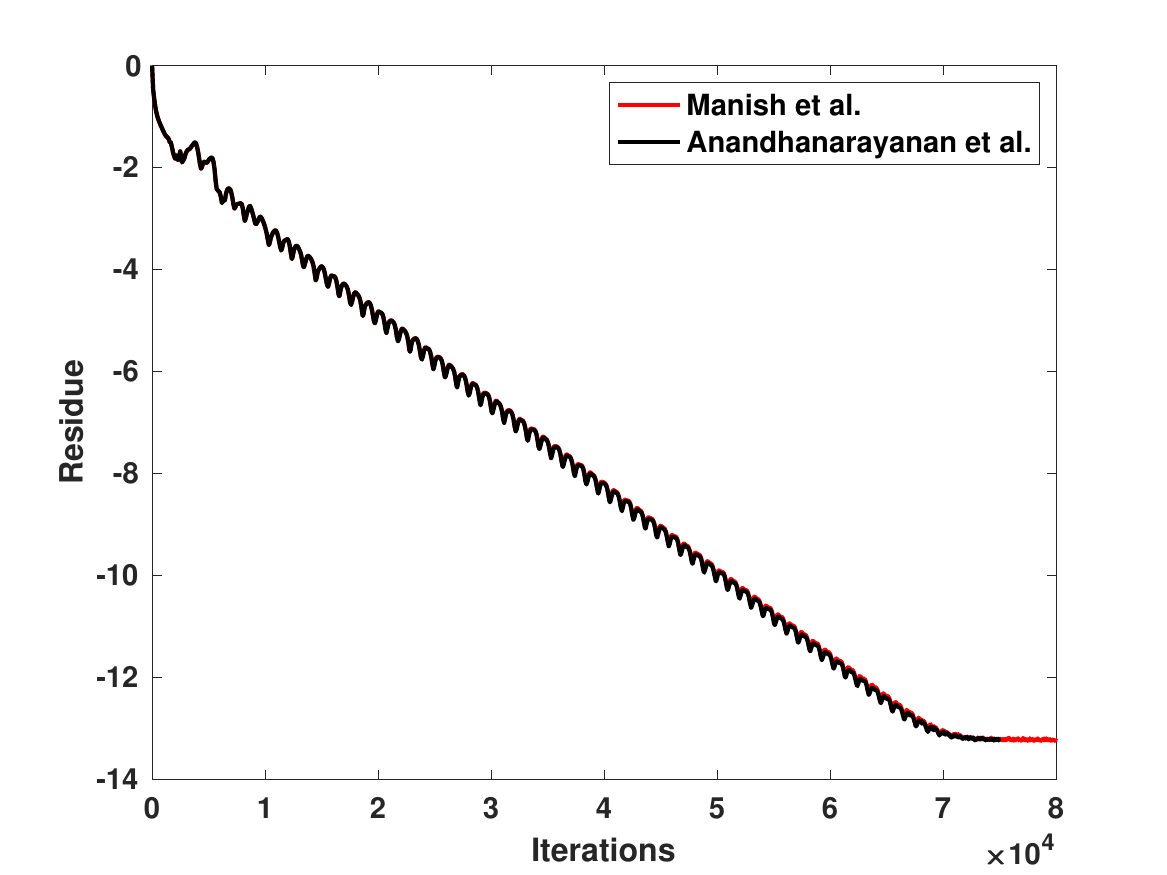} }\label{transonic-naca-0012-coarse-distribution-residues-plot-b}} \\
         \subfloat[\centering ]{{\includegraphics[width=0.48\textwidth,trim={5mm 0mm 5mm 5mm},angle=0,clip]{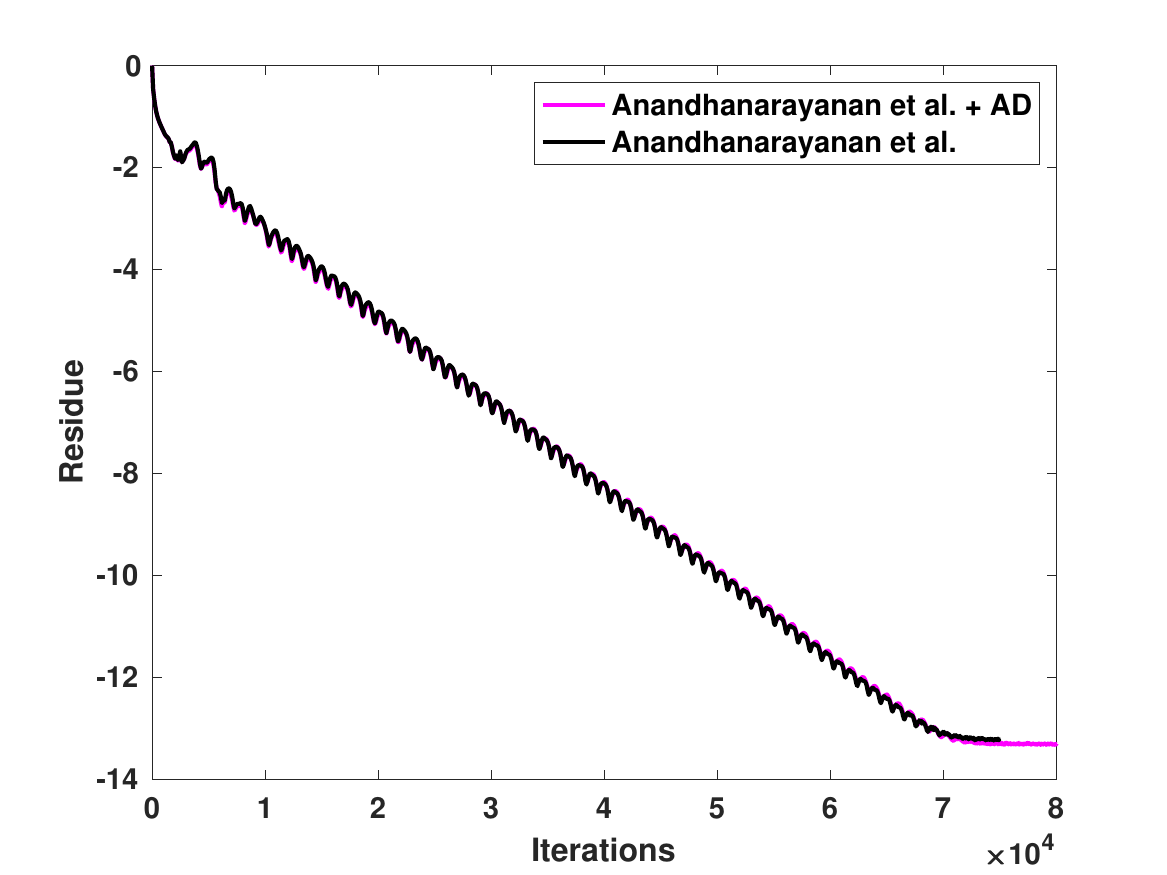} }
         \label{transonic-naca-0012-coarse-distribution-residues-plot-c}}
        \subfloat[\centering ]{{\includegraphics[width=0.48\textwidth,trim={5mm 0mm 5mm 5mm},angle=0,clip]{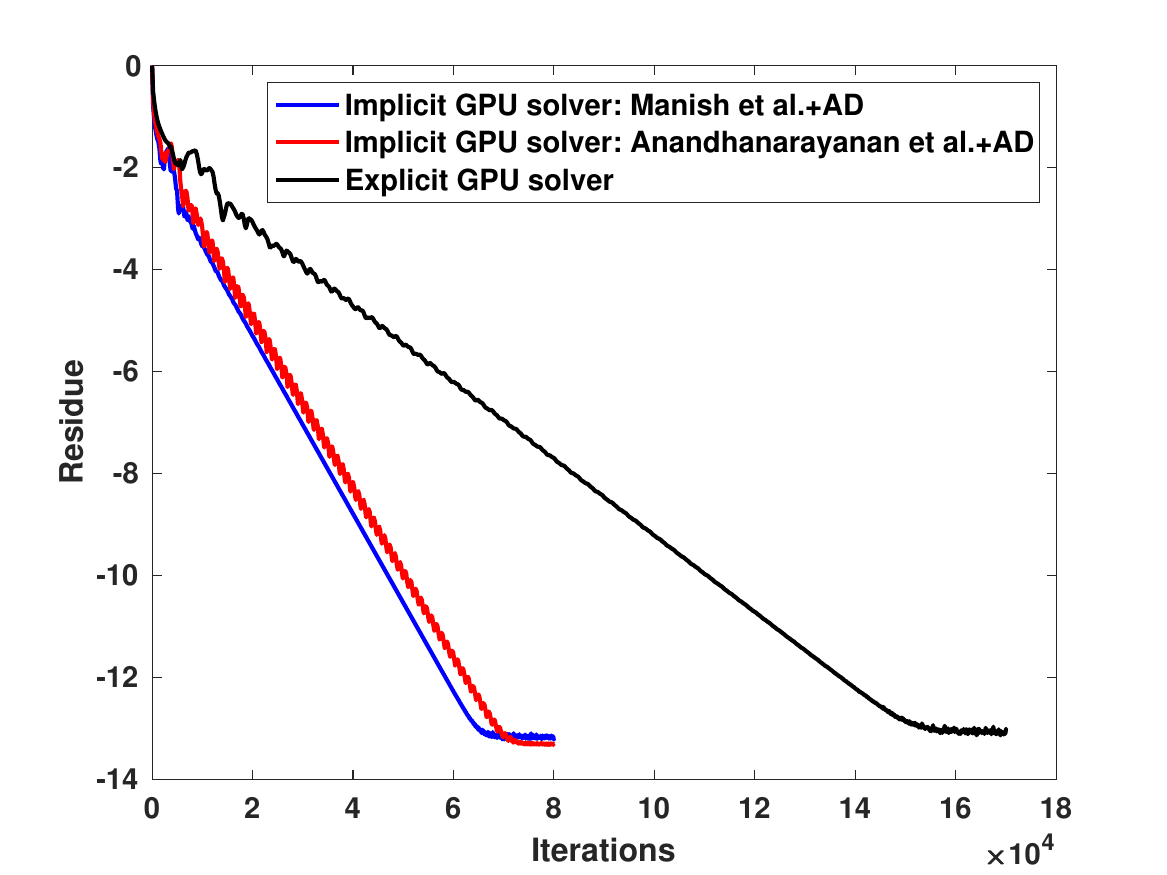} }\label{transonic-naca-0012-coarse-distribution-residues-plot-d}}
\caption{Transonic flow over the NACA 0012 airfoil at $M_{\infty} = 0.85$ and $AoA = 1^o$. Comparison of residues on the coarse distribution with $38,400$ points. }
\label{transonic-naca-0012-coarse-distribution-residues}
 \end{figure}

\begin{figure}[htbp]
\centering
    \subfloat[\centering  ]{{\includegraphics[width=0.48\textwidth,trim={5mm 0mm 5mm 5mm},angle=0,clip]{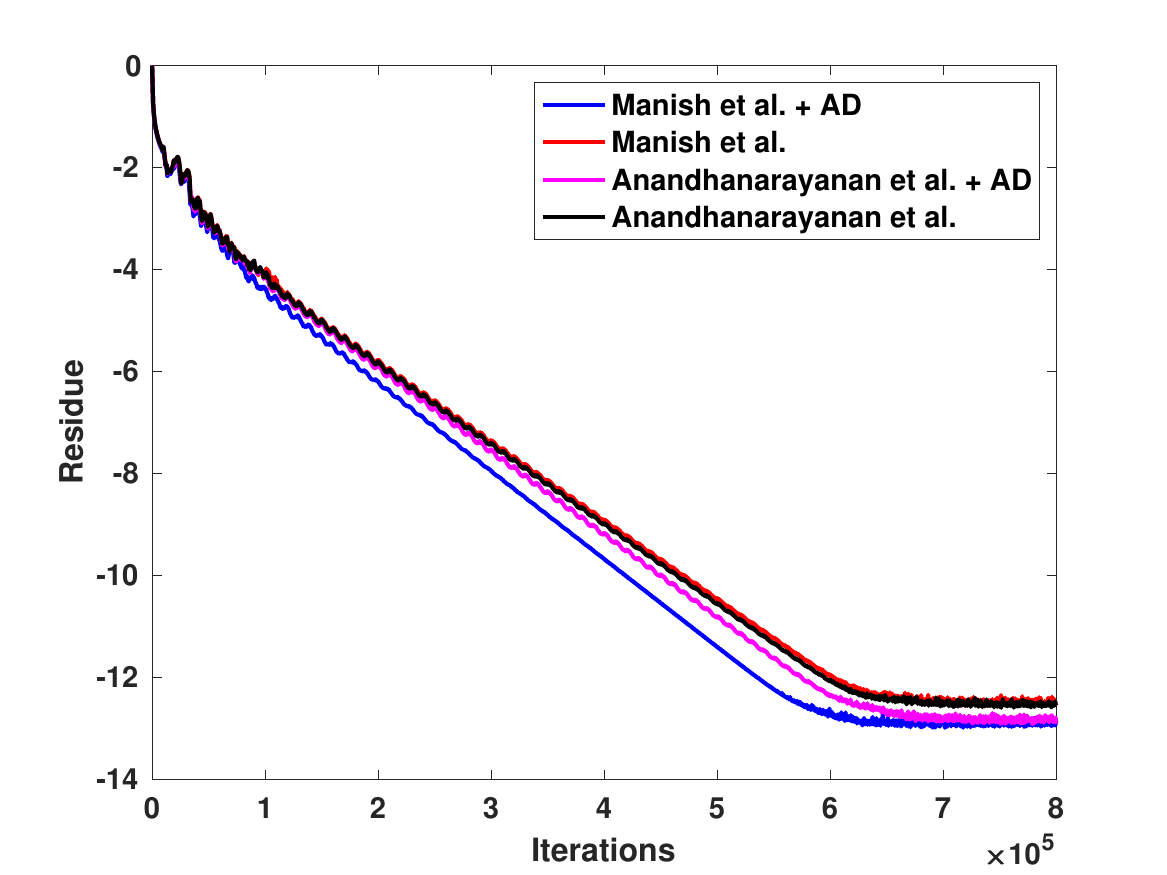} }\label{transonic-naca-0012-fine-distribution-residues-plot-a}}
        \subfloat[\centering ]{{\includegraphics[width=0.48\textwidth,trim={5mm 0mm 5mm 5mm},angle=0,clip]{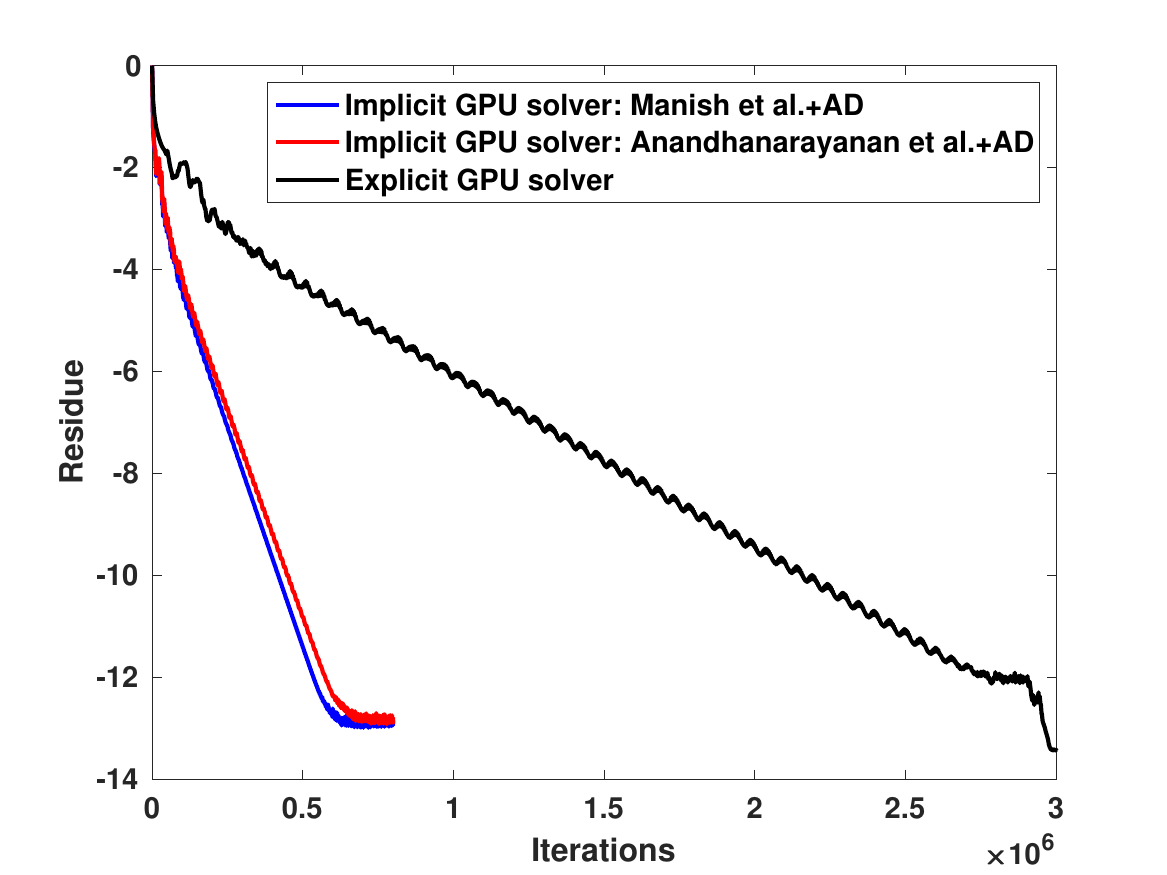} }\label{transonic-naca-0012-fine-distribution-residues-plot-b}} 
\caption{Transonic flow over the NACA 0012 airfoil at $M_{\infty} = 0.85$ and $AoA = 1^o$. Comparison of residues on the fine distribution with $1,228,800$ points. }
\label{transonic-naca-0012-fine-distribution-residues}
 \end{figure}

\begin{figure}[htbp]
\centering
    \subfloat[\centering  Surface pressure distribution]{{\includegraphics[width=0.48\textwidth,trim={5mm 0mm 5mm 5mm},angle=0,clip]{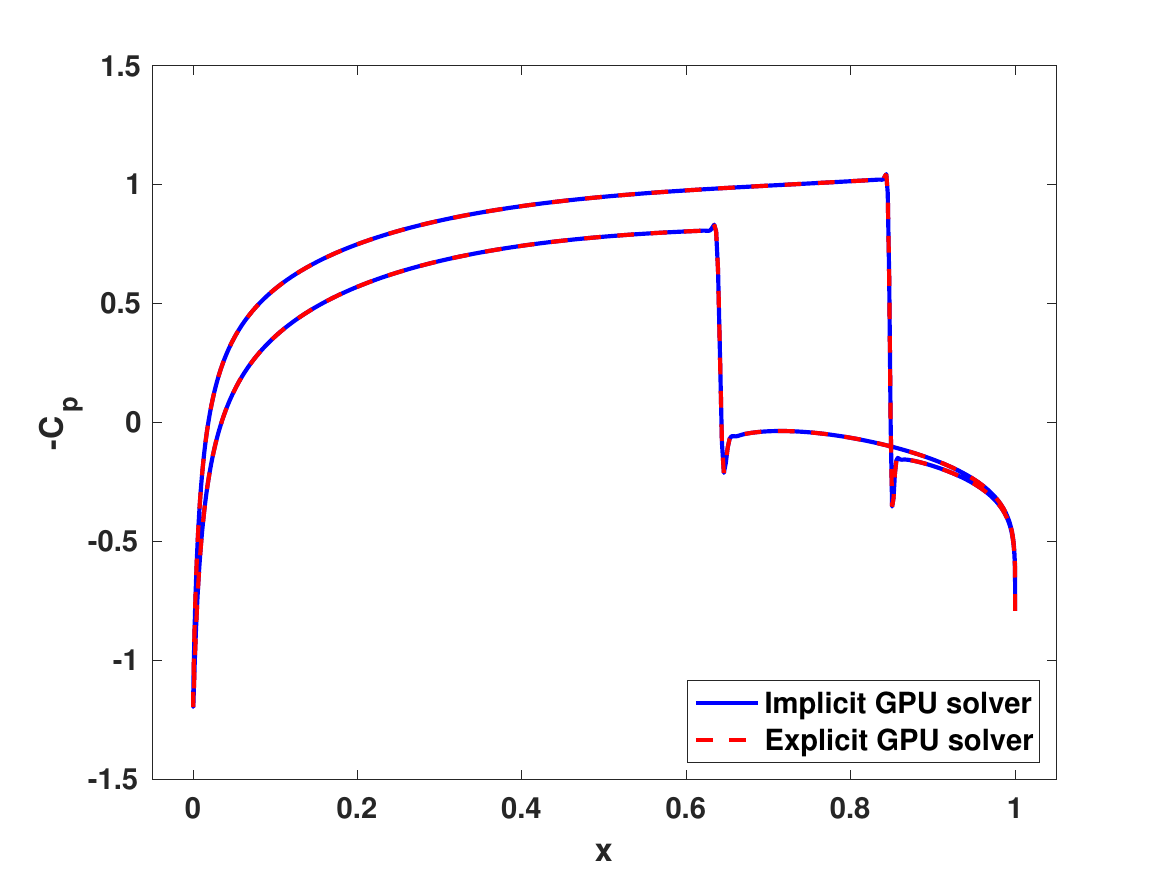}} \label{transonic-naca-0012-fine-distribution-cp-plot}}
        \subfloat[\centering Pressure contours]{{\includegraphics[width=0.48\textwidth,trim={15mm 50mm 0mm 50mm},angle=0,clip]{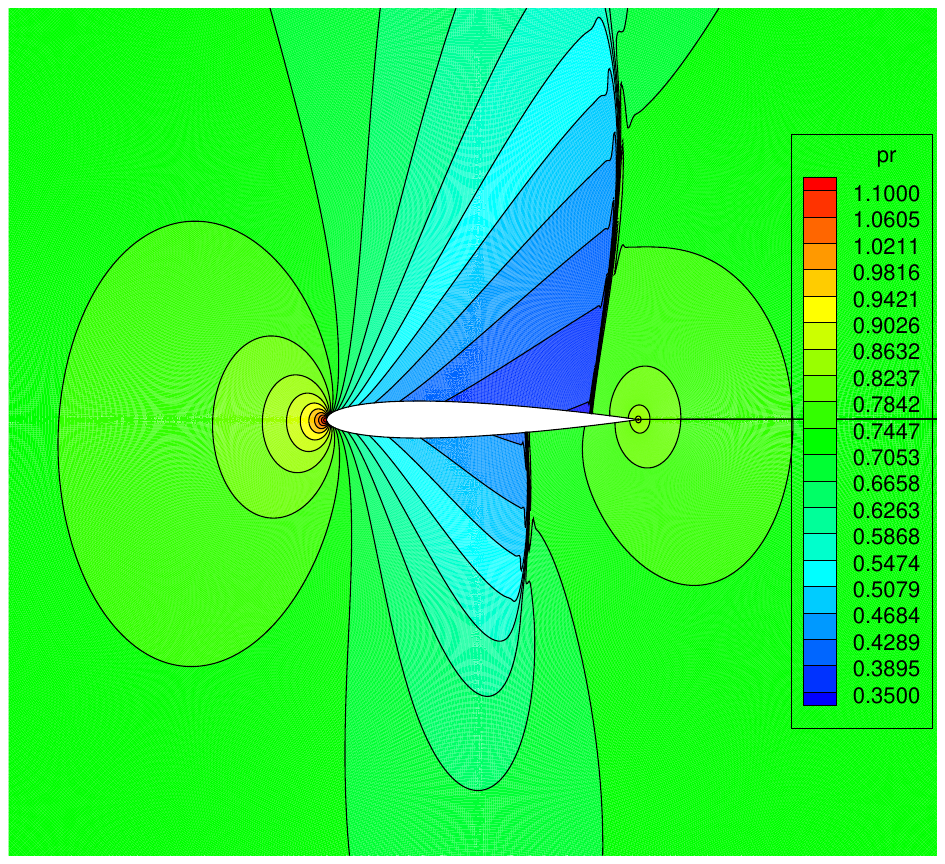}} \label{transonic-naca-0012-fine-distribution-pr-contours} }
\caption{Transonic flow over the NACA 0012 airfoil at $M_{\infty} = 0.85$ and $AoA = 1^o$. Surface pressure distribution and pressure contours on the fine distribution with $1,228,800$ points. }
\label{transonic-naca-0012-fine-distribution-cp-mach-contours}
 \end{figure}

This test case represents a transonic flow over the NACA 0012 airfoil with a freestream Mach number, $M_{\infty}= 0.85$, and angle of attack, $AoA=1^o$. The simulations are performed on coarse and fine point distributions with $38,400$ and $1,228,800$ points. \\ \\ 
Figures \ref{transonic-naca-0012-coarse-distribution-residues} and \ref{transonic-naca-0012-fine-distribution-residues} show a comparison of the residual histories using the implicit and explicit GPU solvers on the coarse and fine distributions, respectively. These plots show that the implicit GPU solver based on the Manish et al.+AD approach yields a faster convergence rate than the other variants of the implicit GPU solver. Furthermore, the residual histories based on the approaches of Manish et al. and Anandhanarayanan et al. are almost identical. Again, we observe that the exact computation of the flux Jacobian conserved vector products improved the diagonal dominance of the linear system in eq. (\ref{short-ldu-equation}), which enhanced the convergence rate. Figures \ref{transonic-naca-0012-coarse-distribution-residues-plot-d} and \ref{transonic-naca-0012-fine-distribution-residues-plot-b} show a good speedup is obtained using the implicit solver compared to the explicit solver. \\ \\
In this test case, the flow results in two shocks of different strengths, one on the suction side and the other on the pressure side of the airfoil. The surface pressure distribution in Figure \ref{transonic-naca-0012-fine-distribution-cp-plot} indicates that the discontinuities are resolved accurately. The pressure contours in Figure \ref{transonic-naca-0012-fine-distribution-pr-contours} show that the shocks are captured crisply. The computed lift and drag force coefficients are compared in Table \ref{transonic-cl-cd}. We observe that the force coefficients agree with the AGARD workshop results \cite{agard-report-AR-211}. 
\begin{table}[htbp]
\centering
\begin{tabular}{lcc}
\toprule
Force coefficient & Implicit GPU solver   & AGARD   \\
\midrule
Lift $\left(C_L\right)$  & $0.3310$ & $0.3300-0.3890$ \\ [0.3em]
Drag $\left(C_D\right)$  & $0.0545$ & $0.0464-0.0590$ \\ [0.1em]
\bottomrule
\end{tabular}
\caption{Transonic flow over the NACA 0012 airfoil at $M_{\infty} = 0.85$ and $AoA = 1^o$. Comparison of the lift and drag coefficients on the fine distribution with $1,228,800$ points. }
\label{transonic-cl-cd}
\end{table}
\subsection{Supersonic flow over the NACA 0012 airfoil}
\begin{figure}[htbp]
\centering
    \subfloat[\centering  ]{{\includegraphics[width=0.48\textwidth,trim={5mm 0mm 5mm 5mm},angle=0,clip]{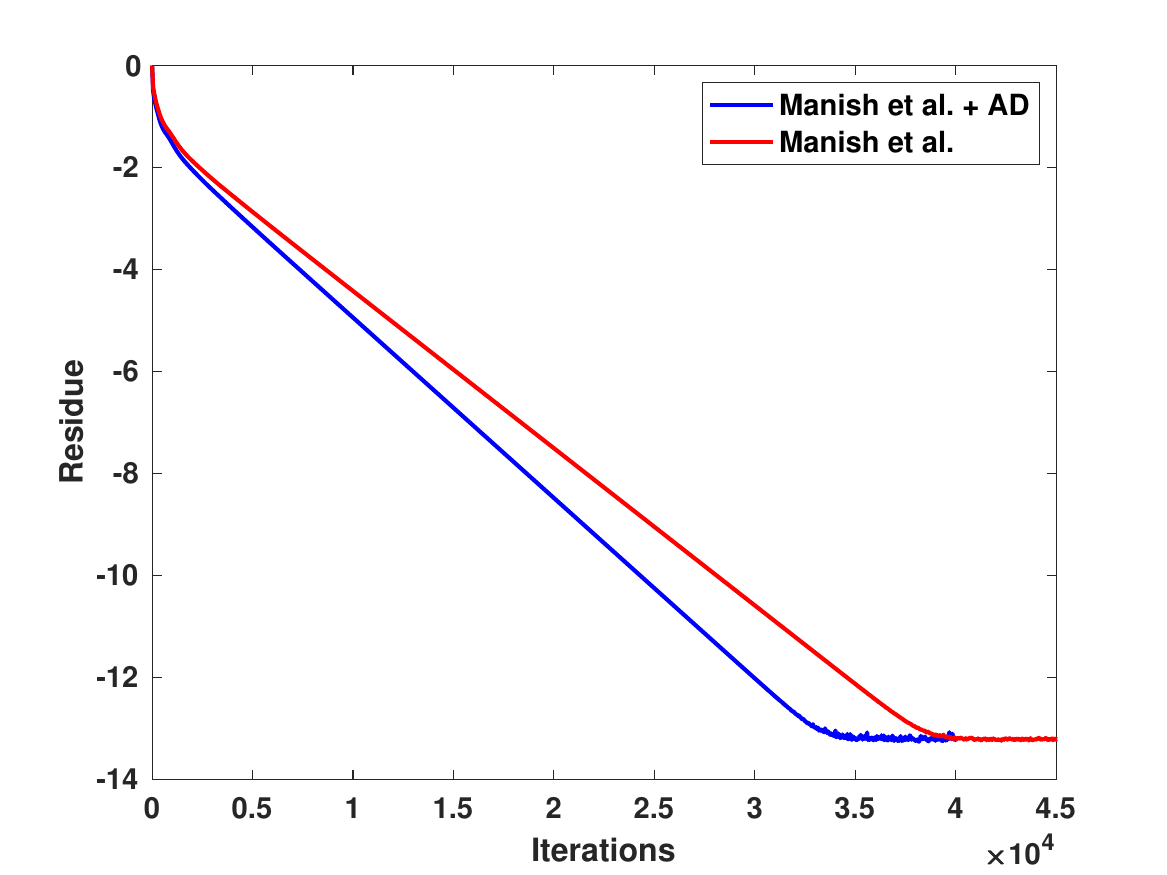} }\label{supersonic-naca-0012-coarse-distribution-residues-plot-a}}
            \subfloat[\centering ]{{\includegraphics[width=0.48\textwidth,trim={5mm 0mm 5mm 5mm},angle=0,clip]{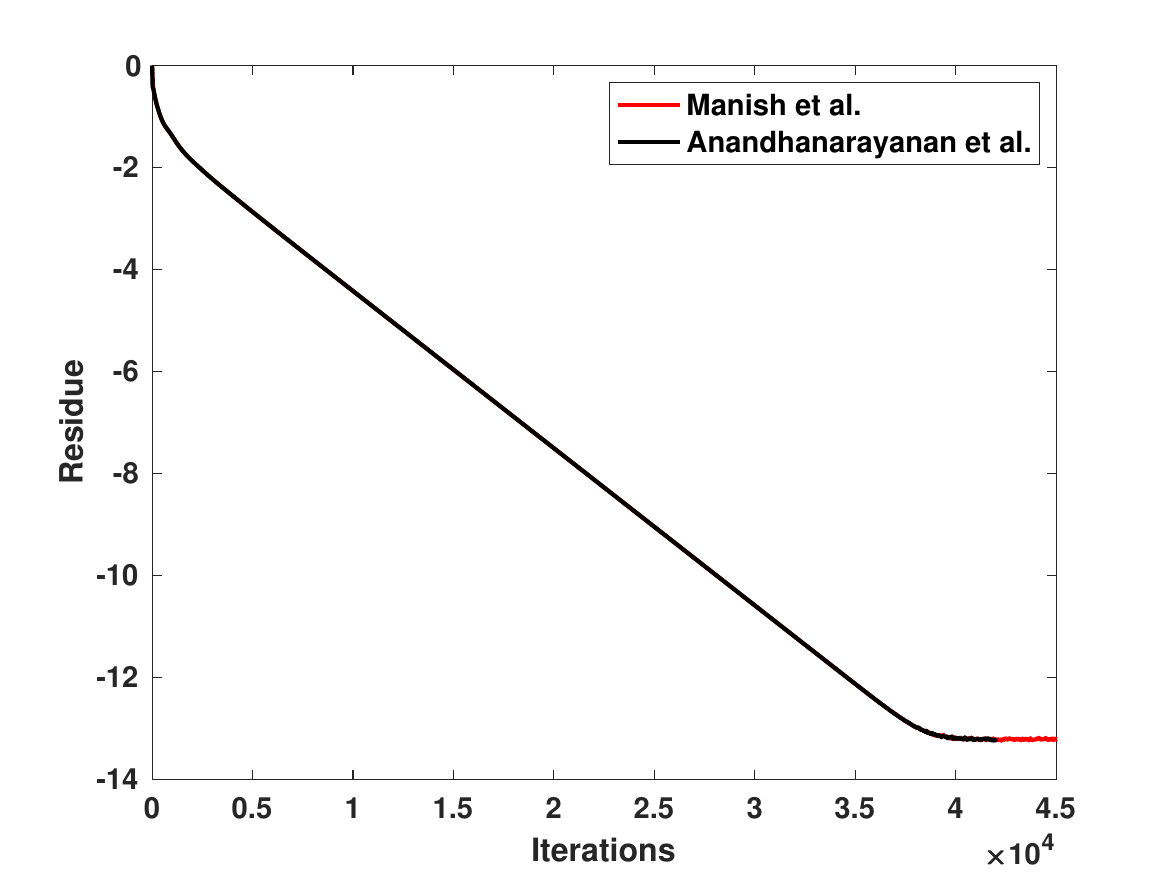} }\label{supersonic-naca-0012-coarse-distribution-residues-plot-b}} \\
         \subfloat[\centering ]{{\includegraphics[width=0.48\textwidth,trim={5mm 0mm 5mm 5mm},angle=0,clip]{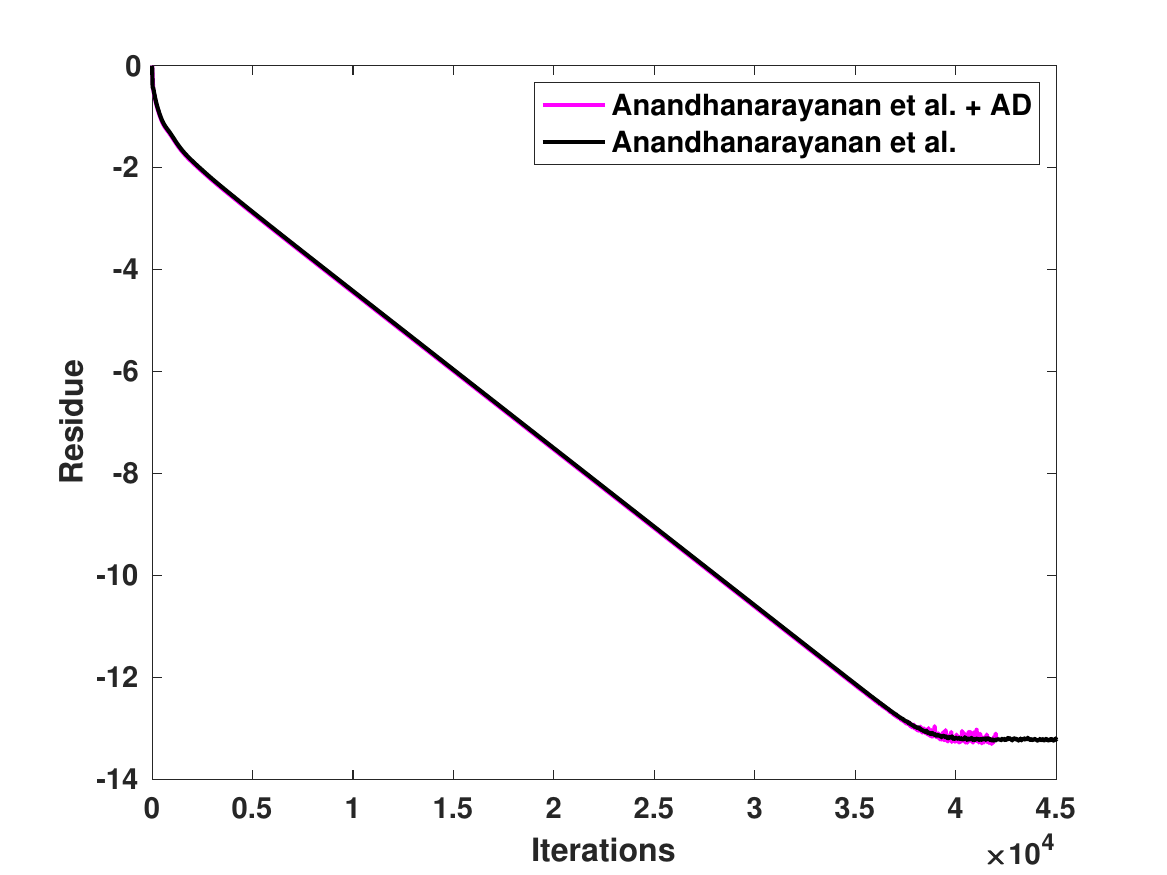} }
         \label{supersonic-naca-0012-coarse-distribution-residues-plot-c}}
        \subfloat[\centering ]{{\includegraphics[width=0.48\textwidth,trim={5mm 0mm 5mm 5mm},angle=0,clip]{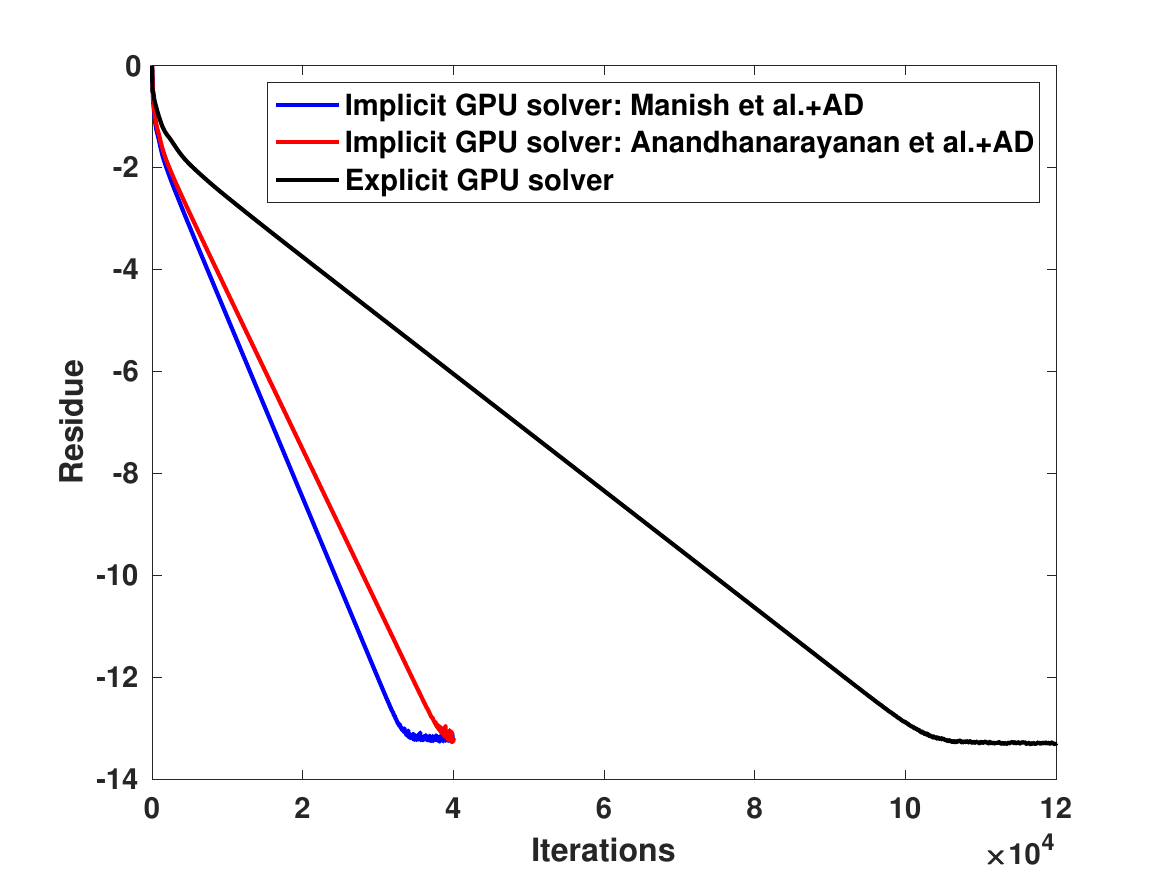} }\label{supersonic-naca-0012-coarse-distribution-residues-plot-d}}
\caption{Supersonic flow over the NACA 0012 airfoil at $M_{\infty} = 1.2$ and $AoA = 0^o$. Comparison of residues on the coarse distribution with $38,400$ points. }
\label{supersonic-naca-0012-coarse-distribution-residues}
 \end{figure}

\begin{figure}[htbp]
\centering
    \subfloat[\centering  ]{{\includegraphics[width=0.48\textwidth,trim={5mm 0mm 5mm 5mm},angle=0,clip]{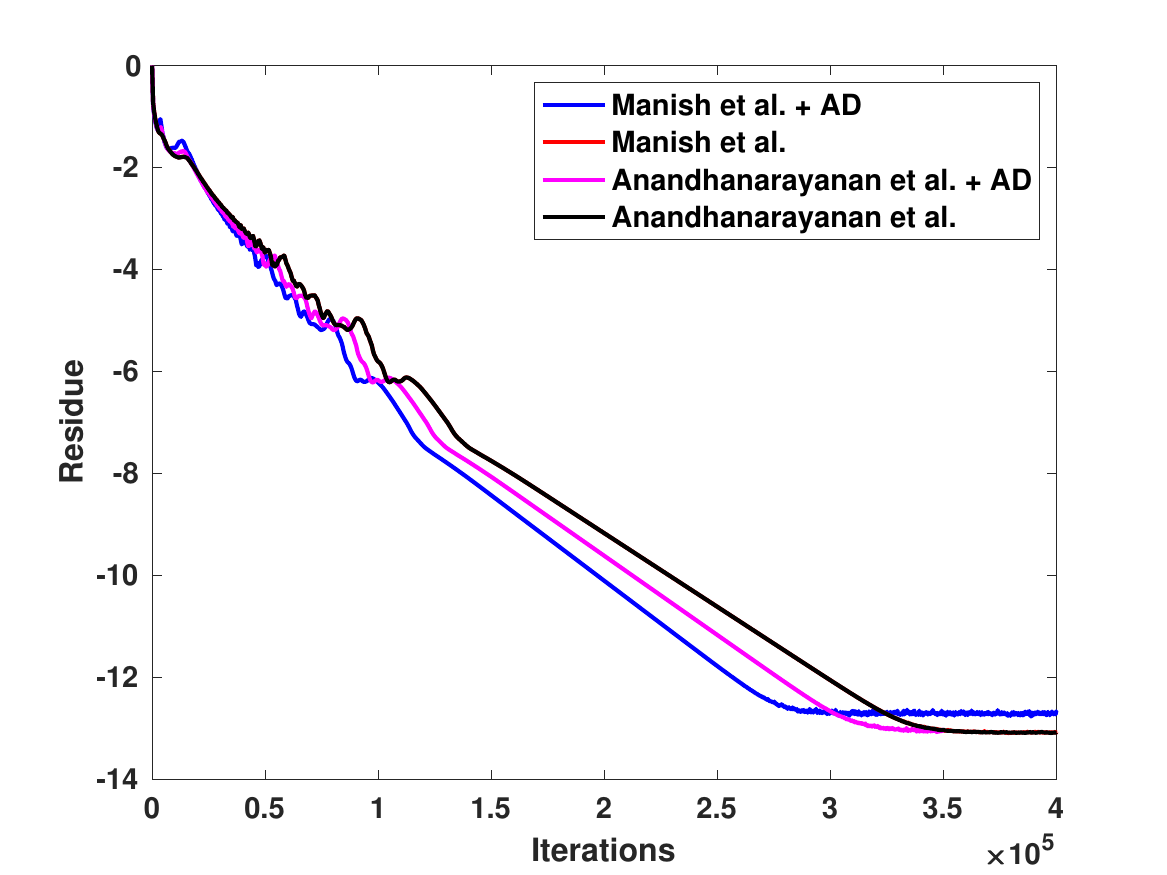} }\label{supersonic-naca-0012-fine-distribution-residues-plot-a}}
        \subfloat[\centering ]{{\includegraphics[width=0.48\textwidth,trim={5mm 0mm 5mm 5mm},angle=0,clip]{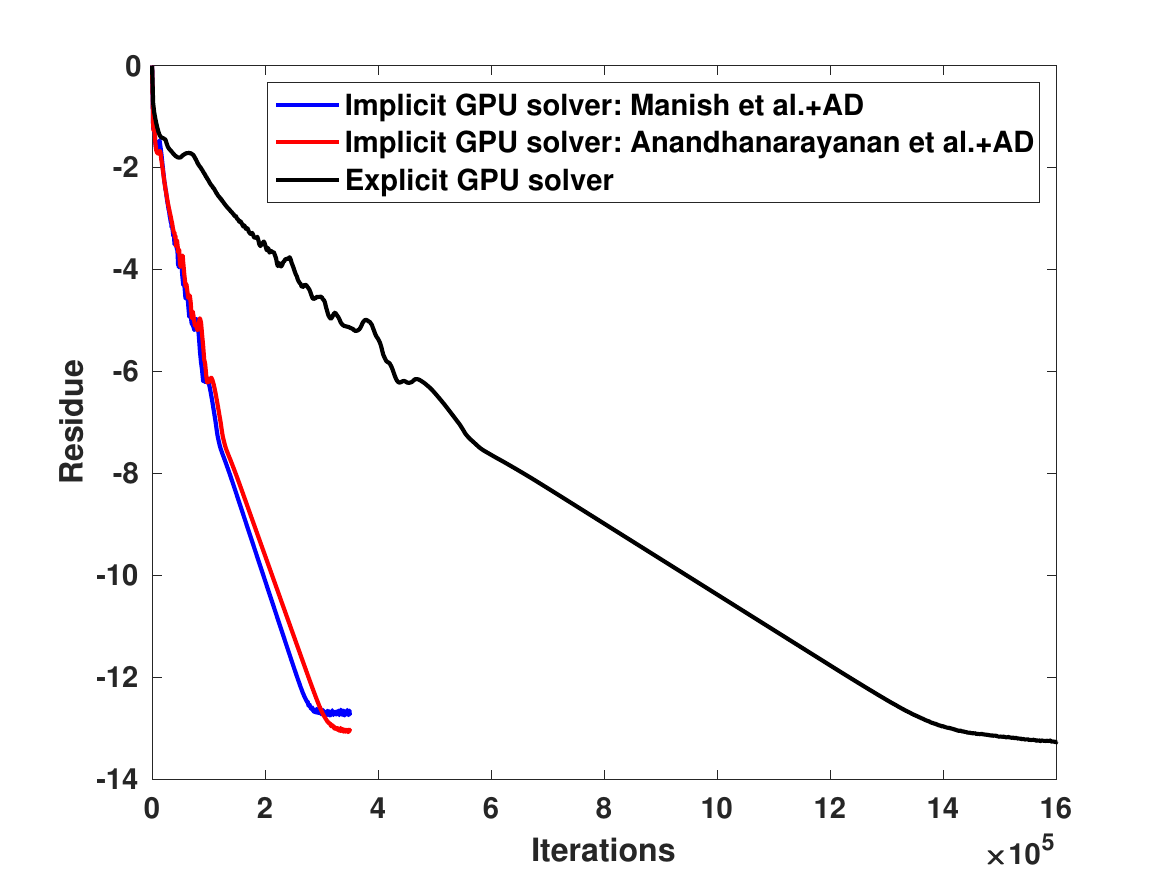} }\label{supersonic-naca-0012-fine-distribution-residues-plot-b}} 
\caption{Supersonic flow over the NACA 0012 airfoil at $M_{\infty} = 1.2$ and $AoA = 0^o$. Comparison of residues on the fine distribution with $1,228,800$ points. }
\label{supersonictzq-naca-0012-fine-distribution-residues}
 \end{figure}

\begin{figure}[htbp]
\centering
    \subfloat[\centering  Surface pressure distribution]{{\includegraphics[width=0.48\textwidth,trim={5mm 0mm 5mm 5mm},angle=0,clip]{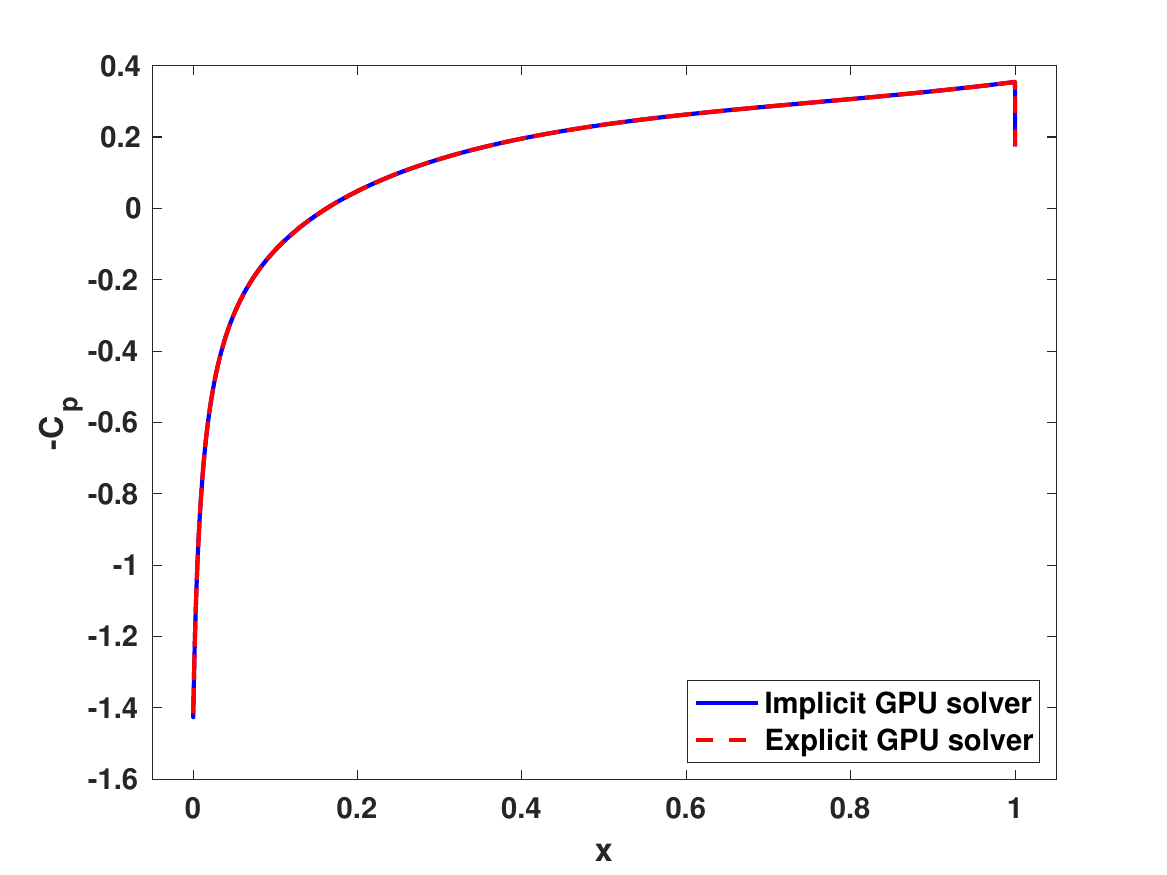}} }\label{supersonic-naca-0012-fine-distribution-cp-plot}
        \subfloat[\centering Pressure contours]{{\includegraphics[width=0.48\textwidth,trim={15mm 50mm 0mm 50mm},angle=0,clip]{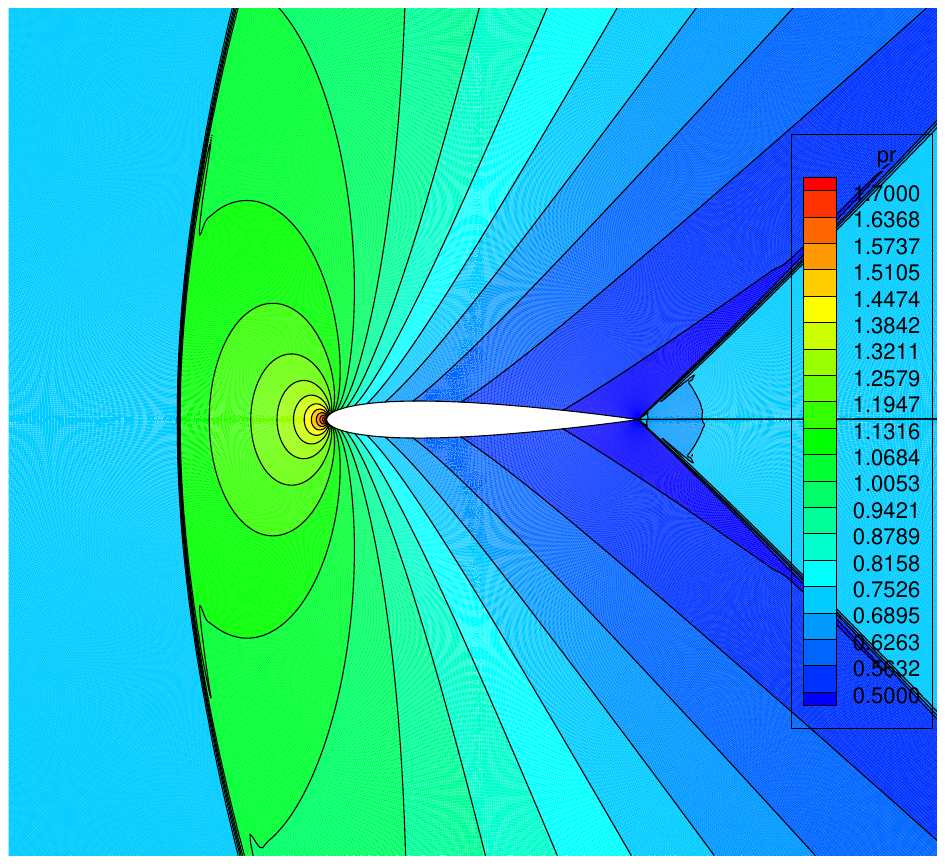}} }\label{supersonic-naca-0012-fine-distribution-pr-contours} 
\caption{Supersonic flow over the NACA 0012 airfoil at $M_{\infty} = 1.2$ and $AoA = 0^o$. Surface pressure distribution and pressure contours on the fine distribution with $1,228,800$ points. }
\label{supersonic-naca-0012-fine-distribution-cp-mach-contours}
 \end{figure}

The supersonic flow over the NACA 0012 airfoil is computed with a freestream Mach number, $M_{\infty}=1.2$ and zero angle of attack. Numerical simulations are performed on coarse and fine point distributions, as used in Section \ref{subsec-num-results-transonic}. \\ \\
Figures \ref{supersonic-naca-0012-coarse-distribution-residues} and \ref{supersonictzq-naca-0012-fine-distribution-residues} show the residue plots based on the implicit and explicit meshfree GPU solvers. Similar to the subsonic and transonic flows, in the supersonic case, the implicit GPU solver based on the Manish et al.+AD approach exhibited a superior convergence rate on the coarse and fine distributions. From Figure \ref{supersonic-naca-0012-coarse-distribution-residues-plot-a}, it is evident that this approach resulted in a better rate of convergence on the coarse distribution compared to subsonic and transonic flows. Again, the residual histories based on the approaches of Manish et al. and Anandhanarayanan et al. are almost the same. Figures \ref{supersonic-naca-0012-coarse-distribution-residues-plot-d} and \ref{supersonic-naca-0012-fine-distribution-residues-plot-b} show the speedup of the implicit GPU solver over the explicit solver. \\ \\
For this test case, the flow results in a bow shock upstream of the leading edge and a fishtail shock at the trailing edge of the airfoil. Since the flow is symmetric, the lift coefficient is zero. Figure \ref{supersonic-naca-0012-fine-distribution-cp-mach-contours} shows the surface pressure distribution $\left(C_p\right)$ and the pressure contours on the finest point distribution. The contour plot shows that the shocks are resolved very crisply. The $C_p$ plot shows the symmetry of the solution. Table \ref{supersonic-cl-cd} compares the lift and drag coefficients on the finest point distribution. We observe that the computed force coefficients agree with AGARD values \cite{agard-report-AR-211}. 
\begin{table}[htbp]
\centering
\begin{tabular}{lcc}
\toprule
Force coefficient & Implicit GPU solver   & AGARD   \\
\midrule
Lift $\left(C_L\right)$  & $1.7414 \times 10^{-5}$ & $0$ \\ [0.3em]
Drag $\left(C_D\right)$  & $0.0975$ & $0.0946-0.0960$ \\ [0.1em]
\bottomrule
\end{tabular}
\caption{Supersonic flow over the NACA 0012 airfoil at $M_{\infty} = 1.25$ and $AoA = 0^o$. Comparison of the lift and drag coefficients on the fine distribution with $1,228,800$ points. }
\label{supersonic-cl-cd}
\end{table}
\subsection{Subsonic flow over the three-element airfoil}
\begin{figure}[htbp]
\centering
\subfloat[\centering  ]{{\includegraphics[width=0.48\textwidth,trim={5mm 0mm 5mm 5mm},angle=0,clip]{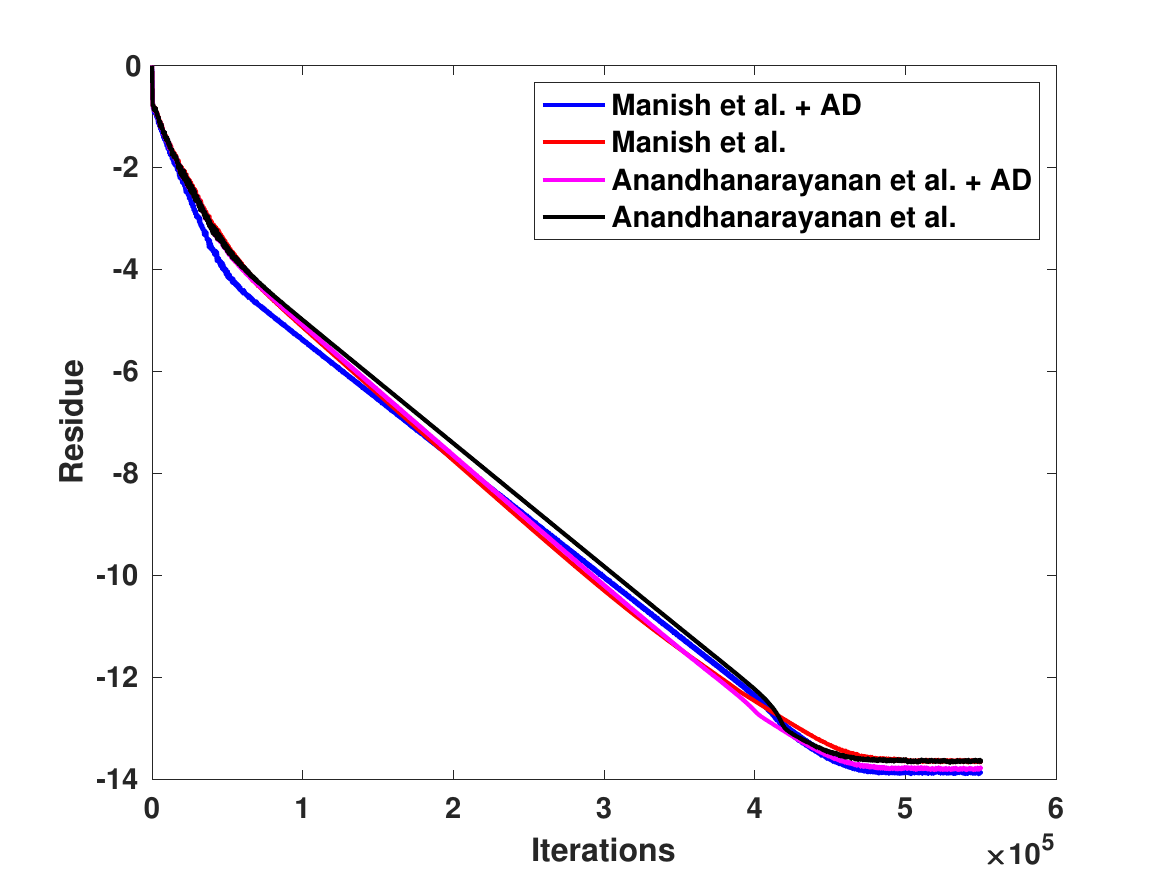}} \label{subsonic-mda-three-element-all-implicit-residues}}
\subfloat[\centering ]{{\includegraphics[width=0.48\textwidth,trim={5mm 0mm 5mm 5mm},angle=0,clip]{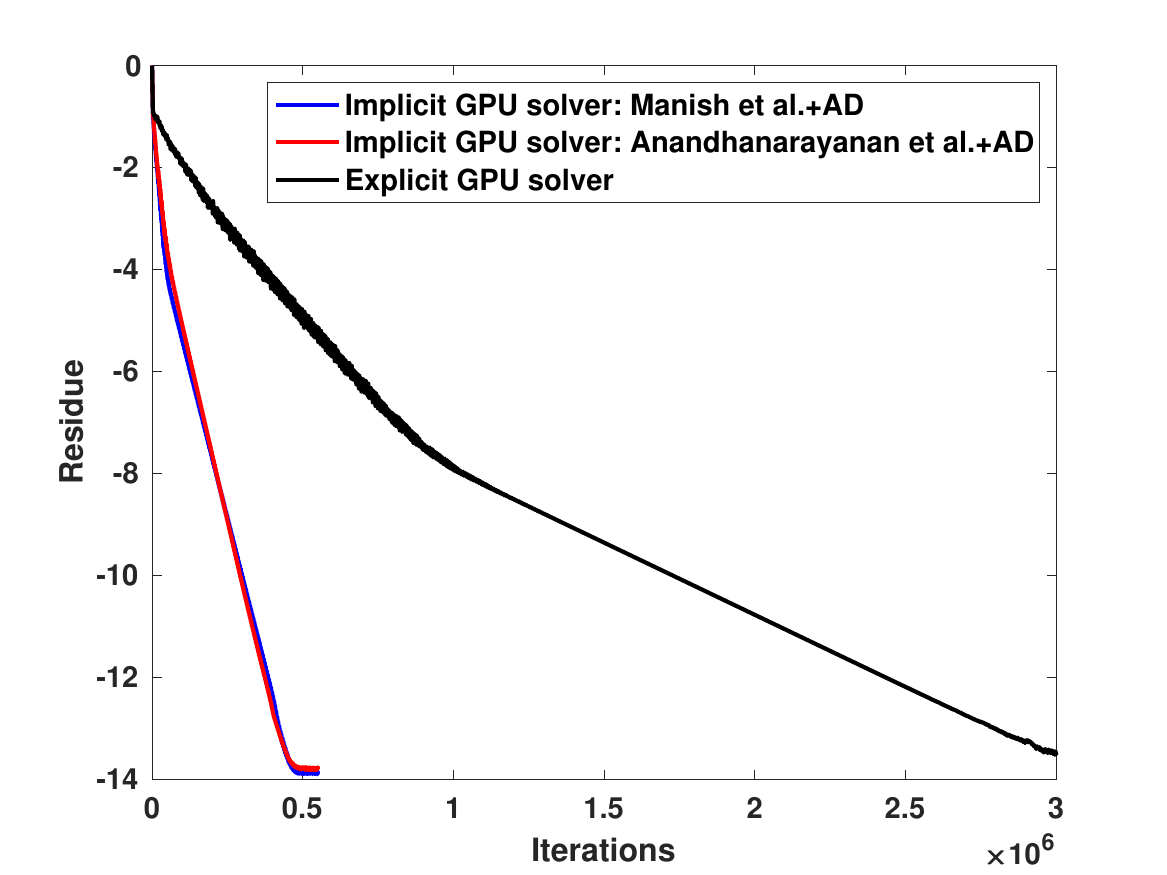}} \label{subsonic-mda-three-element-implicit-explicit-residues} }
\caption{Subsonic flow over the MDA three lement airfoil at $M_{\infty} = 0.2$ and $AoA = 16^o$. Comparison of residues based on implicit and explicit GPU solvers. }
\label{subsonic-mda-three-element-residues}
 \end{figure}

\begin{figure}[htbp]
\centering
    \subfloat[\centering  Surface pressure distribution]{{\includegraphics[width=0.48\textwidth,trim={5mm 0mm 5mm 5mm},angle=0,clip]{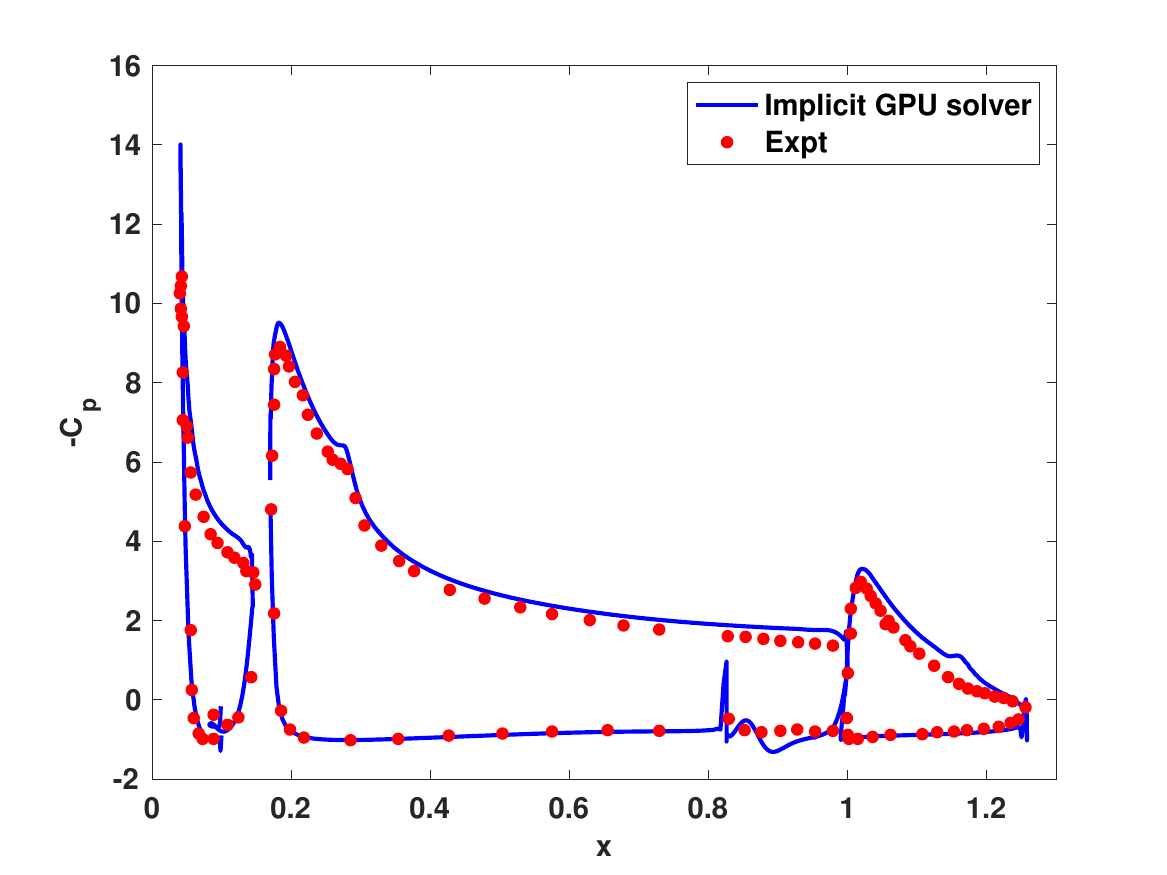}} }\label{subsonic-mda-three-element-cp}
        \subfloat[\centering Mach contours]{{\includegraphics[width=0.48\textwidth,trim={15mm 50mm 0mm 50mm},angle=0,clip]{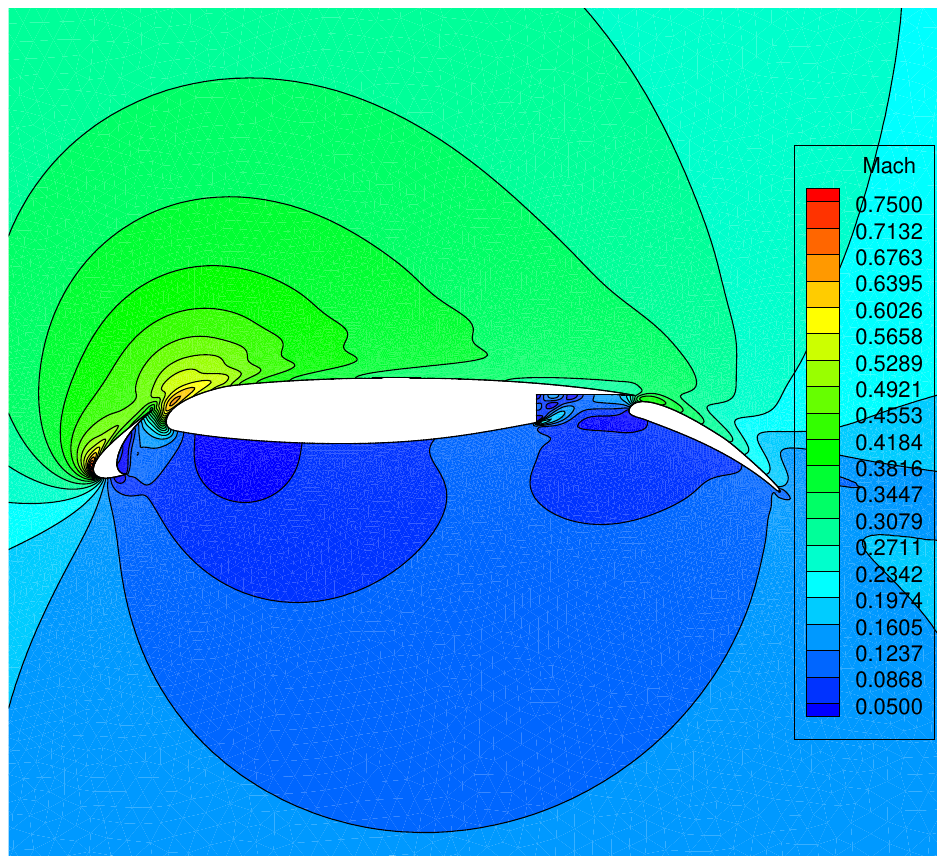}} }\label{subsonic-mda-three-element-mach-contours} 
\caption{Subsonic flow over the MDA three lement airfoil at $M_{\infty} = 0.2$ and $AoA = 16^o$. Surface pressure distribution and pressure contours. }
\label{subsonic-mda-three-element-cp-plot-mach-contours}
 \end{figure}

The next test case under investigation is the subsonic flow over the McDonnell Douglas Aerospace (MDA) $30P$-$30N$ three-element high-lift configuration \cite{mda-three-element-airfoil}. Numerical simulations are performed with a freestream Mach number, $M_{\infty} = 0.2$, and angle of attack, $\alpha=16^{o}$. The computational domain consists of $52,997$ points. The main airfoil has $696$ points, while the slat and flap contain $298$ points each. This test case is chosen to see the effect of AD based modified LU-SGS approaches on the residue fall at low Mach numbers. 
\\ \\
Figure \ref{subsonic-mda-three-element-all-implicit-residues} shows the residue plots based on the implicit GPU solvers. We observe that the implicit solver based on Manish et al.+AD approach exhibited a slightly superior convergence rate up to $6$ decade residue fall. Figures \ref{subsonic-mda-three-element-implicit-explicit-residues} shows a good speedup in convergence is achieved using the implicit GPU solver over the explicit solver. Figure \ref{subsonic-mda-three-element-cp-plot-mach-contours} shows the surface pressure distribution $\left(C_p\right)$ and the Mach contours. The computed $C_p$ is in good agreement with the experimental values \cite{mda-three-element-airfoil}. \\ \\
\section{Performance Analysis of GPU Solvers}
\label{sec-performance-analysis}
\begin{figure}[H]
\centering
    \subfloat[\centering  Speedup of the implicit GPU solvers]{{{\includegraphics[width=0.48\textwidth,trim={5mm 0mm 5mm 0mm},angle=0,clip]{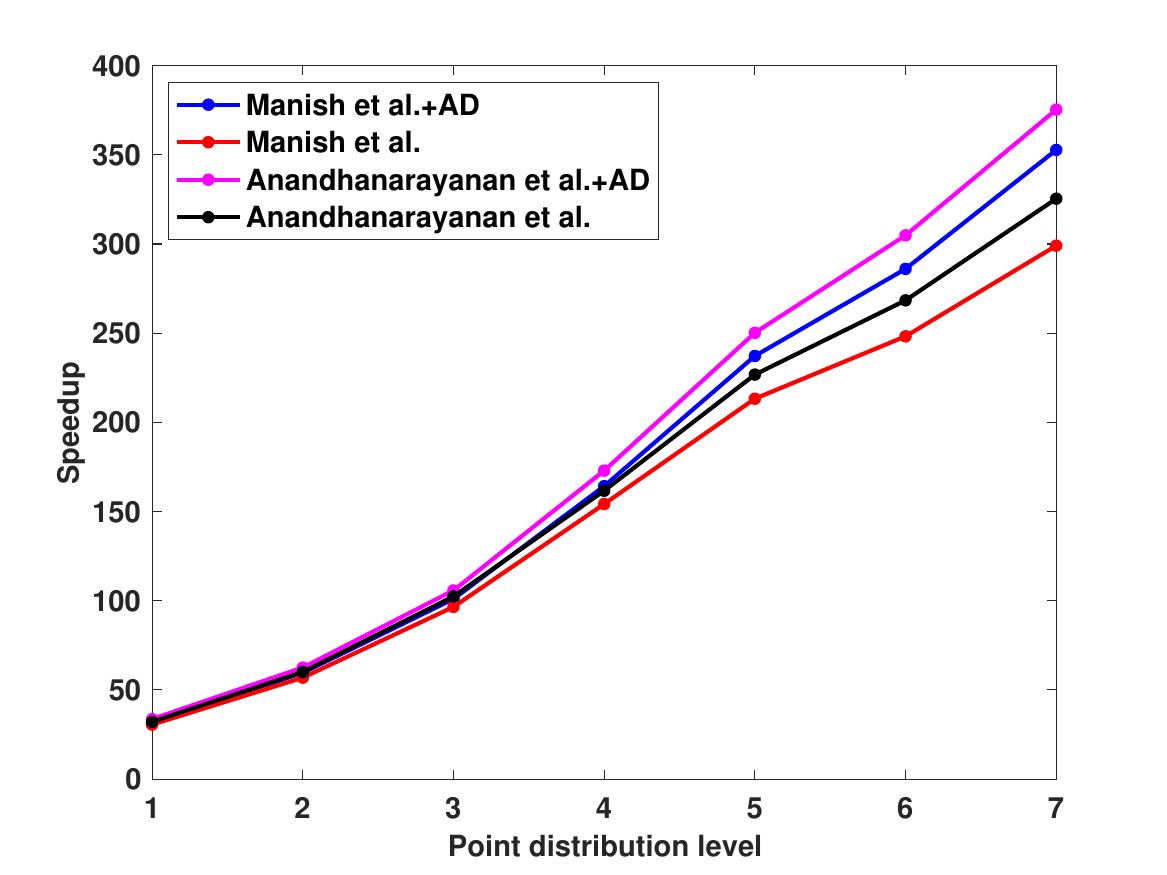}} }\label{speedup-of-implicit-gpu-solvers}}
        \subfloat[\centering Speedup of the explicit GPU solver]{{{\includegraphics[width=0.48\textwidth,trim={5mm 0mm 5mm 0mm},angle=0,clip]{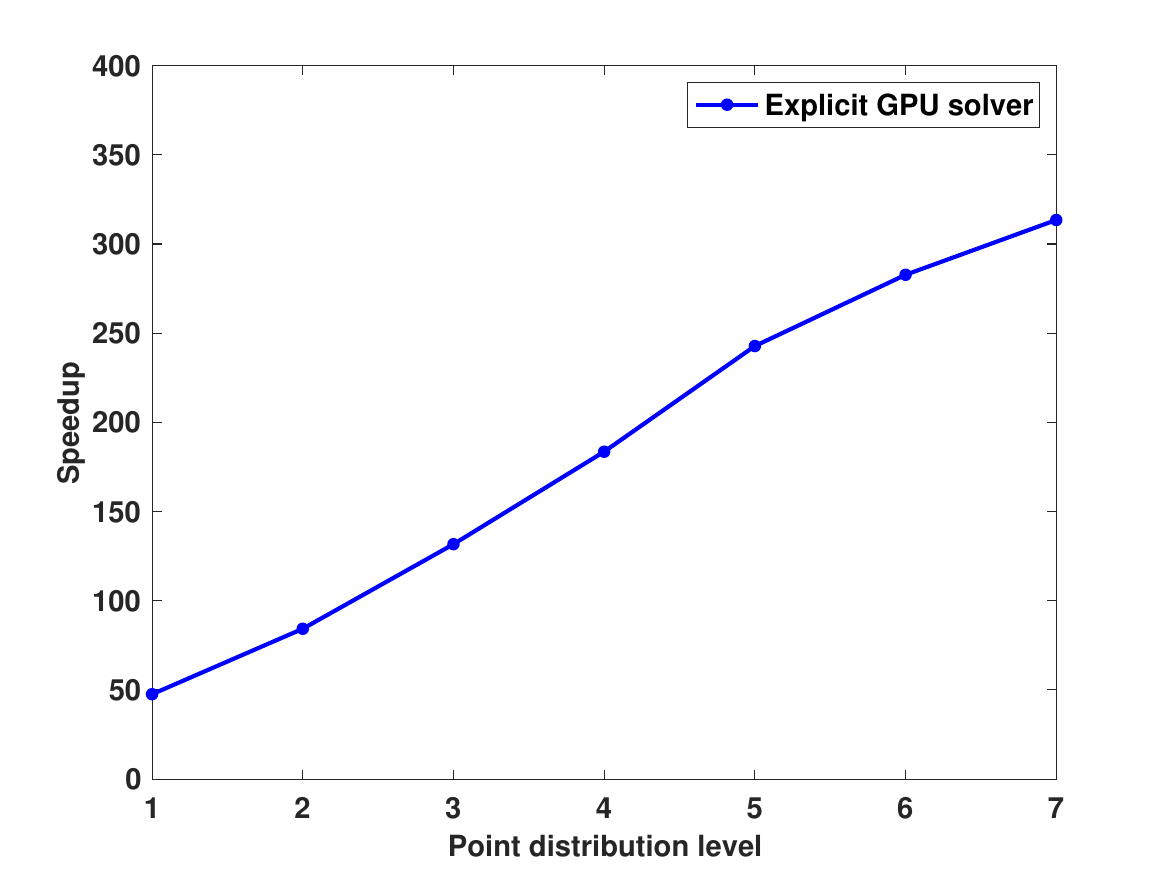}} }\label{speedup-of-explicit-gpu-solvers} }
\caption{Speedup achieved by the implicit and explicit LSKUM GPU solvers. }
\label{speedup-implicit-explicit-gpu-solvers}
 \end{figure}

In this section, we assess the performance of the implicit and explicit meshfree GPU solvers. For the benchmarks, the test case of the subsonic flow over the NACA 0012 airfoil is considered. Numerical simulations are performed on seven levels of point distributions ranging from $38,400$ to $2,457,600$ points. \\ \\
%
%
In order to measure the performance of the GPU solvers, we define a cost metric known as the Rate of Data Processing (RDP). The RDP of a GPU solver is defined as the total wall clock time in seconds per iteration per point in the computational domain. This research evaluates the RDP values by specifying the number of fixed point iterations $N$ in Algorithm \ref{algo-lskum-lusgs-gpu-solver} to $20,000$. Table \ref{rdp-comparison-gpu-codes} compares the RDP values on all levels of point distributions. \\ \\ 
The table shows that the RDP of the GPU codes decreases with continuous refinement in the point distribution. We also observe that the implicit solvers with the exact computation of the products of flux Jacobians and increments in conserved vectors are computationally more efficient as their RDP values are smaller than those with approximations to the Jacobian vector products. Moreover, the efficiency increases with the size of the point distribution. In particular, the efficiency of the implicit solver based on the Anandhanarayanan et al.+AD approach increased from $1.05$ times on the coarse distribution to $1.15$ times on the finest distribution over Anandhanarayanan et al.'s approach. It is worth noting that the exact computation of the flux Jacobian conserved vector product requires only one {\tt CUDA} kernel. On the other hand, approximating it with an increment in the flux function requires two kernel calls. From eqs. (\ref{jacobian-vector-product-approximations}) and (\ref{full-jacobian-vector-product-approximations}), we observe that one of the kernels computes the flux function with the conserved vector $\boldsymbol{U}$ as the input. The other kernel computes the flux function with $\boldsymbol{U}+\delta \boldsymbol{U}$ as input. Repeatedly calling these two kernels over a large number of times increased the overhead costs and thus resulted in higher RDP values for the implicit GPU solvers based on Anandhanarayanan et al. and Manish et al. approaches. \\ \\ 
Table \ref{rdp-comparison-gpu-codes} indicates that the RDP of the implicit GPU solvers is higher than the RDP of the explicit GPU solver. This can be attributed to the additional {\tt CUDA} kernels in implicit solvers that perform the forward and backward sweeps of the LU-SGS algorithm. From Algorithm \ref{algo-cuda-kernels-efficient-forward-backward-sweeps}, it is clear that the more the number of color groups, the more the calls to the {\tt CUDA} kernels and thus an increase in the overheads of the implicit GPU solvers. \\ \\ 
\begin{table}[t]
\centering
\begin{center}
\scalebox{0.77}{
\begin{tabular}{llccccccc}
\toprule
& \multicolumn{7}{c}{\centering{Implicit GPU solver with LU-SGS scheme based on }}  & \\ [0.2em]
\cline{4-7}
Level & Points & Colors & Manish+AD & Manish  & Anandhanarayanan+AD  & Anandhanarayanan & Explicit\\[0.2em]
\midrule
\multicolumn{7}{c}{\centering{RDP $\times$ $10^{-8}$ (Lower is better)}}  \\ [0.2em]
\midrule
$1$ & $38,400$ & $8$ & $20.8373$ & $21.4989$  & $19.5199$ & $20.4602$ & $6.0998$\\[0.5em]
$2$ & $76,800$ & $7$ & $11.0308$ & $11.5563$ & $10.5074$  & $10.9451$ & $3.4449$ \\[0.5em]
$3$ & $1,53,600$ & $7$ & $6.5088$ & $6.8088$ & $6.2158$ & $6.4191$ & $2.2051$ \\[0.5em]
$4$ & $3,07,200$ & $7$ & $4.0029$  & $4.2636$ & $3.8017$ & $4.0663$ & $1.5822$\\[0.5em]
$5$ & $6,14,400$ & $7$ & $2.7715$ & $3.0830$ & $2.6278$ & $2.8991$ & $1.1963$ \\[0.5em]
$6$ & $12,28,800$ & $8$ & $2.2984$ & $2.6486$ & $2.1567$  & $2.4495$ & $1.0273$\\[0.5em]
$7$ & $24,57,600$ & $7$ & $1.8639$ & $2.1985$ & $1.7518$ & $2.0205$ & $0.9266$\\[0.1em]
\bottomrule
\end{tabular}
}
\end{center}
\caption{{\small RDP comparison of implicit and explicit meshfree LSKUM GPU solvers.  }}
\label{rdp-comparison-gpu-codes}
\end{table} 
We also observe that the ratio of the RDP of implicit and explicit solvers decreases from course to fine point distributions. To investigate this behaviour in the RDP ratio, we analyse the streaming multiprocessors (SM) utilisation of the forward and backward sweeps kernels. Tables \ref{sm-utilisation-forward-backward-sweeps-coarse-distribution+Anandh+AD} and \ref{sm-utilisation-forward-backward-sweeps-fine-distribution-anandh+AD} compare the SM utilisation on level $1$ and level $6$ point distributions for the implcit GPU solver based on Anandhanarayanan et al.+AD approach. We observe that the SM utilisation depends on the number of points in a color group. The more the points in a color group the higher the SM utilisation. Although the point groups $1$ and $3$ on the coarse distribution have the same number of points, their SM utilisation is different. This is due to the point types (wall, interior, or outer boundary point) in each group and the number of neighbours in the connectivity set of a point. Higher values of SM indicate an efficient usage of GPU resources \cite{hipc-2022-nischay} and thus resulted in smaller values of RDP ratio. \\ \\
We want to find an AD based implicit solver that effectively utilises the GPU resources. Note that the implicit schemes based on Anandhanarayanan et al.+AD and Manish et al.+AD differ only in their forward sweeps, while having the same backward sweep algorithms. Therefore, it is enough to compare the SM utilisation of the {\tt CUDA} kernels for forward sweep in both the solvers. Table \ref{sm-utilisation-forward-sweep-fine-distribution-manish+AD} shows the SM utilisation of the forward sweep kernel in the Manish et al.+AD implicit solver on level $6$ point distribution. From Tables \ref{sm-utilisation-forward-backward-sweeps-fine-distribution-anandh+AD} and \ref{sm-utilisation-forward-sweep-fine-distribution-manish+AD}, we observe that the SM utilisation is more for the implicit solver based on Anandhanarayanan et al.+AD, indicating better utilisation of GPU resources. 
\\ \\
To assess the overall performance of the implicit and explicit GPU solvers, we define another metric called speedup. The speedup of a GPU code is defined as the ratio of the RDP of the serial code to the RDP of the GPU code. Note that the serial version of the implicit LSKUM solver is based on Algorithms \ref{algo-lskum-lusgs-solver} and \ref{algo-forward-backward-sweeps}. The serial code simulations are performed on a computer node consisting of $2$ {\tt AMD} {\tt EPYC} $7542$ processors with $64$ cores and $256$ GB RAM. \\ \\ 
Figure \ref{speedup-implicit-explicit-gpu-solvers} shows the speedup achieved by the GPU solvers. We observe that the implicit GPU solvers using AD achieved more speedup than the solvers with approximations to the Jacobian vector products. Furthermore, the rate of speedup increases as the number of points in the cloud increases. On the finest distribution, the AD based solvers achieved around $50$ times more speedup compared to the solvers without AD. Among the AD based GPU solvers, the implicit approach based on Anandhanarayanan et al.+AD yielded around $25$ times more speedup than the Manish et al.+AD's approach. We also observe that the explicit GPU solver shows a better speedup on the coarse and medium point distributions than the implicit GPU solvers. However, on the fine distributions, due to the efficient utilisation of GPU resources by the {\tt CUDA} kernels for forward and backward sweeps, the AD based implicit solvers achieved more speedup than the explicit GPU solver. 
%
%
%
%
%
\begin{table}[htbp]
\centering
\begin{center}
\scalebox{0.85}{
\begin{tabular}{lccccccccc}
\toprule
{\tt CUDA} & \multicolumn{7}{c}{\centering{Point color group (Number of points in each color group)}}  & \\ [0.2em]
\cline{2-9}
Kernel & $1(9441)$ & $2(9383)$ & $3(9441)$ & $4(9382)$ & $5(433)$  & $6(316)$ & $7(3)$ & $8(1)$ \\[0.1em]
\midrule
 & \multicolumn{7}{c}{\centering{SM utilisation (shown in percentange) }}  & \\ [0.2em]
\cline{1-9} 
Forward sweep & $11.59$ & $7.02$ & $11.04$ & $7.26$ & $0.50$ & $0.37$ & $0.07$ & $0.16$ \\
[0.5em]
Backward sweep & $7.21$ & $9.90$ & $8.07$ & $9.51$ & $0.59$ & $0.31$ & $0.14$ & $0.19$ \\
[0.1em]

\bottomrule
\end{tabular}
}
\end{center}
\caption{{\small Implicit GPU solver based on Anandhanarayanan et al.+AD: Streaming multiprocessor (SM) utilisation of the forward and reverse sweep kernels on the coarse distribution with $38,400$ points. }}
\label{sm-utilisation-forward-backward-sweeps-coarse-distribution+Anandh+AD}
\end{table} 
\begin{table}[htbp]
\centering
\begin{center}
\scalebox{0.85}{
\begin{tabular}{lccccccccc}
\toprule
{\tt CUDA} & \multicolumn{7}{c}{\centering{Point color group (Number of points in each color group)}}  & \\ [0.2em]
\cline{2-9}
Kernel & $1(303491)$ & $2(302393)$ & $3(300101)$ & $4(296968)$ & $5(22526)$  & $6(2912)$ & $7(396)$ & $8(13)$ \\ 
\midrule
 & \multicolumn{7}{c}{\centering{SM utilisation (shown in percentange) }}  & \\ [0.1em]
\cline{1-9} 
Forward sweep& $32.60$ & $31.09$ & $30.69$ & $27.64$ & $15.98$ & $2.76$ & $2.11$ & $2.09$ \\
[0.5em]
Backward sweep & $33.60$ & $33.97$ & $35.98$ & $36.54$ & $17.42$ & $2.60$ & $3.35$ & $3.45$ \\
[0.1em]
\bottomrule
\end{tabular}
}
\end{center}
\caption{{\small Implicit GPU solver based on Anandhanarayanan et al.+AD: Streaming multiprocessor (SM) utilisation of the forward and reverse sweep kernels on the fine distribution with $12,28,800$ points. }}
\label{sm-utilisation-forward-backward-sweeps-fine-distribution-anandh+AD}
\end{table} 
\begin{table}[htbp]
\centering
\begin{center}
\scalebox{0.85}{
\begin{tabular}{lccccccccc}
\toprule
{\tt CUDA} & \multicolumn{7}{c}{\centering{Point color group (Number of points in each color group)}}  & \\ [0.2em]
\cline{2-9}
kernel & $1(303491)$ & $2(302393)$ & $3(300101)$ & $4(296968)$ & $5(22526)$  & $6(2912)$ & $7(396)$ & $8(13)$ \\ 
\midrule
 & \multicolumn{7}{c}{\centering{SM utilisation (shown in percentange) }}  & \\ [0.1em]
\cline{1-9} 
Forward sweep & $29.48$ & $28.29$ & $28.06$ & $25.19$ & $14.91$ & $2.64$ & $2.03$ & $1.92$ \\
[0.1em]
\bottomrule
\end{tabular}
}
\end{center}
\caption{{\small Implicit GPU solver based on Manish et al.+AD: Streaming multiprocessor (SM) utilisation of the forward sweep kernel on the fine distribution with $12,28,800$ points. }}
\label{sm-utilisation-forward-sweep-fine-distribution-manish+AD}
\end{table} 
\section{Conclusions}
\label{sec-conclusions}
This report presented the GPU acceleration of implicit kinetic meshfree methods for the inviscid compressible flows. The meshfree scheme is based on the Least Squares Kinetic Upwind Method (LSKUM). The implicit schemes were based on modifications to the meshfree LU-SGS approaches proposed by Anandhanarayanan et al. and Manish et al. In the modified LU-SGS algorithms, instead of approximating the product of split flux Jacobians and incremental conserved vector with the increments in the flux function, they are computed exactly using algorithmic differentiation (AD). The programming model {\tt CUDA} was used to develop the GPU solvers. \\ \\ 
%
%
The implicit GPU solvers with the exact and approximate computation of Jacobian vector products in the LU-SGS algorithms for LSKUM were applied to the standard $2D$ test cases for subsonic, transonic, and supersonic flows. Numerical results showed that the exact computation of Jacobian vector products yielded slightly better convergence rate than its approximation. In particular, the implicit GPU solver based on Manish et al. with the modified LU-SGS approach exhibited slightly superior convergence rate (in the sense of slightly better residue fall). In all the test cases, the implicit GPU solvers achieved a good convergence speed over the explicit GPU solver. \\ \\
Two metrics, the RDP and speedup, were defined to assess the computational performance of the implicit GPU solvers. The RDP of a solver was defined as the total wall clock time in seconds per iteration per point in the cloud. The speedup was defined as the ratio of the RDP of serial implicit solver to the RDP of GPU solver. Benchmark simulations on seven levels of point distributions showed that the implicit GPU solvers with AD based exact computation of the Jacobian vector products are computationally more efficient as their RDP values were smaller than the solvers with approximate Jacobian vector products. On the finest distribution, the AD based implicit GPU solvers achieved around $50$ times more speedup than those with approximations. Among the AD based solvers, the implicit approach based on Anandhanarayanan et al.+AD yielded around $25$ times more speedup than the Manish et al.+AD's approach. \\ \\
In summary, the implicit meshfree scheme based on Anandhanarayanan et al. with the modified LU-SGS approach is preferred due to its better utilisation of GPU resources and superior computational efficiency (in terms of much less RDP). We presume that this approach can yield significant speedups in three-dimensional flows with more points in the computational domain.
%
Research is in progress to develop implicit GPU solvers for three-dimensional flows. We also plan to extend these solvers to laminar and turbulent flows. In the future, we plan to develop multi-node and multi-GPU versions of these solvers. 
\newpage
\section{Appendix-A}
\label{appendix}
\begin{lstlisting}[language=Fortran, caption= Fortran code to compute the flux vector $\boldsymbol{Gx}$ along the $x$-coordinate direction. ,  label=routine-for-flux-Gx]
SUBROUTINE flux_Gx(U, Gx)
!
    IMPLICIT NONE
!
    real*8 :: U(4), Gx(4)
    real*8 :: rho, u1, u2, pr
!  
    rho = U(1)
    u1 = U(2) / rho
    u2 = U(3) / rho
    pr = 0.4d0 * (U(4) - 0.5d0 * rho * (u1 * u1 + u2 * u2))
!
    Gx(1) = rho*u1
    Gx(2) = pr+rho*u1*u1
    Gx(3) = rho*u1*u2
    Gx(4) = (pr+U(4))*u1
!
  END SUBROUTINE
\end{lstlisting}
\newpage
\begin{lstlisting}[language=Fortran, caption= The black-box tangent linear code to compute the Jacobian vector product $\boldsymbol{Ax} \delta \boldsymbol{U}$. ,  label=blackbox-routine-for-flux-Gxd]
  SUBROUTINE FLUX_GX_D(U, Ud, Gx, Gxd)
    IMPLICIT NONE
!
    real*8 :: U(4), Gx(4)
    real*8 :: Ud(4), Gxd(4)
    real*8 :: rho, u1, u2, pr
    real*8 :: rhod, u1d, u2d, prd
    real*8 :: temp
!
    rhod = Ud(1)
    rho = U(1)
    temp = U(2)/rho
    u1d = (Ud(2)-temp*rhod)/rho
    u1 = temp
    temp = U(3)/rho
    u2d = (Ud(3)-temp*rhod)/rho
    u2 = temp
    temp = u1*u1 + u2*u2
    prd = 0.4d0*(Ud(4)-0.5d0*(temp*rhod+rho*(2*u1*u1d+2*u2*u2d)))
    pr = 0.4d0*(U(4)-0.5d0*(rho*temp))
!
    Gxd = 0.0d0
    Gxd(1) = u1*rhod + rho*u1d
    Gx(1) = rho*u1
    Gxd(2) = prd + u1**2*rhod + rho*2*u1*u1d
    Gx(2) = pr + rho*u1*u1
    Gxd(3) = u2*(u1*rhod+rho*u1d) + rho*u1*u2d
    Gx(3) = rho*u1*u2
    Gxd(4) = u1*(prd+Ud(4)) + (pr+U(4))*u1d
    Gx(4) = (pr+u(4))*u1
!
  END SUBROUTINE FLUX_GX_D
\end{lstlisting}
\newpage
\begin{lstlisting}[language=Fortran, caption= The optimised tangent linear code to compute the Jacobian vector product $\boldsymbol{Ax} \delta \boldsymbol{U}$. ,  label=optimised-routine-for-flux-Gxd]
  SUBROUTINE Jacobian_vector_product_AxdU(U, Ud, AxdU)
    IMPLICIT NONE
!
    real*8 :: U(4), Ud(4), AxdU(4)
    real*8 :: rho, u1, u2, pr
    real*8 :: rhod, u1d, u2d, prd
    real*8 :: temp
!
    rhod = Ud(1)
    rho = U(1)
    temp = U(2)/rho
    u1d = (Ud(2)-temp*rhod)/rho
    u1 = temp
    temp = U(3)/rho
    u2d = (Ud(3)-temp*rhod)/rho
    u2 = temp
    temp = u1*u1 + u2*u2
    prd = 0.4d0*(Ud(4)-0.5d0*(temp*rhod+rho*(2*u1*u1d+2*u2*u2d)))
    pr = 0.4d0*(U(4)-0.5d0*(rho*temp))
!
    AxdU(1) = u1*rhod + rho*u1d
    AxdU(2) = prd + u1**2*rhod + rho*2*u1*u1d
    AxdU(3) = u2*(u1*rhod+rho*u1d) + rho*u1*u2d
    AxdU(4) = u1*(prd+Ud(4)) + (pr+U(4))*u1d
!  
  END SUBROUTINE 
\end{lstlisting}
\newpage
\bibliographystyle{unsrt}
\bibliography{report}

\begin{thebibliography}{10}

\bibitem{openfoam}
H.~G. Weller, G.~Tabor, H.~Jasak, and C.~Fureby.
\newblock A tensorial approach to computational continuum mechanics using object-oriented techniques.
\newblock {\em Computers in Physics}, 12(6):620–631, November 1998.

\bibitem{su2}
Thomas~D. Economon, Francisco Palacios, Sean~R. Copeland, Trent~W. Lukaczyk, and Juan~J. Alonso.
\newblock Su2: An open-source suite for multiphysics simulation and design.
\newblock {\em AIAA Journal}, 54(3):828–846, March 2016.

\bibitem{gpus-general-purpose-processors}
Mark~J. Harris, William~V. Baxter, Thorsten Scheuermann, and Anselmo Lastra.
\newblock Simulation of cloud dynamics on graphics hardware.
\newblock In {\em HWWS 03: Proceedings of the ACM SIGGRAPH/EUROGRAPHICS conference on Graphics hardware}, pages 92--101, 2003.

\bibitem{pyfr}
F.D. Witherden, A.M. Farrington, and P.E. Vincent.
\newblock {PyFR}: An open source framework for solving advection{\textendash}diffusion type problems on streaming architectures using the flux reconstruction approach.
\newblock {\em Computer Physics Communications}, 185(11):3028--3040, nov 2014.

\bibitem{srikanth-bernardini-gpus}
Matteo Bernardini, Davide Modesti, Francesco Salvadore, Srikanth Sathyanarayana, Giacomo Della~Posta, and Sergio Pirozzoli.
\newblock Streams-2.0: Supersonic turbulent accelerated navier-stokes solver version 2.0.
\newblock {\em Computer Physics Communications}, 285:108644, April 2023.

\bibitem{deshpande-meshfree-lskum}
S.M. Deshpande, J.~C. Mandal, and A.K. Ghosh.
\newblock Least squares weak upwind method for {E}uler equations.
\newblock {\em FM Report No. 1989-FM-4, Dept. of Aerospace Engg., Indian Institute of Science, Bangalore}, 1989.

\bibitem{batina-meshfree}
J.T. Batina.
\newblock A gridless {E}uler/{N}avier-{S}tokes solution algorithm for complex-aircraft applications.
\newblock {\em AIAA Paper 1993-0333}, 1993.

\bibitem{meshfree-lohner}
Rainald L\"{o}hner, Carlos Sacco, Eugenio Oñate, and Sergio Idelsohn.
\newblock A finite point method for compressible flow.
\newblock {\em International Journal for Numerical Methods in Engineering}, 53(8):1765–1779, December 2001.

\bibitem{meshfree-morinishi}
K.~Morinishi.
\newblock An implicit gridless type solver for the navier-stokes equations.
\newblock {\em Computational Fluid Dynamics Journal}, pages 551--560, 2001.

\bibitem{meshfree-sridar}
D.~Sridar and N.~Balakrishnan.
\newblock An upwind finite difference scheme for meshless solvers.
\newblock {\em Journal of Computational Physics}, 189(1):1–29, July 2003.

\bibitem{meshfree-katz}
Aaron Katz and Antony Jameson.
\newblock Multicloud: Multigrid convergence with a meshless operator.
\newblock {\em Journal of Computational Physics}, 228(14):5237–5250, August 2009.

\bibitem{meshfree-Chiu}
Edmond Kwan-yu Chiu, Qiqi Wang, Rui Hu, and Antony Jameson.
\newblock A conservative mesh-free scheme and generalized framework for conservation laws.
\newblock {\em SIAM Journal on Scientific Computing}, 34(6):A2896–A2916, January 2012.

\bibitem{lskum-ghosh-aiaa}
A.~K. Ghosh and S.~M. Deshpande.
\newblock Least squares kinetic upwind method for inviscid compressible flows.
\newblock {\em AIAA paper 1995-1735}, 1995.

\bibitem{lskum-ghosh-journal}
S.~M. Deshpande, P.~S. Kulkarni, and A.~K. Ghosh.
\newblock New developments in kinetic schemes.
\newblock {\em Computers Math. Applic.}, 35(1):75--93, 1998.

\bibitem{qlskum}
S~M. Deshpande, K.~Anandhanarayanan, C.~Praveen, and V.~Ramesh.
\newblock Theory and application of {3-D} {LSKUM} based on entropy variables.
\newblock {\em Int. J. Numer. Meth. Fluids}, 40:47--62, 2002.

\bibitem{meshfree-unsteady-Ramesh}
V.~Ramesh and S.M. Deshpande.
\newblock Unsteady flow computations for flow past multiple moving boundaries using lskum.
\newblock {\em Computers {\&} Fluids}, 36(10):1592–1608, December 2007.

\bibitem{mahendra-lskum-2011}
A.K. Mahendra, R.K. Singh, and G.~Gouthaman.
\newblock Meshless kinetic upwind method for compressible, viscous rotating flows.
\newblock {\em Computers {\&} Fluids}, 46(1):325--332, July 2011.

\bibitem{anandh-lskum-sepdynamic-jaircraft}
K.~Anandhanarayanan, Konark Arora, Vaibhav Shah, R.~Krishnamurthy, and Debasis Chakraborty.
\newblock Separation dynamics of air-to-air missile using a grid-free euler solver.
\newblock {\em Journal of Aircraft}, 50(3):725--731, 2013.

\bibitem{anandh-lskum-viscous-aiaa-journal}
K.~Anandhanarayanan, R.~Krishnamurthy, and Debasis Chakraborty.
\newblock Development and validation of a grid-free viscous solver.
\newblock {\em AIAA Journal}, 54(10):3312--3315, 2016.

\bibitem{qlskum-flutter}
V.~Ramesh and S.~M. Deshpande.
\newblock Kinetic mesh-free method for flutter prediction in turbomachines.
\newblock {\em Sadhana}, 39(1):149--164, 2014.

\bibitem{lusgs-jameson}
Seokkwan Yoon and Antony Jameson.
\newblock Lower-upper symmetric-gauss-seidel method for the euler and navier-stokes equations.
\newblock {\em AIAA Journal}, 26(9):1025–1026, September 1988.

\bibitem{lusgs-Sharov-1998}
Dmitri Sharov and Kazuhiro Nakahashi.
\newblock Low speed preconditioning and lu-sgs scheme for 3-d viscous flow computations on unstructured grids.
\newblock In {\em 36th AIAA Aerospace Sciences Meeting and Exhibit}. American Institute of Aeronautics and Astronautics, January 1998.

\bibitem{lusgs-anandh-icas-2004}
K.~Anandhanarayanan, M.~Nagarathinam, and S.M. Deshpande.
\newblock An entropy variable based gridfree solver with {LU-SGS} accelerator.
\newblock {\em ICAS paper 2004-2.4.3}, 2004.

\bibitem{lusgs-NAL-Manish-2015}
Manish~K. Singh, V.~Ramesh, and N.~Balakrishnan.
\newblock Implicit scheme for meshless compressible euler solver.
\newblock {\em Engineering Applications of Computational Fluid Mechanics}, 9(1):382–398, January 2015.

\bibitem{ad-book-andreas}
A.~Griewank and A.~Walther.
\newblock {\em Evaluating derivatives: Principles and techniques of Algorithmic Differentiation}.
\newblock SIAM, 2008.

\bibitem{hipc-2022-nischay}
Nischay~Ram Mamidi, Dhruv Saxena, Kumar Prasun, Anil Nemili, Bharatkumar Sharma, and S.~M. Deshpande.
\newblock Performance analysis of gpu accelerated meshfree q-lskum solvers in fortran, c, python, and julia.
\newblock In {\em 2022 IEEE 29th International Conference on High Performance Computing, Data, and Analytics (HiPC)}, pages 156--165, 2022.

\bibitem{cir-splitting}
R.~Courant, E.~Issacson, and M.~Rees.
\newblock On the solution of nonlinear hyperbolic differential equations by finite differencesew developments in kinetic schemes.
\newblock {\em Comm. Pure Appl. Math.}, 5:243--255, 1952.

\bibitem{kfvs}
J.~C. Mandal and S.~M. Deshpande.
\newblock Kinetic flux vector splitting for {E}uler equations.
\newblock {\em Comp. \& Fluids}, 23(2):447--478, 1994.

\bibitem{q-lskum}
S.~M. Deshpande, K.~Anandhanarayanan, C.~Praveen, and V.~Ramesh.
\newblock Theory and application of {3-D} {LSKUM} based on entropy variables.
\newblock {\em Int. J. Numer. Meth. Fluids}, 40:47--62, 2002.

\bibitem{smd-nasa-report-entropy-variables}
S.~M. Deshpande.
\newblock On the maxwellian distribution, symmetric form, and entropy conservation for the euler equations.
\newblock {\em NASA-TP-2583}, 1986.

\bibitem{tapenade}
L.~Hasco{\"e}t and V.~Pascual.
\newblock The {T}apenade {A}utomatic {D}ifferentiation tool: {P}rinciples, {M}odel, and {S}pecification.
\newblock {\em {ACM} {T}ransactions {O}n {M}athematical {S}oftware}, 39(3), 2013.

\bibitem{nvidia-documentation}
NVIDIA Corporation.
\newblock {D}eveloper {T}ools {D}ocumentation.
\newblock 2021.

\bibitem{cuda}
David Kirk.
\newblock {NVIDIA} {CUDA} software and {GPU} parallel computing architecture.
\newblock volume~7, pages 103--104, 01 2007.

\bibitem{agard-report-AR-211}
H~Viviand.
\newblock Test cases for inviscid flow field methods.
\newblock {\em AGARD AR-211}, 1985.

\bibitem{mda-three-element-airfoil}
Christopher~L Rumsey, Elizabeth~M Lee-Rausch, and Ralph~D Watson.
\newblock Three-dimensional effects in multi-element high lift computations.
\newblock {\em Computers \& Fluids}, 32(5):631–657, June 2003.

\end{thebibliography}
\end{document}